\documentclass[twoside,11pt]{article}

\usepackage[]{jmlr2e}

\usepackage{amsmath}
\usepackage{psfrag,epsf}
\usepackage{enumerate}
\usepackage{url} \usepackage{float}
\usepackage{etoolbox}
\usepackage[table,xcdraw]{xcolor}

\usepackage{hyperref}

\usepackage{enumitem}

\usepackage[boxruled]{algorithm2e}

\usepackage{multirow}
\usepackage{lipsum} 
\usepackage{amsfonts}

\usepackage{hyphenat} \usepackage{mathtools} \usepackage{thmtools}
\renewcommand\thmcontinues[1]{Continued}

\usepackage[utf8]{inputenc}

\usepackage{amsfonts}

\usepackage[tight]{subfigure}
\usepackage{transparent}
\usepackage{multirow}
\usepackage{color}
\usepackage{blkarray}
\usepackage{booktabs}

\hypersetup{hidelinks}
\usepackage{bm}
\usepackage{cleveref}

\newtheorem{defn}{Definition}
\newtheorem{exa}{Example}
\newtheorem{assumption}{Assumption}

\newcommand{\E}{\mathbb{E}}
\renewcommand{\d}{\mathrm{d}}
\renewcommand{\P}{\mathbb{P}}

\newcommand{\mX}{\mathcal{X}}

\newcommand{\wt}[1]{\widetilde{#1}}

\newcommand{\1}{\mathbb{I}}

\def\C {\,|\:}

\def\C {\,|\:}

\def\mF{\mathcal{F}}

\def\X{\bm{X}}

\def\bmu{\bm{\mu}}

\newcommand{\e}{\mathrm{e}}

\newcommand{\norm}[1]{\left\Vert#1\right\Vert}
\newcommand{\abs}[1]{\left\vert#1\right\vert}

\newcommand{\N}{\mathbb{N}}
\newcommand{\M}{\mathbb{M}}
\newcommand{\R}{\mathbb{R}}

\newcommand{\mH}{\mathcal{H}}
\newcommand{\mN}{\mathcal{N}}

\newcommand{\mP}{\mathcal{P}}

\newcommand{\mD}{\mathcal{D}}
\newcommand{\mO}{\mathcal{O}}

\usepackage{lastpage}
\jmlrheading{23}{2022}{1-\pageref{LastPage}}{4/22; Revised
10/22}{10/22}{22-0383}{Yuexi Wang, Tetsuya Kaji and Veronika Ro\v{c}kov\'{a}}

\ShortHeadings{ABC via Classification}{Wang, Kaji and Ro\v{c}kov\'{a}}
\firstpageno{1}

\begin{document}

\title{ Approximate Bayesian Computation via Classification}

\author{\name Yuexi Wang \email yuexi.wang@chicagobooth.edu 
\AND Tetsuya Kaji \email tetsuya.kaji@chicagobooth.edu 
\AND Veronika Rockova \email veronika.rockova@chicagobooth.edu\\
       \addr Booth School of Business\\
       University of  Chicago\\
       Chicago, IL 60637, USA}

\editor{Sayan Mukherjee}

\maketitle

\begin{abstract}%
Approximate Bayesian Computation (ABC) enables statistical inference in simulator-based models whose likelihoods are difficult to calculate but easy to simulate from. ABC constructs  a kernel-type approximation to the posterior distribution through an accept/reject mechanism which compares summary statistics of real and simulated data.  To obviate the need for summary statistics, we directly compare empirical distributions  with a Kullback-Leibler (KL) divergence estimator obtained via contrastive learning. In particular, we blend flexible machine learning classifiers within ABC to automate fake/real data comparisons.
We consider the traditional accept/reject kernel as well as  an exponential weighting scheme which does not require the ABC acceptance threshold.
Our theoretical results show that the rate at which our ABC posterior distributions concentrate  around the true parameter depends on the estimation error of the classifier. We derive  limiting posterior shape results and find that, with a properly scaled exponential kernel, asymptotic normality holds.  We demonstrate
the usefulness of our approach on simulated examples as well as real data in the context of stock volatility estimation.

\end{abstract}


\section{Introduction}

We consider  a collection of  i.i.d.\ observations $\X=(X_1, \ldots, X_n)'$ where each $X_i\in \mX$ is realized from a parametric model $\{P_{\theta_0}: \theta_0\in \Theta\subset\R^d\}$.  We assume that $P_\theta$, for each $\theta\in \Theta$, admits a density $p_\theta$. We are interested in Bayesian inference about $\theta_0$ based on the posterior distribution 
\begin{equation}\label{eq:posterior}
\pi_n(\theta\mid \X)\propto p_\theta^{(n)}(\X)\pi(\theta)
\end{equation}
prescribed by the likelihood $p_\theta^{(n)}(\X)$ and the prior density $\pi(\theta)$.
We are particularly interested in simulator-based models whose the likelihood function cannot be directly expressed/evaluated (such as discretely observed diffusions \citep{sorensen2004parametric} or generative models) but can be sampled from.

Simulator-based models are often called implicit models because the the likelihood function $p_\theta$ cannot be numerically evaluated \citep{diggle1984monte}.
 Fortunately, it may still be possible to simulate synthetic datasets from the model. The ability to simulate from the likelihood has opened up new opportunities for simulating from the posterior. For example, Approximate Bayesian Computation (ABC) \citep{pritchard1999population, beaumont2002approximate} emerged as a default  likelihood-free Bayesian inferential tool. It is an accept/reject posterior sampling mechanism which obviates likelihood evaluations. Each iteration proceeds by (1) simulating prior parameter guesses and fake data from the likelihood, and then (2) accepting  those parameter values whose fake data were close to the observed data. A big challenge with ABC has been gauging the similitude between observed and fake data.

Measures of similarity between data sets have traditionally been based on summary statistics (see \citet{blum2013comparative} for an overview within the ABC context). In other words, two datasets are considered similar if their summary statistics are close.
In the absence of expert knowledge, however, 
constructing effective summary statistics can be challenging  \citep{joyce2008approximately, nunes2010optimal,blum2013comparative} and one may need to resort to  automated strategies.
One possibility is regressing   parameter values onto   (functionals of) fake data in a pilot ABC run to train a flexible mapping which can be substituted for summary statistics \citep{fearnhead2011constructing,jiang2017learning, akesson2020convolutional}. Another possibility, related to indirect inference,  is to construct summary statistics from an auxiliary model \citep{drovandi2011approximate, wood2010statistical}.  One can also choose a subset of candidate summary statistics that satisfy some optimality criterion  \citep{joyce2008approximately,nunes2010optimal} or find an optimal projection of a set of summary statistics onto a lower-dimensional map \citep{fearnhead2011constructing}.
Alternatively, one can directly use a discrepancy between the empirical distributions of the observed and synthetic data sets inside ABC (such as Kullback-Leibler \citep{jiang2018approximate} or Wasserstein \citep{bernton2019approximate} or Maximum Mean (MM) discrepancy \citep{park2016k2}) or Energy Statistics (ES) \citep{nguyen2020approximate}). See \citet{drovandi2021comparison} for a nice comprehensive review of the distribution-style ABC discrepancies. Our work fortifies this ABC point of view by  focusing on the Kullback-Leibler discrepancy estimated via classification.

The KL divergence is one of the most widely used discrepancy metrics. It expresses the average information per observation to discriminate between two probabilistic models \citep{kullback1958information}.
 In large deviations, for example,  it characterizes the exponential decay rate  at which  empirical measures  converge to their probabilities (see Sanov's theorem in \citet{den2008large}) and the rate of decay of the probability  of error in a binary hypothesis testing problem (see Stein's Lemma in \citet{cover1991entropy}). KL also naturally connects to maximum likelihood estimation through its interpretation as the expectation of the log-likelihood ratio. There exist many methods for estimating the KL divergence.
For example,  \citet{wang2009divergence} proposed nearest-neighbor techniques to obtain a mean-square consistent estimator.  \citet{wang2005divergence} proposed a histogram-based KL estimator based on partitioning of the space into statistically equivalent intervals. \citet{silva2007universal} and \citet{silva2010information} went a step further and proposed using data-driven partitions (including multivariate recursive partitioning) and formulated sufficient consistency conditions. Alternatively, \citet{nguyen2007estimating} proposed a variational approach  by turning KL estimation into a penalized convex risk minimization problem. Our work is different from the approaches above as we  adopt a KL estimator based on classification. 
 
We suggest embedding a machine learning classifier inside ABC to determine whether or not  fake and observed data are similar and, thereby, whether or not the underlying parameter value should be kept in the ABC reference table. The fundamental premise of this proposal is as follows: parameter values that yield indistinguishable simulated datasets can be deemed close. Bayesian inference via classification has been suggested before.
\citet{kaji2021mh} developed a version of the Metropolis-Hastings algorithm, called MHC,  based on classification-based estimators of likelihood ratios. \citet{thomas2022likelihood} derived a marginal approach by contrasting two fake datasets generated from the marginal and conditional likelihoods. \citet{gutmann2018likelihood} proposed a   classification strategy related to ours using a different discrepancy metric. Our paper reframes the method of \citet{gutmann2018likelihood} as a genuine ABC algorithm with a KL divergence discriminator and provides supporting theory which justifies its inferential potential.

In particular, we study statistical properties of the approximate posterior which, in part, depend on the properties of the KL divergence estimator. 
 We consider the traditional accept/reject ABC version (with a uniform kernel) as well as an exponential kernel variant which does not require the ABC tolerance threshold. 
Similar to  \citet{frazier2018asymptotic}, we show that the choice of  the ABC acceptance threshold $\epsilon$ plays a critical role in the convergence rate and in the limiting posterior shape. In practice, it is often not obvious what the optimal threshold $\epsilon$  should be. 
 Motivated by the connections with the MHC algorithm of Kaji and Rockova (2021), we propose an exponential kernel which yields ABC posteriors that correspond to the stationary distribution of MHC.  
 Our ABC kernel method can be thus regarded as a parallelizable counterpart to the sequential MHC sampling, targeting the same posterior approximation. 
The concentration and asymptotic shape behavior of the ABC posterior, which can be derived  from \citet{kaji2021mh},  theoretically justify our exponential weighting scheme.
{Finally, our classification-based ABC approach provides a viable computational strategy for obtaining coarsened  posteriors   for   Bayesian robust inference \citep{miller2018robust}. Our ABC approach leverages machine learning but does so in a perhaps more traditional way than  the recent sequential neural likelihood and mixture density network approaches for learning posteriors \citep{papamakarios2016fast, papamakarios2019sequential}.}

The remaining of the paper is structured as follows. In Section 2, we flesh out the basic idea of ABC and introduce our framework with classification. In Section 3, we investigate the posterior concentration and limiting shape behaviors of the ABC posteriors.  Section 4 shows performance  on simulated datasets and Section 5 further highlights the practical value of our approach on real data. In Section 6, we conclude with a discussion.

\smallskip
{\em Notation.} We use the shorthand notation $p_0=p_{\theta_0}$ and $P_0=P_{\theta_0}$.  We employ the operator notation for expectation, e.g., $P_0 f=\int f dP_0$. The $\epsilon$-bracketing number $N_{[]}(\epsilon, \mF, d)$ of a set $\mF$ with respect to a premetric $d$ is the minimal number of $\epsilon$-brackets in $d$ needed to cover $\mF$\footnote{A premetric on $\mF$ is a function $d: \mF\times \mF \to \R$ such that $d(f,f)=0$ and $d(f,g)=d(g,f)\geq 0$.}. The $\delta$-bracketing entropy integral of $\mF$ with respect to $d$ is $J_{[]}(\delta, \mF, d)=\int_0^{\delta} \sqrt{1+\log N_{[]}(\epsilon, \mF,d)} d\epsilon$. Next, $K(f,g)=\int f\log(f/g)d\mu$ denotes the Kullback-Leibler divergence between two density functions and $V_2(f,g)=\int f\abs{\log(f/g)}^2 d\mu$.  For real-valued sequences $\{a_n\}_{n\geq 1}$ and $\{b_n\}_{n\geq 1}$, $a_n\lesssim b_n$ means that $a_n\leq C\,b_n$ for some generic constant $C>0$, $a_n\asymp b_n$ means that $a_n\lesssim b_n\lesssim a_n$, and $a_n\gg b_n$ indicates a greater order of magnitude. For a sequence of random variables $x_n$, $x_n=o_P(a_n)$  if $\lim_{n\to\infty }P(\abs{x_n/a_n}\geq C)=0$ for every $C>0$, and $x_n=O_P(a_n)$ if for every $C>0$ there exists a finite $M>0$ and a finite $N$ such that $P(\abs{x_n/a_n}\geq M)\leq C$ for all $n>N$. All limits are taken as $n\to \infty$. Take $\norm{\cdot}$ to be the Euclidean norm. 

\vspace{-0.2cm}
\section{ABC without Summary Statistics}

\iffalse
Our framework consists of a collection of  i.i.d. observations $\X=(X_1, \ldots, X_n)'$ where each $X_i\in \mX$ is realized from a parametric model $\{P_{\theta_0}: \theta_0\in \Theta\subset\R^d\}$.  We assume that $P_\theta$, for each $\theta\in \Theta$, admits a density $p_\theta$. 
Hereafter, we will use the shorthand notation $p_0=p_{\theta_0}$ and $P_0=P_{\theta_0}$. 
We are interested in Bayesian inference about $\theta_0$ based on the posterior distribution 
\begin{equation}\label{eq:posterior}
\pi_n(\theta\mid \X)\propto p_\theta^{(n)}(\X)\pi(\theta)
\end{equation}
prescribed by the likelihood $p_\theta^{(n)}(\X)$ and the prior density $\pi(\theta)$.
We are particularly interested in scenarios when the likelihood function cannot be directly expressed/evaluated (such as discretely observed diffusions \citep{sorensen2004parametric} or generative models) but can be sampled from.
 \fi
 
The now default ABC method for Bayesian likelihood-free inference  constructs a nested kernel-type approximation to the posterior distribution. The first approximation occurs when the data is distilled into summary statistics to obtain $\pi(\theta\C S_X) \propto \pi(S_X\C\theta)\pi(\theta)$, where $S_X= S(\X)$ is a vector of summary statistics.  The quality of this approximation  depends crucially on the informativeness of $S_X$.  The actual ABC approximation to the posterior \eqref{eq:posterior}
  is then constructed via a kernel function as
$\pi_{ABC}(\theta\C S_X) =\int \pi(\theta,S\C S_X) dS$ with
$\pi(\theta, S)\propto T_\epsilon(\| S-S_Y\|) \pi (S\C\theta)\pi(\theta)$, where $\norm{\cdot}$ is a general norm to be specified later by the user and where $T_\epsilon(\|u\|)=T(\|u\|/\epsilon)$  
is a standard (smoothing) kernel with a scale parameter $\epsilon>0$. 
The key two challenges with ABC  are (1)   deriving low-dimensional summary statistics with a minimal loss of information and  (2)  selecting the kernel  and its the tolerance level $\epsilon$.
To remediate the reliance of ABC on summary statistics, we focus on viewing the observed and fake data as empirical distributions and gauge the discrepancy between them. See \citet{drovandi2021comparison} for an overview of other discrepancy-based ABC methods. Regarding the choice of aggregation kernels, we consider the traditional uniform kernel yielding an accept/reject algorithm and a smoothing kernel free from $\epsilon$-tuning. 

We are  interested in regimes  where ABC is  most effective   \citep{jiang2017learning}, i.e. settings where the sample size $n$ of $\X$  is moderately high and the dimension of $\Theta$ is low to ensure that we can hit  the high ABC-posterior region with a reasonable prior probability.  The number of observations $n$ has an impact on  the effectiveness of the embedded classifier. The nonparametric  neural  network classifiers usually demand  $n$ to be somewhat large so that the estimation errors are manageable. 

\vspace{-0.2cm}

\subsection{ABC with KL Divergence}
Instead of summary statistics, we use the estimated KL divergence inside the ABC algorithm. 
 Our interest in the KL divergence as a discrepancy metric stems partially from the following connection to the generalized Bayesian inference \citep{bissiri2016general}.  The posterior distribution \eqref{eq:posterior} can be rewritten as a generalized posterior 
$ 
\pi_n(\theta\mid  \X) \propto \pi(\theta)\exp( - n\times  KL(p_0^{(n)},p_\theta^{(n)}))
$
where  the parameter $\theta$ is linked to data through  the empirical Kullback-Liebler (KL) divergence $KL(p_0^{(n)},p_\theta^{(n)})\equiv \frac{1}{n}\sum_{i=1}^n \log ({p_0}/{p_\theta})(X_i)$. 
For when the KL divergence cannot be easily evaluated, we consider various estimators in the next section.
 We denote a generic  KL divergence estimator obtained from observed data $\X \sim P_0^{(n)}$ and pseudo-data $\tilde \X^\theta\sim P_\theta^{(n)}$ as  $\hat K(\X, \tilde \X^\theta)$.  
Adopting $\hat K(\cdot, \cdot)$ as the ABC discrepancy, we consider a simple accept/reject ABC mechanism  detailed in Algorithm \ref{alg:abc_plain} below. While \citet{jiang2017learning} used a nearest-neighbor estimator of the KL divergence, we devise a different estimator based on classification in Section \ref{sec:clas_KL}.

 \begin{figure}[!t]
\begin{algorithm}[H] 
{\small
\vspace{0.1cm}
For a pre-determined tolerance level   $\epsilon>0$ repeat for $j=1,\dots, N$:
\begin{enumerate}[noitemsep]
\item Simulate $\theta_j$  from $\pi(\theta)$.
\item Simulate $\tilde \X^{\theta_j} =(\tilde X_1^{\theta_j}, \ldots ,\tilde X_m^{\theta_j})'$ through i.i.d.\ sampling from the model $p_{\theta_j}$.
\item Construct $\hat K(\X, \tilde \X^{\theta_j})$ by training a classifier distinguishing $\X$ and $\tilde \X^{\theta_j}$ as in \eqref{eq:KL_dnn}.
\item Accept $\theta_j$ when $\hat K(\X, \tilde \X^{\theta_j})\leq \epsilon$.
\end{enumerate}}
\caption{KL-ABC with Accept-Reject}\label{alg:abc_plain}
\vspace{-0.07in}
\end{algorithm}
\vspace{-0.2in}
\end{figure}

Algorithm \ref{alg:abc_plain} simulates  pairs of parameter values and pseudo-data $\{\theta, \tilde \X^\theta\}$ from the joint posterior density
\begin{equation}\label{eq:abc_post}
\hat \pi^{AR}\left(\theta, \tilde \X^\theta\mid \hat K(\X, \tilde \X^\theta)\leq \epsilon\right)=\frac{\pi(\theta)p_\theta^{(n)}(\tilde \X^\theta)\1\big(\hat K(\X, \tilde \X^\theta)\leq \epsilon\big)}{\int \pi(\theta)p_\theta^{(n)}(\tilde \X^\theta)\1\big(\hat K(\X, \tilde \X^\theta)\leq \epsilon\big)\d\tilde \X^\theta \d\theta},
\end{equation}
which margins towards the  following approximate  (Accept/Reject) posterior density 
\begin{align}
\hat \pi^{AR}_\epsilon(\theta \mid  \X)&=\int \hat \pi^{AR}\left(\theta, \tilde \X^\theta\mid \hat K(\X, \tilde \X^\theta)\leq \epsilon\right) \d \tilde \X^\theta \nonumber \\
& \equiv\frac{\pi(\theta)P_\theta^{(n)}  \big(\hat K(\X, \tilde \X^\theta)\leq \epsilon\big)}{\int \pi(\theta)P_\theta^{(n)} \big(\hat K(\X, \tilde \X^\theta)\leq \epsilon\big) \d\theta}\label{eq:alg_post1}. 
\end{align}
The inferential potential of the approximation \eqref{eq:alg_post1}  will be scrutinized theoretically later in \Cref{sec:post_rate}. 
In particular, we will later see  that the convergence rate of  \eqref{eq:alg_post1} around $\theta_0$ depends on the choice of $\epsilon$   \citep{frazier2018asymptotic} as well as the quality of the discriminator.
{It is interesting to note that the ABC posterior \eqref{eq:alg_post1} is mathematically equivalent to the  $c$-posterior proposed by \cite{miller2018robust}  for robust inference in mis-specified (tractable) models.
The computation of the $c$-posteriors has relied on powered-likelihood approximations and MCMC sampling. While we instead view \eqref{eq:alg_post1} as an approximate posterior in models with intractable likelihoods,  our ABC algorithm can be nevertheless used to compute  coarsened posteriors in broader scenarios when MCMC sampling may not be available (see Remark \ref{rem:robust} below).}
Algorithm \ref{alg:abc_plain} uses the uniform kernel which corresponds to the indicator function $T_\epsilon(\|u\|)=\mathbb I(\|u\|\leq \epsilon)$. 
In practice, it is difficult to balance out conflicting demands of smaller  $\epsilon$ (yielding good approximability) and larger acceptance rates (yielding  more posterior samples).
 As a remedy,  we propose a way to aggregate the ABC samples through a scaled exponential kernel
motivated by the connection between KL and the log-likelihood ratio. This ABC variant requires no ad-hoc thresholding and is summarized in Algorithm \ref{alg:abc_exp}.

\begin{figure}[!b]

\begin{algorithm}[H]
{\small
\vspace{0.1cm}
Repeat for $j=1,\dots, N$:
\begin{enumerate}[noitemsep]
\item Simulate $\theta_j$  from $\pi(\theta)$.
\item Simulate $\tilde \X^{\theta_j} =(\tilde X_1^{\theta_j}, \ldots ,\tilde X_m^{\theta_j})'$ through i.i.d.\ sampling from the model $p_{\theta_j}$.
\item Construct $\hat K(\X, \tilde \X^{\theta_j})$ by training a classifier distinguishing $\X$ and $\tilde \X^{\theta_j}$ as in \eqref{eq:KL_dnn}.
\item Assign $\theta_j$  a weight proportional to $\exp\big(-n \hat K(\X, \tilde \X^\theta)\big)$.
\end{enumerate}
\caption{KL-ABC with Exponential Weighting}\label{alg:abc_exp}
\vspace{-0.07in}}
\end{algorithm}
\vspace{-0.15in}
\end{figure}

 Algorithm \ref{alg:abc_exp} generates draws for the pair $\{\theta, \tilde \X^\theta\}$ from a joint posterior density
\begin{equation}
\hat \pi^{EK}(\theta, \tilde \X^\theta\mid \X)=\frac{\pi(\theta)p_\theta^{(n)}(\tilde \X^\theta) \exp\big(-n\hat K(\X, \tilde \X^\theta)\big)}{\int \pi(\theta)p_\theta^{(n)}(\tilde \X^\theta)\exp\big(-n\hat K(\X, \tilde \X^\theta)\big)\d\tilde \X^\theta \d\theta},
\end{equation}
which leads to the approximated Bayesian posterior as
\begin{align}
\hat \pi^{EK}(\theta \mid \X)& = \int  \hat \pi^{EK}(\theta, \tilde \X^\theta\mid \X) \d\tilde\X^\theta=\frac{\pi(\theta)P_\theta^{(n)}\exp\big(-n\hat K(\X, \tilde \X^\theta)\big)}{\int \pi(\theta)P_\theta^{(n)} \exp\big(-n\hat K(\X, \tilde \X^\theta)\big)\d\theta}. \label{eq:alg_post2}
\end{align}

\begin{remark}[Generating Fake Data]\label{rem:generate}
We assume $\tilde \X^\theta=g_\theta(\tilde\X)$, where $\tilde \X\in\R^m$ are random variables arriving from $\tilde P^{(m)}$   and where  $g_\theta:\R^m\rightarrow\R^m$ is a deterministic mapping.
Generating random variable draws by passing $\tilde \X$ through some mapping is commonly done in practice, {also known as the reparameterization  trick \citep{kingma2013auto}}. For example, Gaussian random variables $\tilde\X=\{\tilde X^\theta_i\}_{i=1}^m$ that follow i.i.d. $N(\mu, \sigma^2)$ distribution can be obtained by transforming  $\{\tilde X_i\}_{i=1}^m\overset{i.i.d.}{\sim} N(0,1)$ via $\tilde  X^\theta_i = \mu+\sigma \tilde X_i$. In other cases, one can use uniform draws $\tilde \X$ and  the inverse transform sampling.  
 \end{remark}

Smooth kernels have been used inside ABC before to rescale the acceptance probability (e.g. \citet{beaumont2002approximate} employed the Epanechnikov kernel).  \citet{wilkinson2013approximate} and \citet{sisson2018overview} provide a thorough overview and comparisons of the commonly used kernels. Our  smoothed weights are directly interpretable due to their linkage between the KL divergence and the log-likelihood ratio. Algorithm  \ref{alg:abc_exp} can be regarded as a version of Importance Sampling ABC (see \citet{nguyen2020approximate} for a variant using energy statistics and \citet{park2016k2} for minimal description length ABC).
We later show in \Cref{sec:post_rate} that, with the scaled exponential kernel, the ABC posterior corresponds to the stationary distribution of the MHC algorithm of \citet{kaji2021mh} and can be regarded as a posterior under a misspecified model.
Computational comparisons of the sequential MHC sampler with our  parallelizable ABC sampler are performed in the Appendix Section \ref{sec:mhc_compare} where we show benefits of the ABC strategy when convergence issues may arise for MHC due to initialization. 
In Section \ref{sec:model_misspec}, we perform comparisons of the accept/reject (AR) and exponential kernels under model misspecification  where we show that the AR kernel is far more robust.

 The classifier needs to be trained for each ABC draw, which may incur additional computational cost compared to traditional ABC where the summary statistics and their distance can be computed without optimization. We provide  comparisons of computation times in Appendix (Section F). Although the computation costs of our methods are higher when the data dimensionality $d$ is relatively small, we are less disadvantageous when $d$ is large compared to other discrepancies like Wasserstein distance or Maximum Mean Discrepancy. In addition, nonparametric discriminator classes such as neural network classifiers can efficiently benefit when there is an inherent low-dimensional structure in the data \citep{kaji2020adversarial}. Additionally, training can be accelerated if one initiates the training at some pre-trained neural networks.

{\subsection{Connection to Robust Bayesian  Inference}\label{rem:robust}
Our algorithms can be regarded as ``robust ABC" algorithms that estimate (relative-entropy coarsened posteriors) $c$-posteriors introduced in  \citet{miller2018robust}.  {Note that our focus of robustness is different from the robustness to different distance measures, but more relevant to data perturbation or model misspecification (discussed later in \Cref{sec:model_misspec}).}
The $c$-posteriors yield robust inference by conditioning on the event that the observed data $\X$ is sufficiently close (in terms of the KL divergence)  to the  data generated by the model. {A similar case as the Huber-type data contamination is considered in $\gamma$-divergence ABC \citet{fujisawa2021gamma}.}
 The proposed computation of $c$-posteriors in \citet{miller2018robust} is made feasible only through asymptotic approximations (Section 3.1 in \citet{miller2018robust}), e.g. with powered posteriors that are computable using conjugate priors.
 Our ABC methods can compute them without any approximation and for a broader class of priors.
In particular, if in Algorithm \ref{alg:abc_plain} we draw $\epsilon\sim \text{Exp}(\alpha)$ for some $\alpha>0$ and accept $\theta$ if 
 $ \hat  K(\X, \tilde \X^\theta)<\epsilon$,  then our ABC posterior coincidentally approximates the  relative-entropy $c-$posterior  proportional to $\pi(\theta)P_\theta^{(n)}\e^{-\eta K_n}$ where $K_n=\P_n\log\frac{p_0}{p_\theta}$. 
Algorithm \ref{alg:abc_exp} corresponds to the case $\alpha=n$ in \citet{miller2018robust}   without any approximation.  Interestingly, the degree of robustness corresponds to the acceptance rate of ABC. For example, if we let $\alpha\ll n$, the $c$-posterior puts larger weight on the prior and robustifies the model, which corresponds to accepting many draws (more than proportional to $n$) and the draws reflecting the shape of the prior. On the contrary, if we let $\alpha\gg n$, the $c$-posterior puts most weight on a narrow neighborhood of the observed data, which corresponds to accepting very few draws for which the Kullback--Leibler divergence is the smallest. From an ABC's perspective, probably the most interesting case is when the acceptance rate is roughly fixed throughout $n\to\infty$, in which case $\alpha$ is comparable with $n$ and our algorithms produce a correct $c$-posterior without utilizing the approximation in \citet{miller2018robust} which stands on $n\gg\alpha$ or $n\ll\alpha$.
In addition, another advantage of our method is that it allows us to calculate the $c$-posterior for different $\alpha$ easily. If we use an MCMC with tempering (i.e. the powered likelihood approximation), we might need to run separate MCMC chains for different $\alpha$. On the other hand, our ABC-based algorithm lets us calculate the $c$-posteriors by filtering out independent samples of candidate draws according to various $\alpha$. This may be advantageous in applications when no ex-ante preference is available on the degree of robustness and when one wants to see how the concentration of the $c$-posterior varies with it.

}

\subsection{Estimating KL Divergence via Classification}\label{sec:clas_KL}
We adopt the `$-\log D$' trick to estimate the KL divergence \citep{goodfellow2014generative}.
 More precisely, a flexible discriminator $D$ (such as a neural network or logistic regression) is trained to maximize 
\begin{equation}\label{eq:bce_obj}
\M_{n,m}^\theta(D)= \P_n \log D+\P_m^\theta \log (1-D),
\end{equation}
where we employ the operator notation for expectation, e.g., $\P_n f=\frac{1}{n}\sum_{i=1}^n f(X_i)$ and $\P_m^\theta f=\frac{1}{m}\sum_{i=1}^m f(\tilde X_i^\theta)$. This can also be regarded as a classification problem where we label $\{X_i\}_{i=1}^n$ (`real' data) with 1 and $\{\tilde X_i^\theta\}_{i=1}^m$ (`fake' data) with 0. The oracle maximizer to \eqref{eq:bce_obj} can be shown to be \citep[Proposition 1]{goodfellow2014generative}\begin{equation}\label{eq:oracle}
D_\theta(X)=\frac{p_0(X)}{p_0(X)+p_\theta(X)}.
\end{equation}
The functional form of the oracle solution in \eqref{eq:oracle} naturally suggests the following KL  estimator obtained  from a trained discriminator $\hat D^\theta_{n,m}$  \citep{thomas2020generalised}
\begin{equation}\label{eq:KL_dnn}
\hat K(\X, \tilde \X^\theta)=\P_n\log\frac{\hat D^\theta_{n,m}}{1-\hat D^\theta_{n,m}}=\frac{1}{n}\sum_{i=1}^n\log\frac{\hat D^\theta_{n,m}(X_i)}{1-\hat D^\theta_{n,m}(X_i)}.
\end{equation}

Later we show that our classification-based KL estimator \eqref{eq:KL_dnn} converges to a well-defined limit counterpart $K(p_0,p_\theta)$ under mild conditions in \Cref{sec:estimation_errors}.

\subsection{Other  KL Estimators} 
Our ABC results, presented later in Section \ref{sec:freq_ABC},  can be extended to other types of KL estimators if similar estimation error results as in \Cref{thm:est_error_conc} can be shown. {Since the rate  $1/(nu^2)$ stems from the estimation error of  the empirical KL divergence, the fundamental difference between our classification-based  KL estimator  and other KL  estimators lies in the rate $\delta_n$.} One example is the k-Nearest Neighbor (kNN) estimator  proposed in \citet{perez2008kullback}.  \citet{wang2009divergence} showed that this estimator is asymptotically unbiased and mean-square consistent and they propose a data-dependent choice of $k$ which can improve the convergence speed. \citet{jiang2018approximate} assess   data discrepancy inside ABC with the special case of 1-nearest neighbor, which is defined as
\begin{equation}\label{eq:KL_knn}
\hat K(\X, \tilde \X^\theta)=\frac{d}{n} \sum_{i=1}^n \log \frac{\min_j \|X_i-\tilde X^\theta_j\|}{\min_{j \neq i} \norm{X_i-X_j}}+\log \frac{m}{n-1}.
\end{equation}
where $d$ is the number of covariates in $\X$. \citet{zhao2020analysis} provide  convergence rate of the  bias for this kNN estimator is bounded by $n^{-2\gamma/(d+2)}\log n$, where $\gamma$ is a parameter characterizing the tail behavior of the target distribution.  {Note that  the kNN  estimator is  not applicable to cases where $\X$ arises from a discrete distribution.}

Another route to estimate the KL divergence is via  (data-dependent) partitioning methods. \citet{wang2005divergence} proposed to estimate the Radon-Nikodym derivative  $\d P_0/ \d P_\theta$ using frequency counts on a statistically equivalent partition of $\R^d$. However, the computational complexity of their method is exponential in $d$ and the estimation accuracy deteriorates quickly as the dimension increases. \citet{silva2010information} further contributed to multivariate data-driven partition schemes by using a Barron-type histogram-based density estimate. They provide sufficient conditions on the partitions scheme to make the estimator strongly consistent. 

Lastly, \citet{nguyen2007estimating} adopted a variational approach to estimate KL by reframing the estimation problem as a penalized convex risk minimization problem, where they restrict the estimate to a bounded subset of a Reproducing Kernel Hilbert Space (RKHS). Convergence rates are then  obtained from empirical process theory on nonparametric M-estimators \citep{geer2000empirical}. In an independent contribution, \citet{ghimire2020reliable} used a discriminator in RKHS to estimate KL using  a similar  approach to ours. They showed that the estimator error bound  is related to the complexity of the discriminator in RKHS.  {We compare computational complexities of these methods and our approach in the Appendix Section \ref{sec:complex}.}

Beyond the forward KL divergence, our classification framework allows us to consider other discrepancy metrics.  Alternatively to \eqref{eq:KL_dnn}, we  could instead estimate the reversed KL divergence  
\[
\hat K_{\text{reverse}}( \tilde \X^\theta,\X )= \frac{1}{m}\sum_{i=1}^m \log \frac{1-\hat D^\theta_{n,m}}{\hat D^\theta_{n,m}}(\tilde X_i^\theta)
\]
which converges to  $K(p_\theta, p_0)$ and which is still uniquely minimized at $p_\theta=p_0$. One can show that the estimation error of this reversed KL estimator is still $O_{P^*}(\delta_n)$ by following the same techniques used in \Cref{thm:conv_moments}. The reversed KL divergence is widely used in variational inference  \citep{jordan1999introduction, wainwright2008graphical}. Forward and reversed KL's could perform differently when the function class inside the variational approach is not rich enough. The reversed KL is zero-forcing/mode-seeking, while the forward KL is mass-covering/mean-seeking \citep{bishop2006pattern}. 
Another related metric, deployed by \citet{gutmann2018likelihood}, is the classification accuracy (CA) defined as
\begin{equation}\label{eq:ca}
\text{CA}(\X, \tilde \X^\theta)=\frac{1}{n+m} \Biggl(\sum_{i=1}^n \hat D^\theta_{n,m}(X_i)+\sum_{i=1}^m \bigl(1-\hat D^\theta_{n,m}(\tilde X^\theta_i) \bigr)\Biggr).
\end{equation}
{Since $\hat D(\cdot)$ and $\log \frac{\hat D}{1-\hat D}(\cdot)$ are linked by a logistic transformation which is Lipschitz-continous, CA can be roughly regarded as a weighted average of the forward KL divergence and the reversed KL divergence.} Our framework is connected to other GAN-style discrepancy metrics which are also used in ABC  literature. We provide a discussion in \Cref{sec:other_gans}.

\section{Frequentist Analysis of the ABC Posterior}\label{sec:freq_ABC}
One way to assess the quality of the posterior distribution is through the speed at which it contracts around the truth $\theta_0$ as $n\rightarrow\infty$. While the ABC posterior is ultimately an approximation, it might still  concentrate about $\theta_0$ at a reasonable rate.  In this section, we look into  theoretical properties of both Algorithm 1 and Algorithm 2. First, we develop a tail bound result quantifying how fast the classification-based estimator $\hat K(\X, \tilde\X^{\theta})$  converges to the true KL divergence $K(p_0, p_\theta)$ conditional on approximability of the discriminator class. The tail bound analysis is crucial in our convergence analysis.  Next, we show that the convergence rate of the accept-reject ABC in Algorithm 1 is determined jointly by the accept-reject threshold $\epsilon_n$, the estimation error  $\delta_n$ (with respect to  $\P_n \log p_0/p_\theta$), and the rate  $n^{-1/2}$ of estimation between $\P_n \log p_0/p_\theta$ and  $K(p_0, p_\theta)$. Further,  the typical posterior distribution converges to a uniform distribution over an ellipse when the acceptance threshold $\epsilon_n$ dominates the other two. On the other hand, the exponentially weighted posterior in Algorithm 2 can be viewed as the posterior under a ``misspecified'' model. The convergence rate is then determined by the contraction rate of the true posterior and the estimation error, where the ABC posterior is asymptotically normal around the KL projection  under  LAN conditions \citep{kaji2021mh}.

\subsection{Convergence Rate of Estimation Errors}\label{sec:estimation_errors}
We assume that the set of considered classifiers $\mD$ resides in a sieve $\mD_n$ that expands with the sample size and its size is measured by bracketing entropy $N_{[]}(\epsilon, \mF, d)$. \citet{kaji2020adversarial} prove the rate of convergence of such a classifier (under assumptions reviewed later) with a Hellinger-type distance defined as
\[
d_\theta(D_1, D_2)=\sqrt{h_\theta(D_1, D_2)^2+ h_\theta(1-D, 1-D_\theta)^2},
\]
where $h_\theta(D_1, D_2)=\sqrt{(P_0+P_\theta)(\sqrt D_1-\sqrt D_2)^2}$.

\begin{assumption} \label{ass:entropy}
Assume that $n/m$ converges and that an estimator $\hat D^\theta_{n,m}$ exists that satisfies $\M_{n,m}(\hat D^\theta_{n,m})\geq  \M_{n,m}^\theta(D_\theta)-O_P(\delta_n^2)$ for a nonnegative sequence $\delta_n$.  Moreover, assume that the bracketing entropy integral satisfies $J_{[]}(\delta_n, \mD^\theta_{n,\delta_n},d_\theta)\lesssim \delta_n^2\sqrt n$ and that there exists $\alpha<2$ such that $J_{[]}(\delta, \mD^\theta_{n,{\color{blue}\delta}},d_\theta)/\delta^\alpha$ has a majorant decreasing in $\delta$. Here $\mD^\theta_{n,\delta_n}= \{D\in \mD_n: d_\theta(D, D_\theta)\leq \delta_n\}$.
\end{assumption}

\Cref{ass:entropy} requires three conditions. First,  the synthetic data sample size $m$ should be at least as large as the actual data size $n$, which can be assured. Second, the discriminator class  needs to be expressive enough so we can find a  sufficiently good maximizer approximating the oracle discriminator $D_\theta$. Lastly, the entropy of the sieve should be moderate to prevent overfitting.

The following theorem states that the sequence $\delta_n$ in \Cref{ass:entropy} determines the convergence rate of $\hat{D}$.
The speed at which $\delta_n$ converges to $0$ depends on the choice of the sieve and smoothness of the model.
When a nonparametric estimator is employed, $\delta_n$ is often slower than $n^{-1/2}$.
In \Cref{sec:nn}, we give a specific expression of $\delta_n$ for a neural network classifier and give a few examples in which $\delta_n$ vanishes faster than $n^{-1/4}$.

\begin{lemma}[{\citealp[Theorem S.1]{kaji2020adversarial}}]\label{lem:delta_n} Under Assumption \ref{ass:entropy}, we have $d_\theta(\hat D_{n,m}^\theta, D_\theta)=O_P^*(\delta_n)$.\footnote{We use $P^*$ to denote outer expectation (see Section 1.2 of \citet{van1996weak}), here is the expectation of a ``a smallest measureable function g that dominates $d_\theta(\hat D_{n,m}^\theta, D_\theta)$".}
\end{lemma}

To establish the rate of convergence of our approximated posterior, we have the following assumption.

\begin{assumption}\label{ass:ratio_bd}
There exists $\Lambda>0$ such that for every $\theta\in \Theta$, $P_0(p_0/p_\theta)$ and $P_0(p_0/p_\theta)^2$ are bounded by $\Lambda$ and
\[
\sup_{D\in \mD^\theta_{n,\delta_n} }P_0\left(\frac{D_\theta}{D}\middle\vert \frac{D_\theta}{D} \geq \frac{25}{16}\right)<\Lambda, \quad \sup_{D\in \mD^\theta_{n,\delta_n} }P_0\left(\frac{1-D_\theta}{1-D}\middle\vert  \frac{1-D_\theta}{1-D} \geq \frac{25}{16}\right)<\Lambda, 
\]
for $\delta_n$ in Assumption \ref{ass:entropy}. The brackets in Assumption \ref{ass:entropy} can be taken so that $P_0(\sqrt{u/l}-1)^2=O(d_\theta(u,l)^2)$ and $P_0(\sqrt{(1-l)/(1-u)}-1)^2=o(d_\theta(u,l))$.
\end{assumption}

\Cref{ass:ratio_bd} constrains the tail behavior of the discriminator so that the residual of the cross-entropy loss in \eqref{eq:bce_obj} can be circumscribed by the bracketing entropy. See \citet{kaji2021mh} for a discussion of how this can be reasonably satisfied for logistic discriminator and neural network discriminators that use sigmoid activation functions.

The following theorem quantifies the rate of convergence of our estimator \eqref{eq:KL_dnn} towards the {\em empirical} KL divergence $\P_n\log \frac{p_0}{p_\theta}$.

\begin{lemma}[Convergence Rate of Estimation Errors]\label{thm:conv_moments} 
 Under Assumptions \ref{ass:entropy} and \ref{ass:ratio_bd}, 
\begin{align}
&\abs{\hat K(\X, \tilde \X^\theta)-\P_n\log \frac{p_0}{p_\theta}}=O_{P^*}(\delta_n).\label{eq:approx_error}
\end{align}
\end{lemma}
\textbf{Proof. }
Since the estimation error can be rewritten as
\begin{align*}
\hat K(\X, \tilde \X^\theta)-\P_n \log \frac{p_0}{p_\theta} = - \P_n\biggl(\log \frac{1-\hat D^\theta_{n,m}}{1-D_\theta}- \log\frac{\hat D^\theta_{n,m}}{D_\theta}\biggr),
\end{align*}
it follows from \citet[Theorem 4.1]{kaji2021mh}.

\vspace{-0.05in}
\begin{remark}[Uniform Convergence Rate]\label{rmk:worst_rate} 
The rate of estimation in \Cref{lem:delta_n} and \Cref{thm:conv_moments}  is characterized as  point-wise. While for each $\theta \in \Theta$ the estimation error is shrinking at the rate $\delta_n$, the multiplication constant in front the rate could potentially depend on $\theta$. Assuming a compact parameter support (which is not a-typical in deep learning models; see e.g. \citet{schmidt2020nonparametric} or \citet{polson2018posterior} and \citet{wang2020uncertainty}) and continuity of the multiplication constant, we can essentially regard the rate as uniform.  Hereafter, we thereby abuse the notation and tacitly assume that $\delta_n$ is the worst rate over all $\theta \in \Theta\subset\R^d$, i.e. the rate with the largest multiplication constant. 
\end{remark}

Next, we  investigate convergence around the {\em actual} KL divergence $K(p_0, p_\theta)$. The next lemma will be utilized later in the proof of  \Cref{thm:post_rate}. However, it is of independent interest as it shows how
the speed at which the joint error probability (accounting   for randomness of  both the observed and fake data  $(\X,\tilde\X)$)  decays in terms of the estimation error. Below, the probability $P$ corresponds to $P_0^{(n)}\otimes \tilde P^{(m)}$, where $\tilde P^{(m)}$ is the measure for $\tilde\X$ (see Remark \ref{rem:generate}).

\begin{theorem}\label{thm:est_error_conc}
 For a given $\theta\in\Theta$, we define for $u>0$ and  $\delta_n>0$ as in Lemma \ref{lem:delta_n} and for an arbitrarily slowly increasing sequence $C_n>0$
\begin{equation}\label{eq:est_error_bigO}
\rho_{n,\theta}(u; C_n;\delta_n)\equiv P\biggl(\abs{\hat K(\X, \tilde \X^\theta)-K(p_0, p_\theta)} >2u, d_\theta(\hat D^\theta_{n,m}, D^\theta)\leq C_n\delta_n \biggr).
\end{equation}
Under Assumptions \ref{ass:entropy} and \ref{ass:ratio_bd}, we then have
$$
\rho_n(u;C_n;\delta_n)\equiv\sup_{\theta\in \Theta}\rho_{n,\theta}(u;C_n;\delta_n)=O\Big(\frac{C_n\delta_n}{u} +\frac{1}{nu^2}\Big).
$$
\end{theorem}
The proof can be found in \Cref{pf:est_error_conc}. {Note that the intersecting event above has probability converging to one, according to  Lemma \ref{lem:delta_n}.}

\begin{remark}[Neural Network Sieve]\label{rmk:nn_sieve}
The results discussed in this section apply to any nonparametric sieve discriminator that satisfies the entropy conditions. To facilitate  understanding of how the rate of convergence $\delta$ can be affected by the dimension of the data space $d$, the smoothness of the density, and the choice of the classifier, we provide a discussion on neural network sieve specifically here.  

Borrowing from the idea of \citet{bauer2019deep}, the convergence rate of the appropriately configured neural network discriminator depends only on the low underlying dimension of the oracle discriminator, however large the ostensible dimension of the data is. Intuitively, the low-dimensional structure can be described as follows. The log-likelihood ratio $\log(p_0/p_\theta)$ takes a $d$-dimensional input $X$ as an argument, which may be large. If this function admits a representation as a nested composition of smooth functions, each of which takes a possibly smaller number $d^\ast$ of arguments, the neural network sieve can adapt to this underlying structure and converges faster than the traditionally proven rate. 

In particular, if $d^*<2p$, we have $\delta_n=o_P(n^{-1/4})$, which is often the desired rate for the nonparametric estimator of a nuisance parameter. Additionally we show that one can obtain $\delta_n \lesssim n^{-2/5}$ for the binary choice model with logistic errors and $\delta_n$ arbitrarily close to $n^{-1/2}$ for the discretely sampled Brownian motion model. The  detailed characterizations of the low underlying dimension $d^\ast$ and two examples can be found in \Cref{sec:nn}. 

Finally, with \Cref{ass:ratio_bd}, the convergence rate $\delta_n$ translates into the convergence rate of the Kullback-Leibler estimator in \Cref{thm:conv_moments}. Thus, in smooth low\hyp{}dimensional hierarchical models, our Kullback-Leibler estimator converges reasonably fast even when the nominal dimension of the data is large.

\end{remark}

\subsection{Posterior Concentration Rate}\label{sec:post_rate}

Concentration rates are typically quantified in terms of a prior concentration (measured in terms of a combination of the KL divergence and the KL variation) and the entropy of the model. We have a similar prior mass condition (see (8.4) in  Section 8.2 in \citet{ghosal2017fundamentals}). We denote the KL-neighborhood of $p_0$ by
\begin{equation}\label{eq:KL_ball}
B_2(p_0, \epsilon)=\{{\theta}:  K(p_0,p_\theta)<\epsilon^2\}.
\end{equation}

\begin{assumption}[Prior Mass]\label{ass:prior_mass} 
There exist some constants $\kappa>0$ and $\xi>0$ such that for every $0<\epsilon<\xi$ and some constant $C>0$, the prior probability satisfies $\Pi[B_2(p_0, \epsilon)]\geq  C\epsilon^\kappa$.
\end{assumption}

Next, we assume that the parameter $\theta$ is identifiable in the sense that the KL divergence is locally compatible with the Euclidean norm. This assumption is adopted from Assumption 3(ii) of \citet{frazier2018asymptotic}.

\begin{assumption}[Identification]\label{ass:identification} 
The density function $p_\theta$ is continuous in $\theta$ and for every $\theta$ in some open neighborhood of $\theta_0$ satisfies
\[
\norm{\theta-\theta_0}\leq L\times K(p_0, p_\theta)^\alpha
\]
 for some $L>0$ and $\alpha>0$.
\end{assumption}

Similarly as in (5.1) in \citet{kleijn2006misspecification},
\Cref{ass:identification} ensures posterior concentration around $\theta_0$ when $K(p_0, p_\theta)\to 0$.    This holds for many distributions. For example, for the exponential distribution with a rate parameter $\theta$, we have $K(p_0, p_\theta)=\frac{\theta}{\theta_0}-\log(\frac{\theta}{\theta_0})-1$. Since $\log(1+x)=x-\frac{x^2}{2}+o(x^2)$ when $x\to 0$, we have 
$K(p_0, p_\theta)\geq \frac{1}{2}\theta_0^{-2}(\theta-\theta_0)^2$. For multivariate normal distribution with a known variance $\Sigma$ and an unknown location $\mu$, we have
$K(p_0, p_\theta)=\frac{1}{2}(\mu-\mu_0)\Sigma^{-1}(\mu-\mu_0)\geq \frac{1}{2}\rho(\Sigma)^{-1}\norm{\mu-\mu_0}^2$, where $\rho(\Sigma)$ is the spectral radius, i.e., the largest eigenvalue, of a matrix $\Sigma$.

First, we focus on the uniform kernel $T_{\epsilon}(x)=\1(\abs{x}\leq \epsilon)$ used in Algorithm \ref{alg:abc_plain}. 
Recall (from Remark \ref{rem:generate}) that $\tilde \X^\theta=g_\theta(\tilde \X)$ for some suitable mapping $g_\theta(\tilde \X)$ where $\tilde\X\sim\tilde P^{(m)}$.
The ABC joint posterior \eqref{eq:abc_post} is a weighted aggregation of uniform kernels, i.e
\begin{equation}
\hat \pi^{AR}\left(\theta,\tilde\X\mid \hat K[\X,  g_\theta(\tilde \X)]\leq \epsilon\right)=\frac{\tilde\pi(\tilde\X)\pi(\theta)\1\big(\hat K[\X,  g_\theta(\tilde \X)]\leq \epsilon\big)}{\int \tilde\pi(\tilde\X) \pi(\theta)\1\big(\hat K[\X,  g_\theta(\tilde \X)]\leq \epsilon\big)\d \tilde\X \d\theta},
\end{equation}
which yields  the following ABC posterior distribution  
\begin{align}\label{eq:ABC_accept_post}
\hat \Pi^{AR}_{\epsilon_n}(A\mid  \X) &=\frac{ \int_{\theta\in A} \pi(\theta)\tilde P^{(m)} \big( \hat K(\X, \tilde \X^\theta)\leq \epsilon_n\big)\d \theta}{\int_{\Theta}  \pi(\theta)\tilde P^{(m)}\big( \hat K(\X, \tilde \X^\theta)\leq \epsilon_n\big)\d \theta}
\quad \text{for a Borel-measurable $A\subset  \Theta$}.
\end{align}

\iffalse
margins towards the  following approximate  (Accept/Reject) posterior density 
\begin{align}
\hat \pi^{AR}_\epsilon(\theta \mid  \X)&=\int \hat \pi^{AR}\left(\theta, \tilde \X \mid \hat K[\X,  g_\theta(\tilde \X)]\leq \epsilon\right) d \tilde \X  \nonumber \\
& =\frac{\pi(\theta) \tilde P^{(m)}\big(\hat K[\X, g_\theta(\tilde \X)]\leq \epsilon\big)}{\int \pi(\theta)\tilde P^{(m)} \big(\hat K[\X, g_\theta(\tilde \X)]\leq \epsilon\big) d\theta}\label{eq:alg_post1}. 
\end{align}
In the fixed design, we would have, conditionally on a given $\tilde\X$ and $\X$,
\begin{equation}
\hat \pi^{AR}_\epsilon\left(\theta\mid \X,\tilde \X \right)=\frac{\pi(\theta)\1\big(\hat K(\X, \tilde \X^\theta)\leq \epsilon\big)}{\int \pi(\theta)\1\big(\hat K(\X, \tilde \X^\theta)\leq \epsilon\big)  d\theta}.
\end{equation}

The corresponding ABC posterior distribution is defined through the following expression implied by \eqref{eq:alg_post1}
\begin{align}
\hat \Pi^{AR}_{\epsilon_n}(A\mid  \X) &=\frac{ \int_{\theta\in A} \pi(\theta)P_\theta^{(n)} \big( \hat K(\X, \tilde \X^\theta)\leq \epsilon_n\big)\d \theta}{\int_{\Theta}  \pi(\theta)P_\theta^{(n)}\big( \hat K(\X, \tilde \X^\theta)\leq \epsilon_n\big)\d \theta}
\quad \text{for a Borel-measurable $A\subset  \Theta$}.
\end{align}
\fi

The following theorem (a modification of Theorem 1 in \citet{frazier2018asymptotic}) quantifies the concentration rate in terms of the tolerance threshold $\epsilon_n$ as well as the rate at which the classification-based KL estimator can estimate $\P_n\log(p_0/p_\theta)$ (as formulated in \eqref{eq:approx_error}).

\begin{theorem} \label{thm:post_rate}
Let Assumptions \ref{ass:entropy}, \ref{ass:ratio_bd} and \ref{ass:prior_mass} hold and take $\delta_n$ as in \eqref{eq:approx_error} in \Cref{thm:conv_moments}. Then, as $n\to \infty$ and with $\epsilon_n=o(1)$ such that $n\epsilon_n^2\rightarrow\infty$ and $C_n\delta_n=o(\epsilon_n)$ for some arbitrarily slowly increasing sequence $C_n>0$ we have
\begin{equation}\label{eq:kl_rate}
P_0^{(n)}\Pi[K(p_0,p_\theta)>\lambda_n \mid\hat K(\X, \tilde \X^\theta) \leq \epsilon_n]=o(1), 
\end{equation}
where $\lambda_n =\epsilon_n+M_nC_n \delta_n\epsilon_n^{-\kappa} +\sqrt{M_n} n^{-1/2}\epsilon_n^{-\kappa/2}$ for some arbitrarily slowly increasing sequence $M_n>0$. Moreover, if Assumption \ref{ass:identification} also holds, as $n\to \infty$, we have
\begin{equation}\label{eq:theta_abc_rate}
 P_0^{(n)}\Pi[\norm{\theta-\theta_0}>L\lambda_n^\alpha \mid\hat K(\X, \tilde \X^\theta)\leq \epsilon_n]= o(1). 
\end{equation}
\end{theorem} 
The proof of the theorem is provided in \Cref{pf:post_rate}. Thus, the convergence rate of our ABC posterior depends on three components: the accept-reject threshold $\epsilon_n$, the estimation error of the KL estimator $\delta_n$ and the rate of discrepancy  $n^{-1/2}$ between the empirical and true KL divergence. Since $\delta_n$ will typically be greater than the parametric rate $n^{-1/2}$, the overall convergence rate is then driven by $\lambda_n =\epsilon_n+ \wt M_n \delta_n\epsilon_n^{-\kappa}$, where $\wt M_n$
is an arbitrarily slowly increasing sequence.

\smallskip

In practice, it is unclear how to properly choose  $\epsilon_n$. In Algorithm \ref{alg:abc_exp}, we proposed to weight the draws using a scaled exponential kernel $\exp(-n\hat K(\X, \tilde \X^\theta) )$. We denote the ABC posterior under the exponential kernel as $\hat \Pi^{EK}(\cdot \mid \X)$ where 
\begin{align}
\hat \Pi^{EK}(A\mid \X) &=\frac{\int_A \tilde P^{(m)} \exp\big(-n\hat K(\X, \tilde \X^\theta)\big)\pi(\theta) \d \theta}{\int_\Theta \tilde P^{(m)} \exp\big(-n\hat K(\X, \tilde \X^\theta)\big)\pi(\theta) \d \theta}. \label{eq:abc_post_exp}
\end{align}

To gain more insights into the  ABC posterior behavior under the exponential kernel, we take a closer look at the ``likelihood function" above
\begin{align*}
\tilde P^{(m)} \exp\big(- n\hat K(\X, \tilde \X^\theta)\big)  &=\frac{p_\theta^{(n)}}{p_0^{(n)}}\tilde P^{(m)} \e^{u_\theta},
\end{align*}
where
$
u_\theta(\X, \tilde\X^\theta)=- n\times \Big(\hat K(\X, \tilde \X^\theta)-\P_n\log\frac{p_0}{p_\theta}\Big).
$
From the equations above, we can write  
\begin{equation}\label{eq:abc_post_exp_u}
\hat \pi^{EK} (\theta\mid \X) \propto p_\theta^{(n)}(\X)\times \e^{\hat u_\theta(\X)} \times \pi(\theta)\quad \text{with}\quad \hat u_\theta (\X)=\log \int  \e^{u_\theta(\X,\tilde \X^\theta)}\d \tilde P^{(m)}(\tilde\X).
\end{equation} 
When $\hat K(\X, \tilde\X^\theta)$ is the classification-based estimator,   $\hat u_\theta(\X)$ can be related to the random generator setting of the Metropolis-Hastings MHC algorithm in \citet{kaji2021mh} which has  \eqref{eq:abc_post_exp_u} as its stationary distribution. Similarly as in  {Appendix Section 5} of \citet{kaji2021mh}, we can regard the posterior approximation in \eqref{eq:abc_post_exp_u} as a posterior $\hat \pi^{EK}(\theta \mid \X) \propto  q_\theta^{(n)} \tilde \pi(\theta)$ under a misspecified likelihood 
\begin{equation}\label{eq:tilde_p}
q_\theta^{(n)}=\frac{p_\theta^{(n)}(\X)\e^{\hat u_\theta(\X)}}{C_\theta} \quad \text{ where }\quad C_\theta=\int_\mX p_\theta^{(n)}(\X)\e^{\hat u_\theta(\X)} \d\X
\end{equation}
and a modified prior  $\tilde \pi(\theta)\propto \pi(\theta)C_\theta$.
Since the likelihood is misspecified, the ABC posterior concentrates around a projection point $\theta^*$ defined as
\begin{equation}\label{eq:theta_star}
\theta^*=\arg\min_{\theta\in \Theta} -P_{0}^{(n)}\log[q_\theta^{(n)}/p_0^{(n)}],
\end{equation}
which corresponds to the mis-specified model that is closest to $P_0^{(n)}$ in the KL sense \citep{kleijn2006misspecification}. 
\citet{kaji2021mh} study posterior concentration of \eqref{eq:abc_post_exp_u}. We recall their Theorem 4.5 in \Cref{sec:exp_kernel_post}.
 Unlike in Theorem \ref{thm:post_rate}, the posterior concentration rate  here depends both on the estimation error $\delta_n$ and the actual concentration rate of the true posterior, not the acceptance threshold.
 
 \iffalse
 \begin{remark}(Extension to i.i.d Case) The likelihood in \eqref{eq:tilde_p} does not separate due to the nested dependence of $u_\theta$ and $\X$. When $n\to \infty$, one can also extend the theorem to  i.i.d. data case as studied in \citet{kleijn2006misspecification}. This can be achieved by splitting $\X$ into two halfs, one for training ($\X^{train}$) and one for prediction ($\X^{pred}$). Then $u_\theta(\cdot)$ only depends on $\X^{train}$, and thus $u_\theta(\X^{pred})$ is separable.
 \end{remark}
\fi

\begin{remark}[Vanishing Bias] To better understand the severity of the centering bias of the misspecified model, we note
\begin{align*}
-P_{0}^{(n)}\log[q_{\theta^*}^{(n)}/p_0^{(n)}] &\leq -P_{0}^{(n)}\log[q_{\theta_0}^{(n)}/p_0^{(n)}] 
= P_0^{(n)} \log \frac{p_0^{(n)}}{p_0^{(n)} \e^{\hat u_{\theta_0}(\X) }/C_{\theta_0}} \\
&= \log C_{\theta_0}- P_0^{(n)} {\hat u_{\theta_0}(\X)} = \log P_0^{(n)} \e^{\hat u_{\theta_0}(\X)} - P_0^{(n)} {\hat u_{\theta_0}(\X)}.
\end{align*}
This is essentially the Jensen gap. If we have this Jensen gap vanishing when $n\to \infty$, then we can conclude that the centering bias is also vanishing, and the ABC posterior in \eqref{eq:abc_post_exp} will eventually concentrate at the right location. 
\end{remark}

\iffalse
\begin{remark}[Connection to Generalized Bayesian Inference]
The exponential kernel has a connection to the generalized posterior within the Generalized Bayesian Inference (GBI) framework \citep{bissiri2016general}. 
 \citet{schmon2020generalized} show that one can view the ABC posterior constructed from the scaled exponential kernel as
$
p_l(\theta\mid\X) \propto \int \exp(-w\cdot l(\X, \tilde \X^\theta)) \pi(\theta)\d\tilde P^{(m)}(\tilde\X),
$
where   $\l(\cdot, \cdot)$ is a ``loss function" which can be any distance between the summary statistics or a data discrepancy metric used in ABC, and where the weight  $w >0$ can be chosen to adjust the influence of the loss on the posterior. 
The exponential kernel can thus be applied more generally  to other summary statistics or data discrepancy measures, as in \citet{park2016k2}. 
\end{remark}
\fi

\subsection{Shape of the Limiting ABC Posterior Distribution}
We now analyze the limiting shape of $\hat \Pi^{AR}_{\epsilon_n}(\cdot\mid \X)$ defined in \eqref{eq:ABC_accept_post}. We focus on  the case when $\epsilon_n\gg \delta_n^{1/(\kappa+1)}$, where $\kappa$ was defined in Assumption \ref{ass:prior_mass}, since the posterior is not guaranteed to converge when the decision threshold $\epsilon_n$ is smaller than the estimation error $\delta_n$ of the KL estimator.

\begin{assumption}\label{ass:KL_taylor}
Assume that for every $\varepsilon>0$, we have $\inf_{\|\theta-\theta_0\|>\varepsilon}K(p_0,p_\theta)>0$. In addition, assume that $\log p_\theta$ is twice differentiable with respect to $\theta$ and that, for every $\theta$ in some neighborhood of $\theta_0$, the remainder of the second order Taylor expansion of $K(p_0, p_\theta)= P_0 \log\frac{p_0}{p_\theta}(x)$ around $\theta_0$  is comparatively small  relative to the second-order term, i.e.
\begin{align*}
K(p_0,p_\theta)&=\nabla_{\theta=\theta_0} K(p_0, p_\theta) (\theta-\theta_0)+\frac{1}{2}(\theta-\theta_0)'\nabla^2_{\theta=\theta_0} K(p_0, p_\theta) (\theta-\theta_0)\{1+o(1)\} \\
&=\frac{1}{2} (\theta-\theta_0)'I(\theta_0) (\theta-\theta_0)\{1+o(1)\},
\end{align*}
where $I(\theta)=\nabla^2_{\theta} K(p_0, p_\theta)= P_0[(\nabla_\theta \log p_\theta)^2]$ is the Fisher information matrix.
\end{assumption}

When \Cref{ass:KL_taylor} is satisfied, the condition in \Cref{ass:identification} is immediately satisfied as well and the identification of $\theta_0$ is guaranteed.  In the open-neighborhood of $\theta_0$, since we have
 \[
 (\theta-\theta_0)'I(\theta_0) (\theta-\theta_0) = 2K(p_0, p_\theta)\{1+o(1)\},
 \]
and $I(\theta_0)$ is positive definite, the convergence in the bilinear form $ (\theta-\theta_0)'I(\theta_0) (\theta-\theta_0)$  ensures the convergence of the parameters.

\begin{theorem}\label{thm:unif_ci}
Assume that the prior function $\pi(\cdot)$ is continuous around $\theta_0$. Then,  under Assumptions \ref{ass:entropy}, \ref{ass:ratio_bd}, \ref{ass:prior_mass} and \ref{ass:KL_taylor}, if  $\lim_n \delta_n/\epsilon_n^{\kappa+1}\to 0$, the average posterior distribution of $\epsilon_n^{-1/2}(\theta-\theta_0)$ converges to the uniform distribution over the ellipse $\{w: w' I(\theta_0) w\leq 2\}$ where $I(\theta)$ is the Fisher information matrix defined in \Cref{ass:KL_taylor}. In particular,  as $n\to 0$, we have
\[
P_0^{(n)}\int f\Big(\epsilon_n^{-1/2}(\theta-\theta_0)\Big) \hat \Pi^{AR}_{\epsilon_n}(\theta \mid \X) \to \int_{u'I(\theta_0)u\leq 2} f(u) \d u/\int_{u'I(\theta_0)u\leq 2} \d u
\]
for every  continuous and bounded function  $f(\cdot):\mX\rightarrow\R$. 
\end{theorem}

The proof is provided in \Cref{sec:proof_unif_ci}.

\begin{remark}
\Cref{thm:unif_ci} is adapted from the  case (i) in Theorem 2 of \citet{frazier2018asymptotic}. We only consider situations when $\epsilon_n \gg\delta_n^{1/(\kappa+1)}$ with the prior shrinkage parameter $\kappa$  defined in Assumption \ref{ass:prior_mass}. In other words, we   assume that the ABC decision threshold $\epsilon_n$ is dominating both the estimation error $\delta_n$ and the asymptotic error $n^{-1/2}$ and, thereby, determines the posterior concentration rate.  It is not entirely obvious how the posterior would behave when the threshold $\epsilon_n$ shrinks faster than the estimation error $\delta_n$.
\iffalse
\Cref{thm:unif_ci} is adapted from  case (i) in Theorem 2 of \citet{frazier2018asymptotic}. We only consider the case where $\epsilon_n \gg\delta_n^{1/(\kappa+1)}$ with the prior shrinkage parameter $\kappa$  from Assumption \ref{ass:prior_mass}. In other words, we   assume that the ABC decision threshold $\epsilon_n$ is dominating both the estimation error $\delta_n$ and the asymptotic error $n^{-1/2}$ and, thereby, determines the posterior concentration rate. It is not entirely obvious how the posterior would behave when the threshold $\epsilon_n$ shrinks faster than the estimation error $\delta_n$. In addition, we study the posterior behavior of $\epsilon_n^{-1/2}(\theta-\theta_0)$ rather than $\epsilon_n^{-1}(\theta-\theta_0)$. This inefficiency is the result of $\nabla_{\theta=\theta_0} K(p_0,p_\theta) =0$ and the use of second-order expansion.
\fi
\end{remark}

For the asymptotic behavior of the ABC posterior \eqref{eq:abc_post_exp_u} induced by the exponential kernel,  we resort to BvM characterizations under misspefication in LAN models \citep{kleijn2012bernstein}. {When the posterior \eqref{eq:abc_post_exp_u} concentrates around $\theta^*$ in \eqref{eq:theta_star} at the rate $\epsilon_n^*$,  one can show under a suitable LAN condition that the approximate posterior converges to a sequence of normal distributions in total variation at the rate $\epsilon_n^*$. The centering and the asymptotic covariance matrix both depend on $\theta^*$. The formal statement and the proof is in \citet{kaji2021mh}. 

 Although we only consider the case where the model is correctly specified in our paper, our results can be  extended to the mis-specified model along the lines of \citet{frazier2019model}. 
}

\subsection{ABC Kernels and Model Misspecification}\label{sec:model_misspec}
Although the exponential kernel ABC (Algorithm \ref{alg:abc_exp}) obviates the need for the threshold $\epsilon_n$ and performs very well in our examples,   the accept/reject kernel ABC (Algorithm \ref{alg:abc_plain})  may perform better under mis-specification (see Remark \ref{rem:robust}).
When the model is mis-specified, the posterior   under Algorithm \ref{alg:abc_plain} will converge to the ``pseudo-true" value, which is the point that minimizes the distance  between summary statistics within the mis-specified class. Since our summary statistic is replaced with KL divergence, this point will coincide with the KL projection in our case.
The  exponential kernel will also concentrate around a certain KL projection but its bias will now be compounded by the  influence of both $P_0\notin \mP=\{\theta \in \Theta: P_\theta\}$ and the exponential tilt  $e^{\hat u_{\theta}(\X)}$ arising from the approximation error of the KL estimator. We would thereby expect a bigger bias from the exponential kernel when the model is misspecified.

We illustrate the intuition above with a  toy example. We use the simple example proposed in \citet{frazier2019model} where the assumed data-generating process (DGP) is i.i.d.  $\mN(\theta,1)$, but the actual DGP for $\X$  is i.i.d. $N(\theta, \sigma^2)$ for $\sigma^2\neq 1$. In other words, the assumed DGP has an incorrect specification  of the variance of the observed data. We consider the oracle logistic classifier built on $\X$ and the quadratic term $\X^2$ for our KL estimator.
We  fix $\theta=1$ and simulate $\X$ (using   $n=100$) with respect to different values of $\sigma^2$, ranging  from 0.5 to 5 with evenly spaced increments of 0.05. Using one common set of latent variables $\nu_i \sim \mN(0,1)$,  the observed data is generated as $X_i = 1+\nu_i \sigma$  for each $\sigma^2$. The prior belief  is  $\theta \sim \mN(0, 25)$, and we implement  ABC methods with $N=100\,000$  pseudo datasets. The parameter draws and the  latent variable datasets are the same across the  different values of $\sigma^2$.  To  explore how the tail behavior  influences the ABC bias, we also include misspecified lognormal distributions using the same setup.

\begin{figure}[!t]
\centering
\subfigure[normal]{
\includegraphics[width=0.45\textwidth]{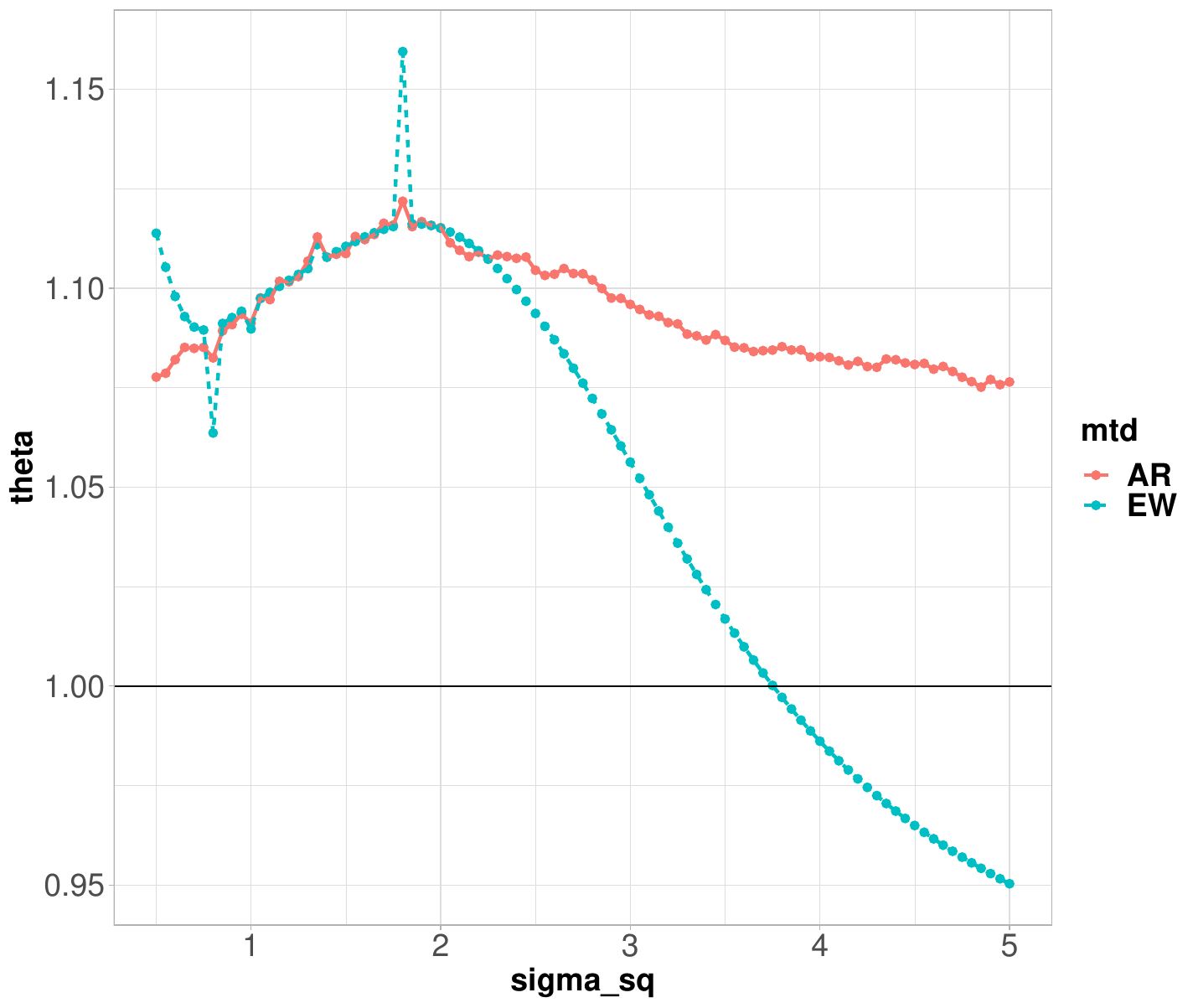}\label{fig:mis_normal}
}
\subfigure[lognormal]{
\includegraphics[width=0.45\textwidth]{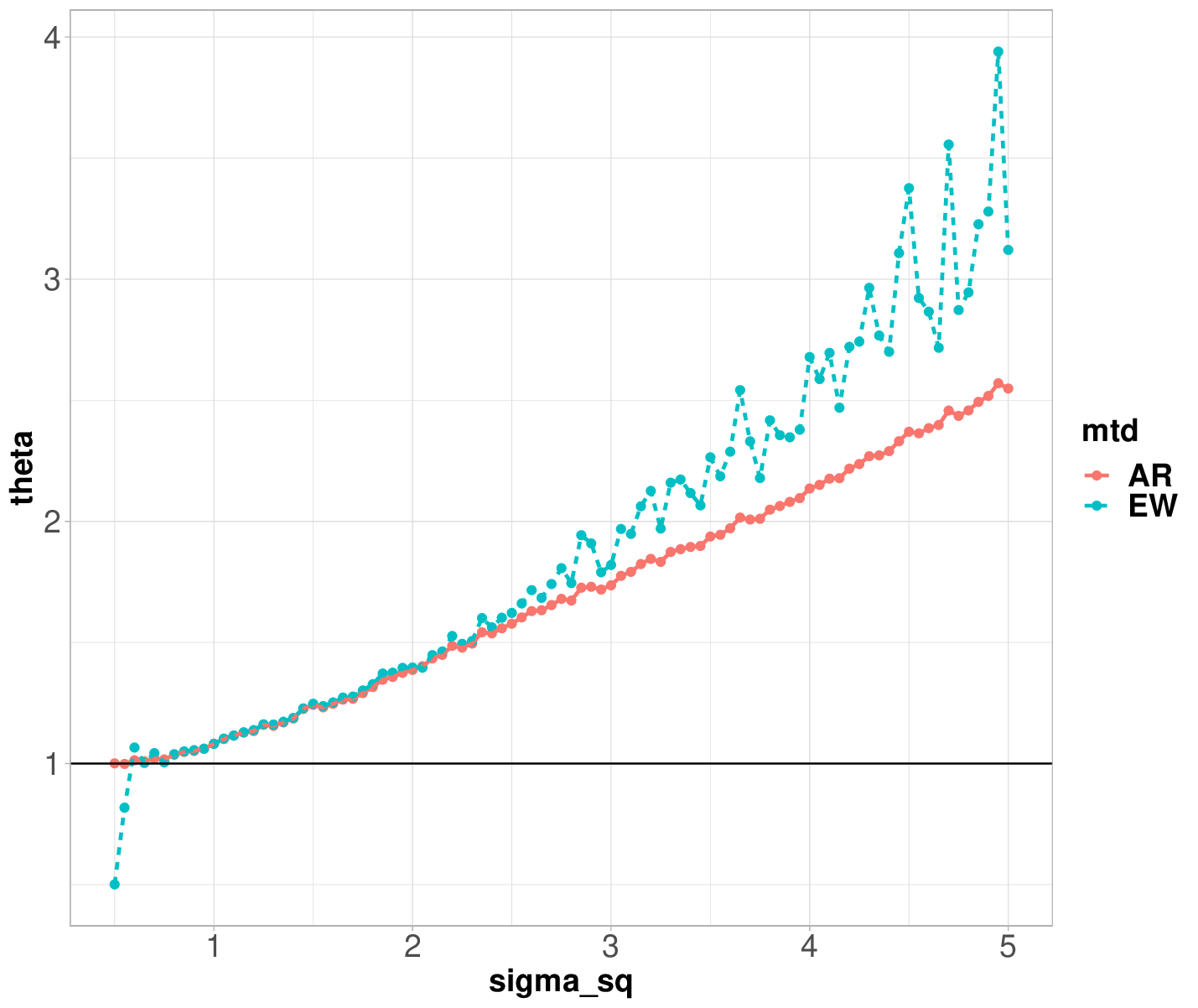}\label{fig:mis_lognormal}
}
\vspace{-0.1in}
\caption{ \footnotesize ABC posterior under misspecified models.}
\end{figure}

\Cref{fig:mis_normal}  compares the posterior mean of the accept-reject ABC (AR)  and the ABC with exponential weighting (EW) across different values of $\sigma^2$. We can see that both AR and EW have a relatively small  bias in  estimating $\theta$ when the misspecification level in $\sigma^2$ varies. Since the logistic regression classifier on  $\X$ and $\X^2$ is almost the ``oracle" discriminator for gaussian distributions, the error of the KL estimator should be minimal.  Nevertheless, the posterior mean of EW exhibits a downward moving trend when the level of misspecification increases. For the heavy-tailed lognormal distribution shown in \Cref{fig:mis_lognormal}, although the posterior mean of AR does shift away from true value $\theta=1$ as the degree of model misspecification increases, the posterior mean of EW shifts away from $\theta=1$ at a faster speed.

\section{Simulations}
In this section, we illustrate our approach and make comparisons with other likelihood-free inference techniques. Within our KL-ABC framework, we include two types of KL estimators. One is obtained with the logit   discriminator score, which we refer to as   KL estimation via classification (KLC), and the other one is estimated via the kNN method (kNN) with $k=1$ \citep{jiang2018approximate}. For both estimators, we aggregate  ABC samples  with the accept-reject kernel as in Algorithm \ref{alg:abc_plain} and the exponential kernel as in Algorithm \ref{alg:abc_exp}. The latter will be denoted with a suffix  `exp', e.g.   KLC-exp or kNN-exp. The discriminator used for each dataset will be specified later. We provide more discussions on   discriminator calibrations  in \Cref{sec:dhat_training}. Besides the two examples presented here, we have another analysis on the g-and-k distribution in \Cref{sec:gk}.

The  ABC discrepancy metrics we choose for comparisons are (1) the classification accuracy (CA) \citep{gutmann2018likelihood} defined as \eqref{eq:ca}; (2) the 2-Wasserstein (W2) distance  under the Euclidean metric \citep{bernton2019approximate} defined as   $\text{W2}(\X,\tilde \X^\theta)= \min_\gamma [\sum_{i=1}^n \sum_{j=1}^m \gamma_{ij}||X_i-\tilde X^\theta_j||^2]^{1/2}$ s.t. $\gamma' {\bf 1}_m={\bf 1}_n, \gamma' {\bf 1}_n={\bf 1}_m$ with  $0\leq \gamma_{ij}\leq 1$; (3) $\ell_2$-distance between summary statistics (SS) and we use the semi-automatic (SA) method \citep{fearnhead2011constructing} if no candidate summary statistics are given; {(4) approximated posterior mean of the  parameters predicted by trained deep neural network (DNN) \citep{jiang2017learning};  (5) Maximum Mean (MM) discrepancy \citep{park2016k2} defined as $MM(\X,  \tilde\X^\theta)=\frac{1}{n(n-1)}\sum_{ i\neq j} k(X_i, X_j)+\frac{1}{m(m-1)}\sum_{ i\neq j} k(\tilde X^\theta_i, \tilde X^\theta_j)-\frac{2}{nm}\sum_{ i,  j} k(X_i, \tilde X^\theta_j)$ where $k(\cdot,  \cdot)$ is a Gaussian kernel with the bandwidth being the median of $\{\norm{X_i - X_j} : i\neq j \}$;(6) a V-statistic estimator of  Energy Statistics (ES) proposed by \citet{nguyen2020approximate};}  (7) auxiliary likelihood $AL=\frac{1}{m} \ln p_A(\tilde\X^\theta\mid \hat \phi(\tilde\X^\theta)) -\frac{1}{m} \ln p_A(\tilde\X^\theta\mid \hat \phi(\X)) $ proposed by \citet{drovandi2011approximate}, where $p_A(x\mid \phi)$ is a $d$-dimensional  Gaussian distribution with $\phi$ being  the sample mean and covariance. For the classification accuracy (CA), we use the same discriminator as the one in KL estimation. {For the DNN approach,  we deploy a 3-layer DNN with  100 neurons  and hyperbolic tangent (tanh) activation on each hidden layer. The model is trained on $10^6$ samples and validated  on $10^5$ samples, with early  stopping  once the validation  error starts to  increase.}  In each experiment, unless otherwise noted,  we set the tolerance threshold $\epsilon$ adaptively such that $1\,000$ of $100\,000$ (i.e. the top 1\%) proposed ABC samples are accepted.

\subsection{M/G/1-Queuing Model}
Because queuing models are usually easy to simulate from, but have no tractable likelihoods, they have been frequently used as test cases in the ABC literature, see e.g. \citet{fearnhead2011constructing} and \citet{bernton2019approximate}. 
Here, we choose the same setup as in \citet{jiang2017learning}. Each datum is a 5-dimensional vector consisting of the first five inter-departure times $x_i=(x_{i1},x_{i2},x_{i3},x_{i4},x_{i5})'$. In the model, the service times $u_{ik}$ follow a uniform distribution $U[\theta_1, \theta_2]$, and the arrival times $w_{ik}$ are exponentially distributed with the rate $\theta_3$. We only observe the interdeparture times $x_i$, given by the process $x_{ik}=u_{ik}+\max(0, \sum_{j=1}^k w_{ij} -\sum_{j=1}^{k-1} x_{ij})$. We perform ABC on $n=500$ observed samples which are generated from the true parameter $\theta_0=(1, 5, 0.2)$. The prior on $(\theta_1, \theta_2-\theta_1, \theta_3)$ is uniform on $[0,10]^2\times [0, 0.5]$. \footnote{We place the uniform prior on $\theta_2-\theta_1$ instead of $\theta_2$, since $\theta_2$ must be larger than $\theta_1$. This is  used in \citet{jiang2017learning}.} 

{Regarding the choice of the discriminator, we consider both the  Random Forest (RF) classifier  and a $\ell_1$-penalized logistic classifier (LRD). For the former, we use the default setting in the \texttt{R} package \texttt{randomForest}. We also denote CA calculated from RF classifier as RF-CA. For the latter, we implement the discriminator with R package \texttt{glmnet},   and the model is built on degree-2 polynomials of the data, including quadratic  and interaction terms, with penalty parameter $\lambda$ is selected via 5-fold cross-validation. We find that RF outperforms LRD. }

\begin{figure}[!t]
\begin{center}
\includegraphics[width=0.9\textwidth]{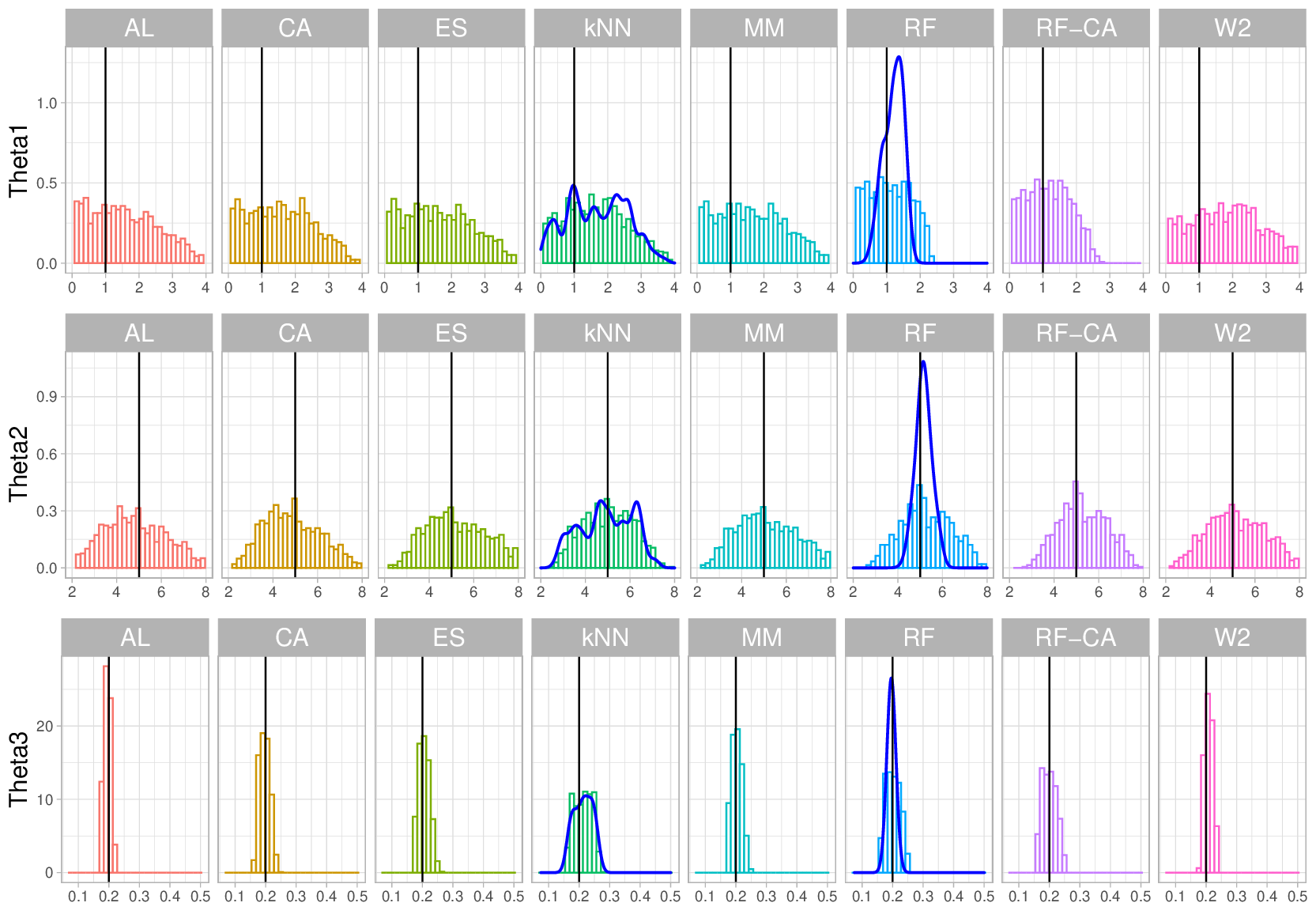}
\end{center}
\vspace{-0.1in}
\caption{ABC reconstructed posteriors under M/G/1-queuing model with $\theta_0=(1, 5, 0.2)'$.  The vertical  black lines mark the true values. The blue curves in kNN and RF boxes mark a smoothed density calculated from the exponential kernel.}\label{fig:mg1_thetas}
\vspace{-0.2in}
\end{figure}

The shape of the ABC posteriors is given in \Cref{fig:mg1_thetas}. The plot reveals that, among the three parameters, $\theta_1$ is the hardest one to estimate where all methods, except for RF  and its  variants, gave relatively flat posterior estimates. Regarding $\theta_2$, all methods seem to center well around the truth. RF-exp provides the tightest estimation. Regarding $\theta_3$, all except  kNN return spiky posterior. Next, we repeat the experiments on 10 different datasets, and summarize the average squared estimation errors and the width of the 95\% credible intervals in  \Cref{tab:mg1_stats}. Overall, we see that  RF and RF-exp are able to correctly identify the right locations of the parameters and outperform  the kNN estimator, with RF-exp providing the tightest credible interval. The method RF-CA tends to give very similar results to  RF, which is not entirely unexpected since they are derived from the same discriminators.

\begin{table}[!t]
\centering
\scalebox{0.9}{
\begin{tabular}{l | *{2}{c} | *{2}{c} |*{2}{c}} \toprule  
   & \multicolumn{2}{c|}{$\theta_1=1$}                                       & \multicolumn{2}{c|}{$\theta_2=5$}                                       & \multicolumn{2}{c}{$\theta_3=0.2$}                         \\
   {Method}     &{$(\hat\theta_1-\theta_1)^2$} &{95\% CI width} & {$(\hat\theta_2-\theta_2)^2$} & {95\% CI width} &{$(\hat\theta_3-\theta_3)^2$}  &{95\%CI width}   \\ 
           (scale) & &  & &     & {\small$(10^{-4})$ }&  \\
          \midrule 
          RF      &  \bf{0.008}  & 2.044 &  0.038  & 4.304  & 0.571 & 0.084 \\
RF -exp & 0.035  &  \bf{0.959} &  0.055  &  \bf{1.473}  & 0.307 &  \bf{0.028} \\
          LRD    & 0.197    & 3.116 & 0.217 &  4.599 & 0.308 & 0.064 \\
LRD-exp & 0.169 &  2.851 & 0.312 & 3.708 &  \bf{0.234} & 0.030 \\  
kNN      & 0.525 & 3.135 & 0.106 & 3.986 & 3.659 &0.094 \\  
kNN-exp  & 1.057 &  2.664 (0.8) & 0.431 &  3.331 (0.9)& 2.634 &  0.072 \\
\vspace{-0.1in}&&&&&&\\
DNN     & 0.150  & 3.350 & 1.100  & 7.361  & 81.931 & 0.328 \\
ES      & 0.255  & 3.321 & 0.087  & 5.280  & 1.176 & 0.070 \\
CA      & 0.191 & 3.107 & 0.235 & 4.602 & 0.280 & 0.064 \\
RF-CA   & 0.014 & 2.194&  \bf{0.036} & 4.104 & 0.404  & 0.084 \\
AL      & 0.259 & 3.423 & 0.057 &  5.651 & 0.575 &  0.044 \\
SA      & 0.180 &  2.457 & 0.355 &  3.514 & 45.297 &  0.446 \\
W2      & 0.595 &  3.631 &  0.039 &  4.871 & 3.846 &0.052 (0.8)\\
\bottomrule    
\end{tabular}
}
\vspace{-0.1in}
\caption{\label{tab:mg1_stats} \footnotesize ABC performance on the M/G/1 queuing model over 10 repetitions with top 1\% ABC samples selected. Most of the 95\% CIs have full coverage with the rest having  their coverage marked in brackets. The bold fonts mark the best model in each metric.}
\vspace{-0.2in}
\end{table}

\subsection{Lotka-Volterra Model}
The Lotka-Volterra (LV) predator-prey model \citep{wilkinson2018stochastic}  describes population evolutions in ecosystems where predators interact with prey. It is one of the classical   stochastic kinetic network model examples. The state of the population is prescribed deterministically via a system of  ordinary differential equations (ODEs). Inference for such models is challenging because the transition density is intractable. However, simulation from the model is possible, which  makes it a natural candidate for ABC methods. 

The model monitors population sizes of  predators $X_t$ and  prey $Y_t$ over time $t$. The changes in states are determined by  four parameters $\theta=(\theta_1, \ldots, \theta_4)'$ controlling: (1) the rate $r_1^t=\theta_1 X_t Y_t$ of a predator being born; (2) the rate $r_2^t=\theta_2 X_t$ of a predator dying; (3) the rate $r_3^t=\theta_3Y_t$ of a prey being born; (4) the rate $r_4^t=\theta_4X_tY_t$ of a prey dying.  Given the initial population sizes  $(X_0, Y_0)$ at time $t=0$, the dynamics can be simulated using the Gilliespie algorithm \citep{gillespie1977exact}. The algorithm samples times to an event from an exponential distribution (with a rate $\sum_{j=1}^4 r_j^t$) and picks one of the four reactions with probabilities proportional to their individual rates $r_j^t$. 

\begin{figure}[!t]
\centering
\subfigure[$(\theta_1, \theta_4) \in (0,0.1)^2$]{
\includegraphics[width=0.4\textwidth]{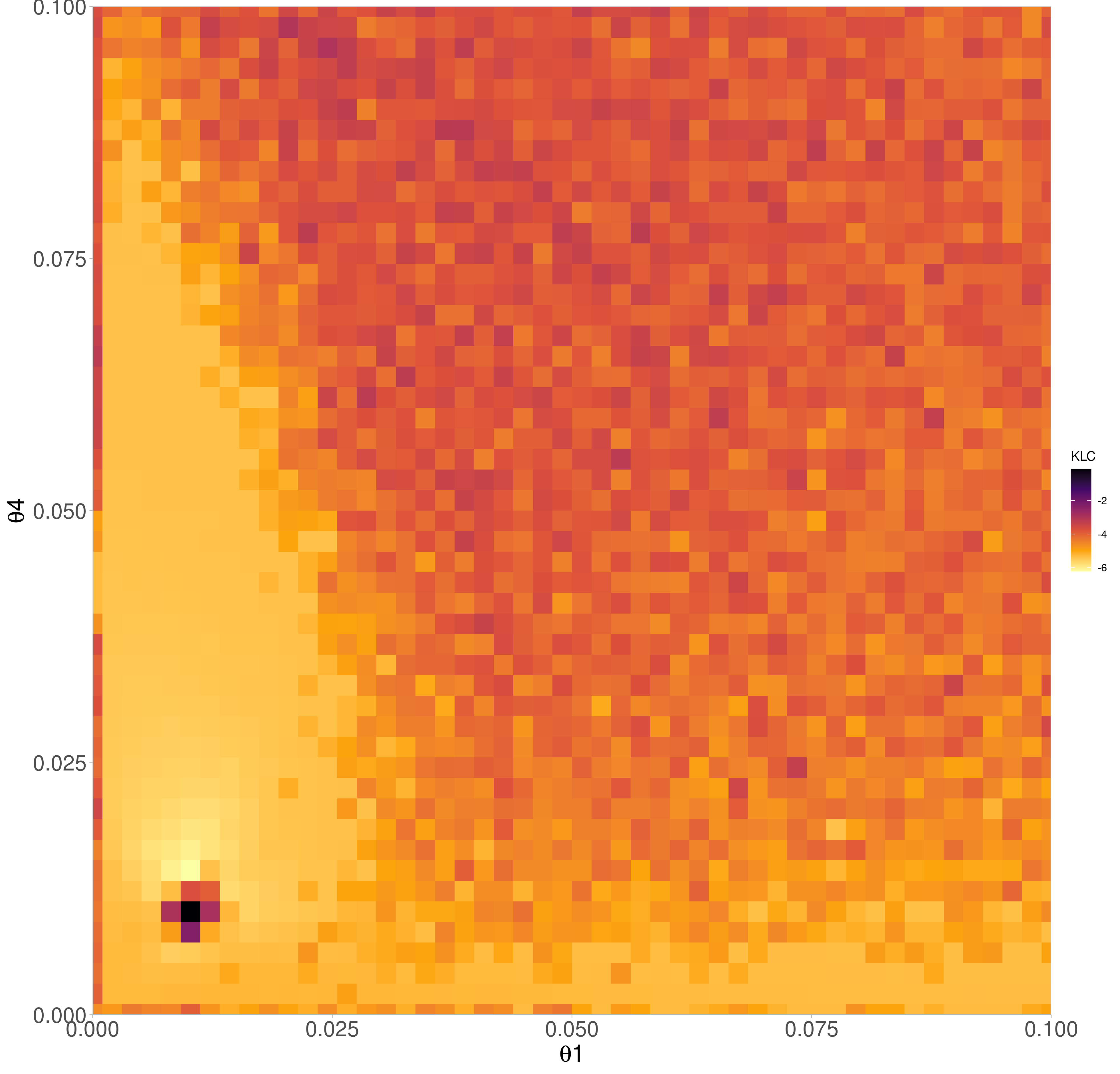}\label{fig:lv_zoomout}
}
\subfigure[$(\theta_1, \theta_4) \in (0,0.02)^2$ ]{
\includegraphics[width=0.4\textwidth]{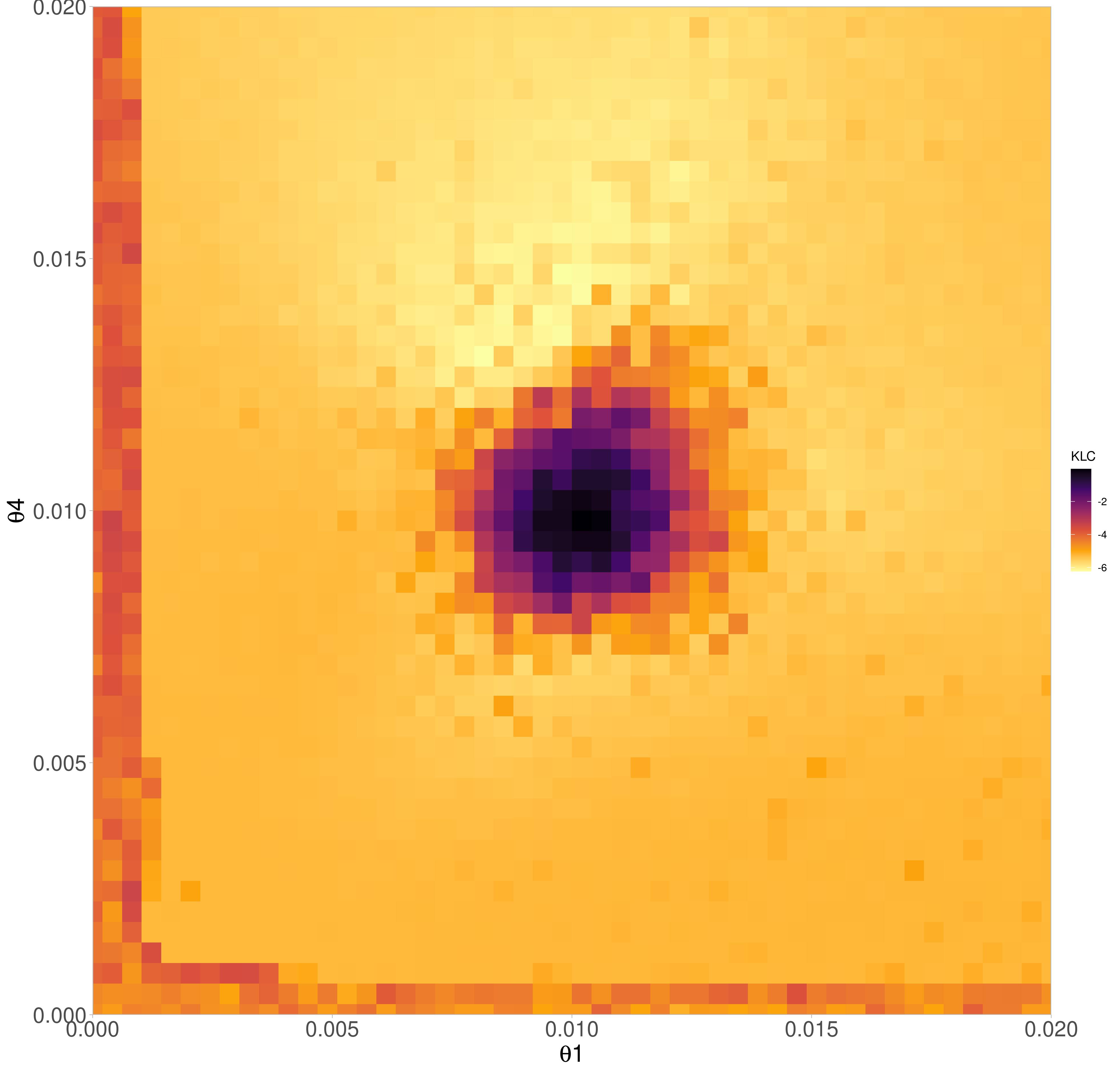}\label{fig:lv_zoomin}
}
\vspace{-0.1in}
\caption{  \footnotesize Estimated $\hat K(\X, \tilde \X^\theta)$ (via Classification) over a grid of $(\theta_1, \theta_4)$ values. The other two parameters are fixed as $\theta_2=0.5, \theta_3=1$.} \label{fig:lv_heatmap}
\vspace{-0.2in}
\end{figure}

We use the same simulation setup as \citet{kaji2021mh}. Each simulation is started at $X_0=50$ and $Y_0=100$ and state observations   are recorded every  0.1 time units for a period of 20 time units, resulting in a series of $T=201$ observations each. The real data ($n=20$ time series) are generated with true values $\theta_0=(0.01, 0.5, 1, 0.01)'$.  The   predator-prey interaction dynamic is  very sensitive to parameter changes. For example, Figure 4 of \citet{kaji2021mh} shows that  a slight perturbation in $\theta_2$ leads to significant changes in the population renewal cycle. We rely on the ability of the discriminator to tell such different patterns apart.  The sensitivity of the model to minor parameter changes is confirmed with  heat-map plots of the estimated KL divergence as a function of $(\theta_1, \theta_4)'$ in \Cref{fig:lv_heatmap}. \Cref{fig:lv_zoomout} provides a plot of the estimated KL over the region $[0,0.1]^2$ where, apparently, the majority of the region is flat and uninformative with a sharp spike around the true values at $\theta_1=\theta_4=0.01$. We thus  narrow the investigation down to a smaller region $[0,0.02]^2$   in \Cref{fig:lv_zoomin}. Again, the curvature in the estimated KL around the truth is quite steep. This may pose some issues for Metropolis-Hasting algorithms, since the majority of the prior region is uninformative and improper initialization could lead to extremely slow convergence.

Previous ABC analyses of this model suggested various summary statistics including the mean, log-variance, autocorrelation (at lag 1 and 2) of each series as well as their cross-correlation \citep{papamakarios2016fast}. For the discriminator of our method, we choose the $\ell_1$-penalized (LASSO) logistic regression classifier (LRD) with $m=n$ and with a penalty $\lambda$ selected via  5-fold cross-validation (as implemented  in the R package \texttt{glmnet}), {as well as a random forest classifier (RF).}

\begin{figure}[!t]
\centering
\includegraphics[width=0.85\textwidth]{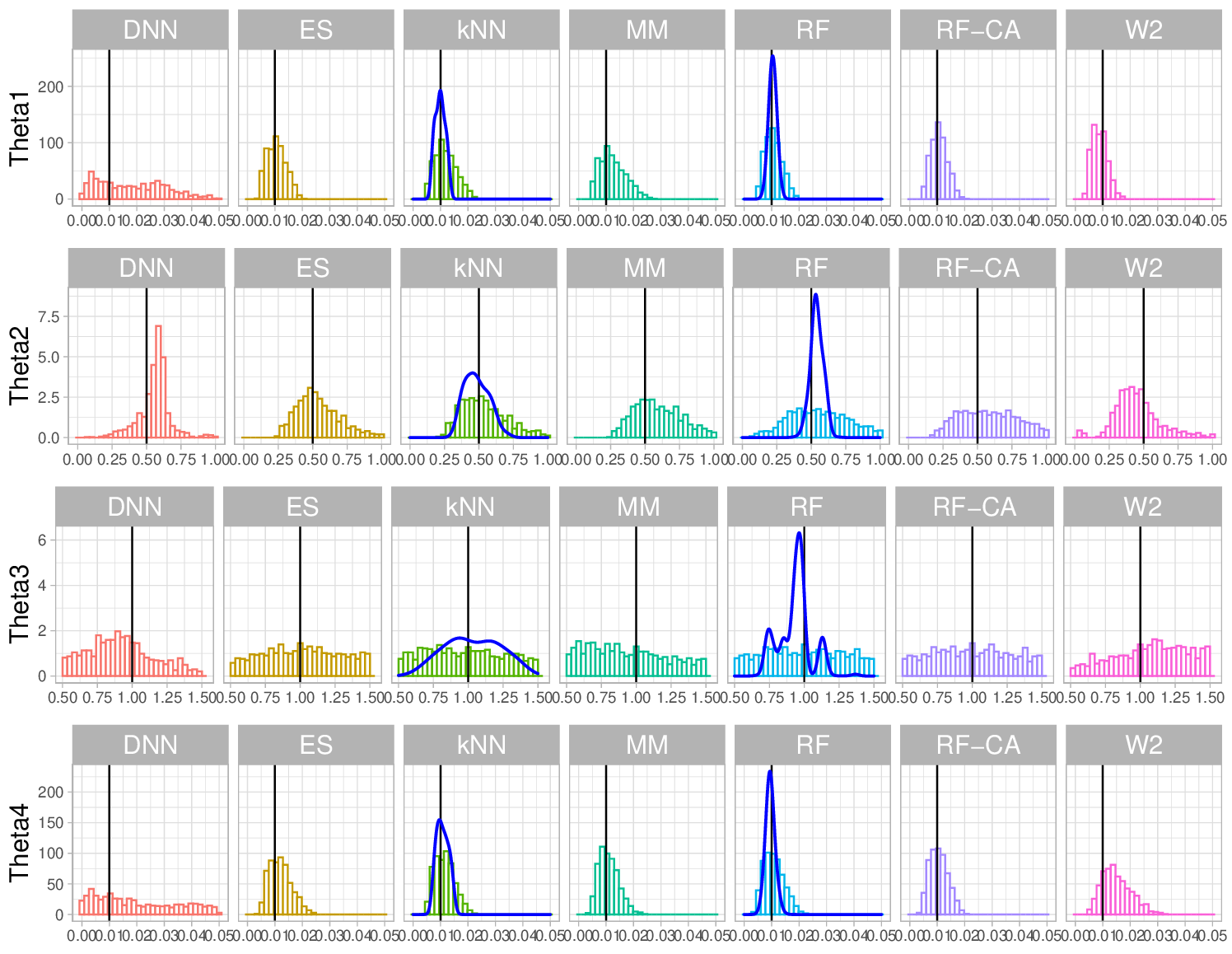}
\vspace{-0.1in}
\caption{  \footnotesize ABC posterior densities for the Lotka-Volterra model with $\theta_0=(0.01, 0.5, 1, 0.01)'$. The black vertical lines mark the true parameter values. The blue curves in kNN and RF boxes represent the smoothed density calculated from the exponential kernel. The ABC posteriors are plotted  from the top 1\% out of $10^5$ samples.}\label{fig:lv_post}
\vspace{-0.2in}
\end{figure}

Similar to \citet{kaji2021mh}, we use an informative prior $\theta \in U(\Xi)$ with a restricted domain $\Xi =[0, 0.1] \times [0, 1] \times [0,2] \times [0, 0.1]$ so that the computation is more economic and efficient. A typical snapshot of the ABC posteriors is plotted in  \Cref{fig:lv_post}. From the plot, we can see that the difficulty in estimating different parameters varies a lot and $\theta_3$ is the most challenging among all. While other methods give relatively flat posteriors for $\theta_3$,  our method (RF) combined with the scaled exponential kernel identifies the correct location of the parameter with a much tighter posterior. 	For the DNN approach, the posteriors seem  to be very flat and the estimation for $\theta_2$ is biased. Considering its heavy  computation costs shown in \Cref{tab:comp_cost}, we exclude this method  in the repetition experiment. The average performance of ABC methods under this model is summarized in \Cref{tab:lv_comp}. We can see that RF with the exponential kernel gives the tightest CI most of the time, while maintaining relatively small estimation errors.

\begin{table}[!ht]
\centering
\scalebox{0.7}{
\begin{tabular}{l | *{2}{c} | *{2}{c}  | *{2}{c} | *{2}{c} }
             \toprule
             & \multicolumn{2}{c|}{$\theta_1=0.01$}                             & \multicolumn{2}{c|}{$\theta_2=0.5$}                             & \multicolumn{2}{c|}{$\theta_3=1$}                             & \multicolumn{2}{c}{$\theta_4=0.01$}                             \\
{Method}     & {$(\hat\theta_1-\theta_1)^2$} & {95\% CI} width & {$(\hat\theta_2-\theta_2)^2$} & { 95\% CI} width & {$(\hat\theta_3-\theta_3)^2$}  & { 95\% CI}  width &  {$(\hat\theta_4-\theta_4)^2$}  & {95\% CI} width\\ 
        (scale) & {\small$(10^{-5})$} &  & &    &   {\small$(10^{-2})$} & & {\small$(10^{-5})$ }&  \\
   \midrule
RF      & 0.118  & 1.116 & 0.389 & 0.719 & \bf{0.037} & 0.938 & 0.051  & 1.290 \\
RF-exp  & \bf{0.015}  & \bf{0.354} & 0.092 & \bf{0.191} & 0.272 & \bf{0.483} & 0.037  & \bf{0.493} \\
LRD     & 42.949 & 4.474 & 1.688 & 0.626 & 0.119 & 0.945 & 17.159 & 4.720 \\
LRD-exp & 0.066  & \bf{0.354} & 0.247 & 0.285 & 0.405 & 0.643 & \bf{0.021}  & 0.626 \\
KL      & 0.563  & 1.462 & 0.107 & 0.602 & 0.054 & 0.934 & 0.203  & 1.327 \\
KL-exp  & 0.049 & 0.544  (0.9)  & \bf{0.044} & 0.243  (0.9)  & 0.346 & 0.599  (0.9) & 0.066  & 0.616 (0.9)  \\
\vspace{-0.1in} &&&&&&&& \\
CA      & 36.477 & 4.375 & 1.602 & 0.629 & 0.118 & 0.945 & 15.684 & 4.622 \\
RF-CA   & 0.141  & 1.061 & 0.699 & 0.700 & 0.042 & 0.933 & 0.056  & 1.181 \\
ES      & 0.101  & 1.233 & 0.203 & 0.573 & 0.045 & 0.923 & 0.341  & 1.443 \\
W2      & 1.265  & 1.666 & 0.478 & 0.599 & 0.011 & 0.917 & 4.080  & 2.051 \\
SS      & 1.639  & 1.777 & 2.260 & 0.721 & 1.019 & 0.901 & 3.767  & 1.660 \\
MM      & 0.824  & 1.727 & 0.959 & 0.601 & 0.414 & 0.932 & 0.074  & 1.294\\
\bottomrule
\end{tabular}}
\vspace{-0.1in}
\caption{\label{tab:lv_comp} \footnotesize ABC performance evaluations on the Lotka-Volterra Model, averaged over 10 repetitions, with the top 1\% ABC samples selected. Most of the 95\% CIs have full coverage  with the rest having their coverage  marked in brackets. The bold fonts mark the best model in each metric.}
\vspace{-0.2in}
\end{table}

\section{Empirical Analysis}
We further demonstrate   our approach on the nontrivial problem of estimating   stock volatility    using  merely daily observations on high, low and closing prices. All of these price observations are typically available to investors. We use a similar  data generating process as in \citet{magdon2003maximum},   assuming that the assets follow a Brownian motion with a constant drift and volatility. In particular, suppose that the log-price processes $X_j(t), i=1, \ldots, d$, are correlated Brownian motions, that is  $E[X_i(s)X_j(t)]=\sigma_{ij}\min\{s,t\}$, 
and that the joint movement of the log-price processes $\X(t)=(X_1(t),  \ldots, X_d(t))'$ follows a multivariate Brownian motion as
\begin{equation}
d\X(t)= \bmu  dt+\Sigma d \bm{W}(t),
\end{equation}
where $\bmu=(\mu_1, \ldots, \mu_d)'$ and $\Sigma=[\sigma_{ij}]_{1\leq i,j\leq d}$ denote the drift and the volatility of the log processes, respectively. We write 
\[
H_j =\max_{0\leq t\leq 1} X_j(t), \qquad L_j=\min_{0\leq t\leq 1} X_j(t),  \qquad S_j=X_j(1),
\]
for the high, low and final log price, respectively, over a fixed time interval $[0,1]$. We  want to estimate the drift $\bmu$ and the volatility matrix $\Sigma$  merely from observing these three prices over a period of time. 

We impose a normal-inverse-Wishart prior $(\bmu,\Sigma )\sim NIW (\bmu_0, \lambda, \Phi, \nu)$. This distribution can be sampled from in two steps: (1) sample $\Sigma$ from an inverse Wishart distribution $\Sigma\mid \Phi, \nu \sim W^{-1} (\Phi, \nu)$; (2) sample $\bmu$ from a multivariate normal distribution $\bmu\mid\bmu_0 , \lambda, \Sigma \sim N(\bmu_0, \frac{1}{\lambda}\Sigma)$. Since $\Sigma$ is a semi-positive definite matrix, we model the parameters through its Cholesky root $\Sigma^{1/2}$. Without loss of generality, we only consider the case $W(0)=(0, \ldots, 0)'$, since the closing prices on the previous day or the opening prices of today are usually known.

\subsection{Synthetic Data}
\begin{figure}[!ht]
\subfigure[$\sigma_{12}=0.05$]{
\includegraphics[width=0.3\textwidth]{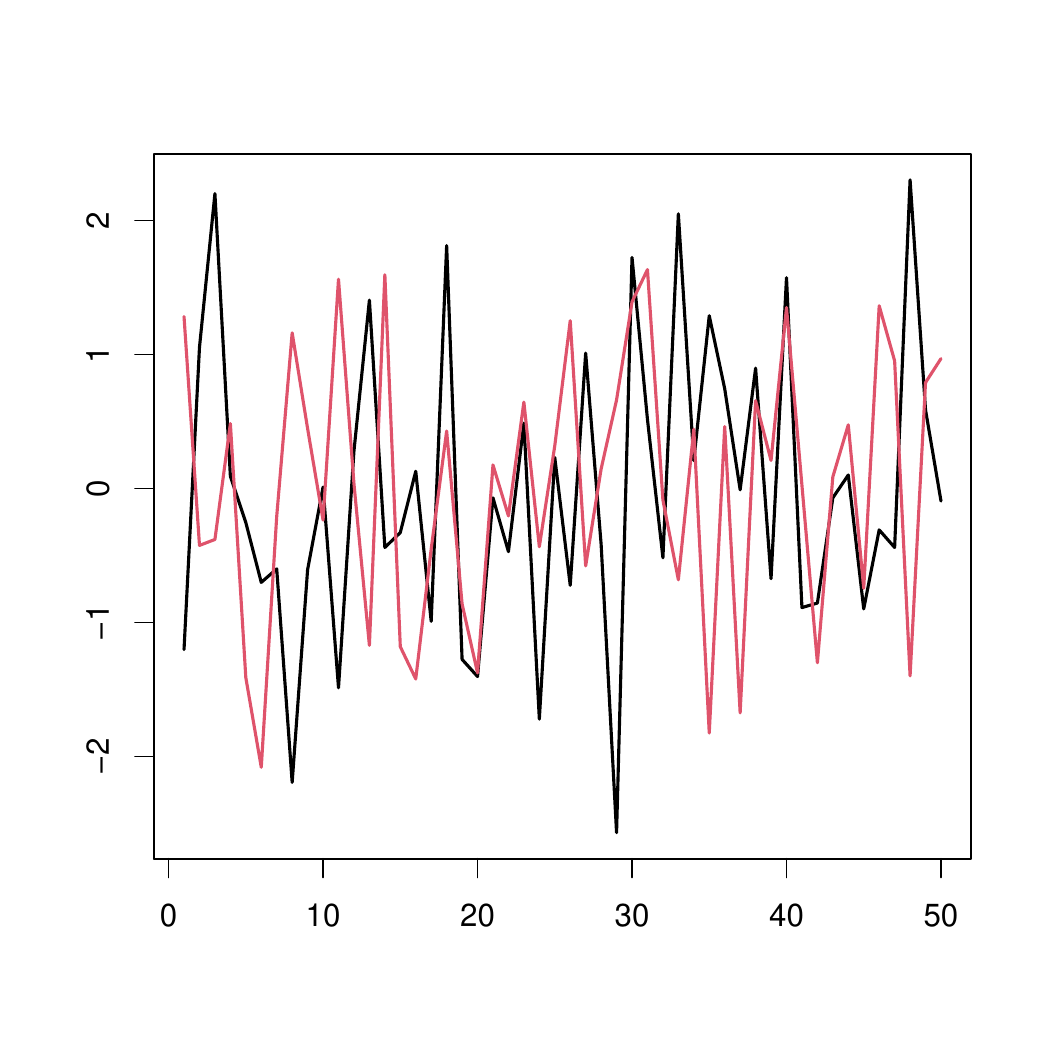}
}
\subfigure[$\sigma_{12}=0.5$]{
\includegraphics[width=0.3\textwidth]{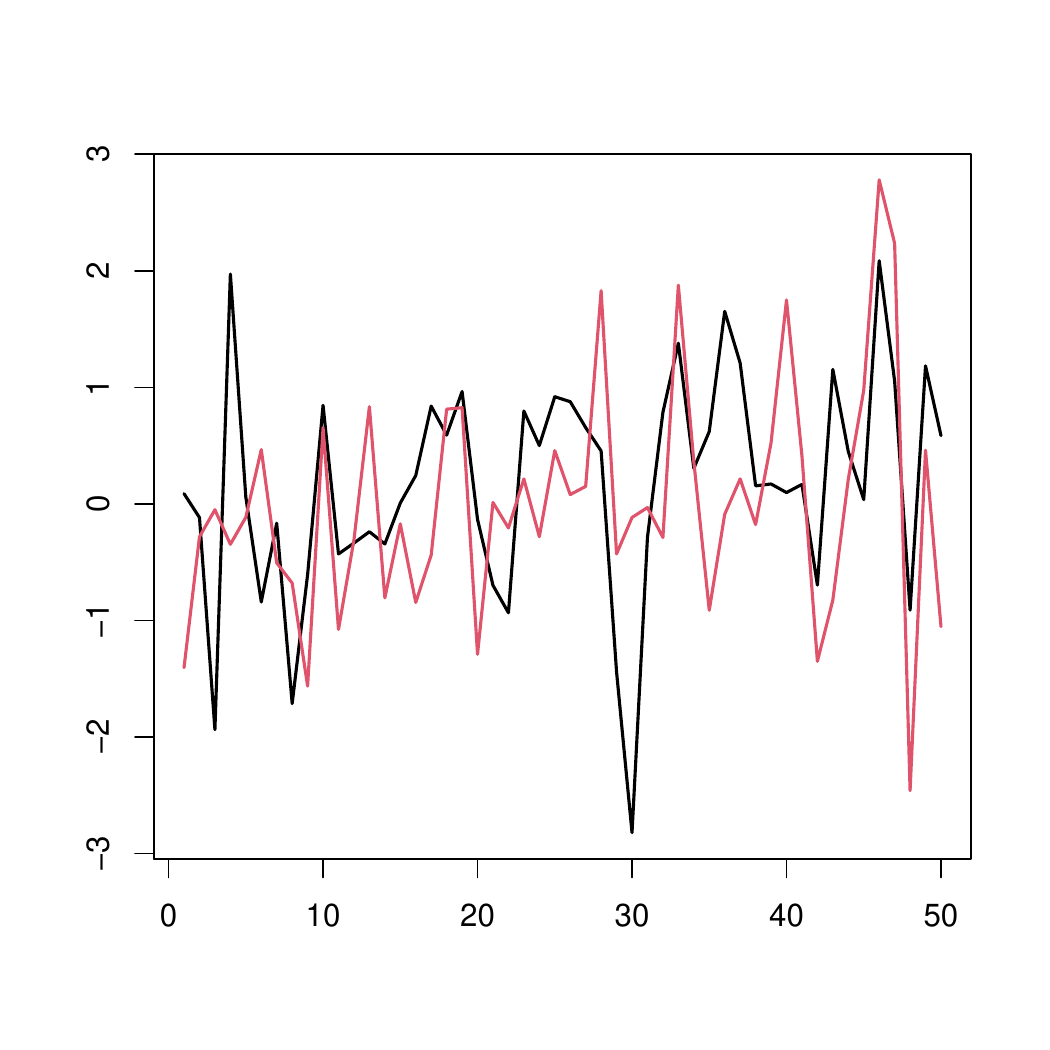}
}
\subfigure[$\sigma_{12}=0.95$]{
\includegraphics[width=0.3\textwidth]{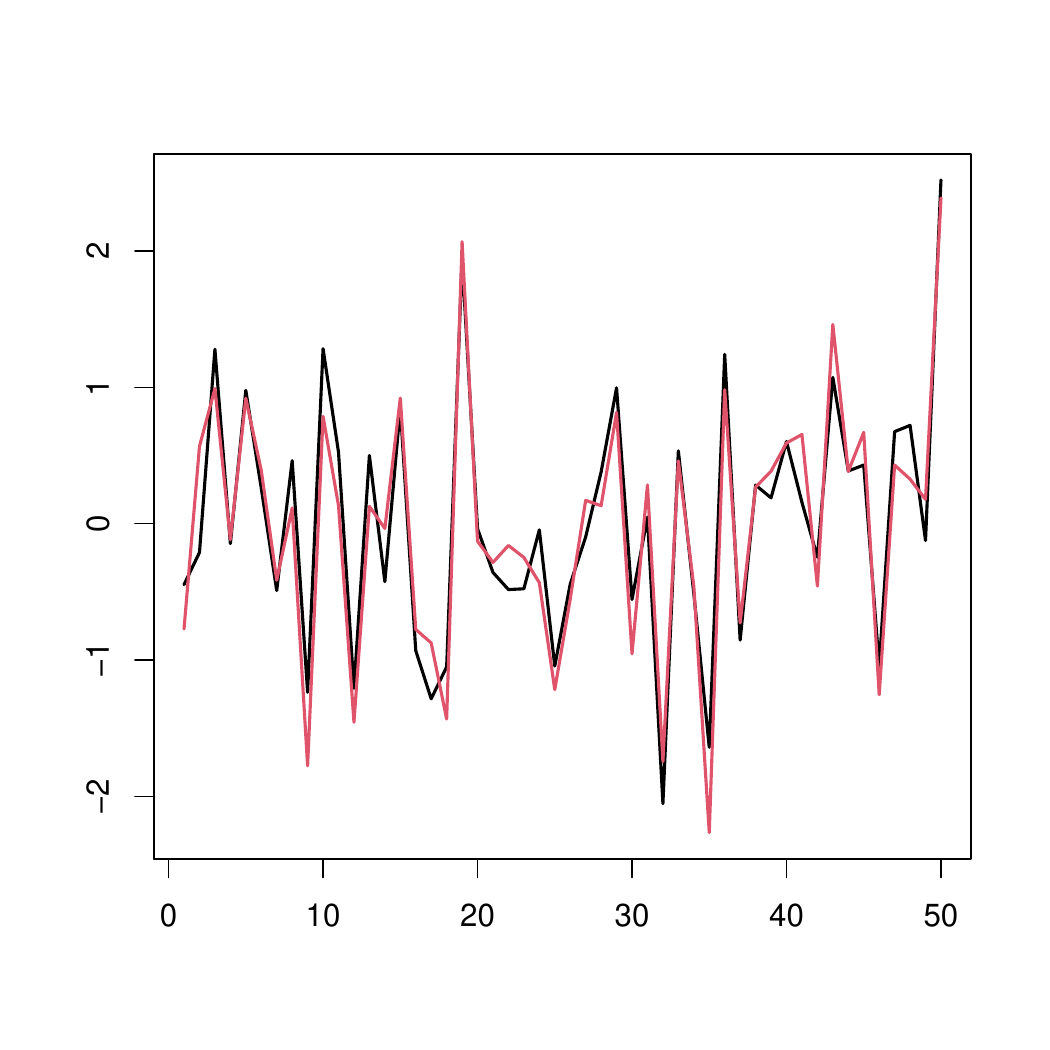}
}
\vspace{-0.1in}
\caption{ \footnotesize Closing prices time series for three choices of $ \sigma_{12}$ ($\mu_1=\mu_2=0, \sigma_{11}=\sigma_{22}=1$).}
\label{fig:AP_s12}
\end{figure}

To compare various likelihood-free estimators, we first generate synthetic data for 1\,000 trading days. For each day (of length $t=1$), we simulate the Brownian motion using 500 time steps to obtain the high, low and closing price data for each particular window. 
We first restrict our attention to the case of just two assets, which leaves us with $5$ parameters to estimate: $\mu_1, \mu_2$ and the upper triangular root of $\Sigma$, denoted with 
$
L={\scriptsize\left[\begin{array}{cc}
l_{11} & 0 \\
 l_{12}& l_{22}
\end{array}\right]}$.
We first illustrate how the covariance parameter $\sigma_{12}$  impacts the co-movement of   asset prices. Holding $\mu_1=\mu_2=0$ and $\sigma_{11}=\sigma_{22}=1$ fixed, we  plot  time series realizations of the closing prices for three particular choices of $\sigma_{12}$ in  \Cref{fig:AP_s12}. The patterns are as expected where the prices tend to co-fluctuate when $\sigma_{12}$ is closer to one. 
The success of our method depends on how well the discriminator can tell apart these trajectories.

\begin{figure}[!ht]
\centering
\includegraphics[width=0.85\textwidth]{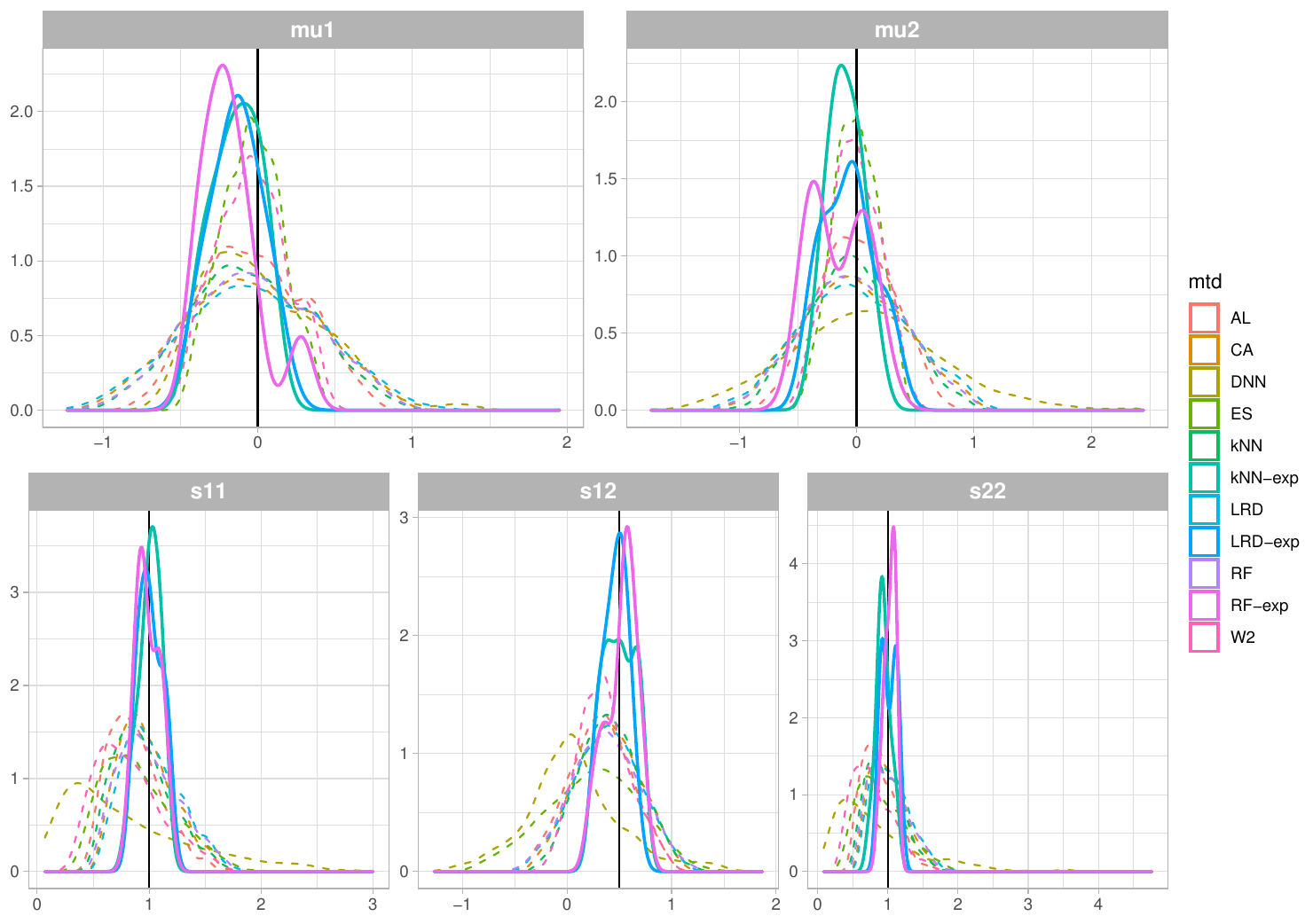}
\vspace{-0.1in}
\caption{  \footnotesize Posterior densities on simulated volatility data ($d=2$)}\label{fig:brownian_simu}
\end{figure}

For our simulation, we set $\mu=(0,0)'$ and  $\sigma_{11}=\sigma_{22}=1, \sigma_{12}=0.5$. We choose relatively non-informative prior hyper-parameters $\bmu_0=(0,0)',\lambda=1, \Phi=I_d$ and  $\nu=d$.
Posterior distributions reconstructed with different ABC methods are given in \Cref{fig:brownian_simu} and the averaged performance over 10 repetitions is summarized in \Cref{tab:brownian_simu}. We explore two discriminators built on the prices  and the quadratic/interaction terms of the prices: (1) a lasso classifier  with penalty term $\lambda$ selected from a 5-fold  cross-validation  (LRD); (2) a random forest classifier (RF). We find out that for the Lotka-Volterra model, the linear classifier performs better than the nonparametric random forest, as reflected in \Cref{fig:brownian_simu}. We observe a smaller estimation bias and the computation of LRD is much shorted than RF. Thus, we only include  LRD in the repetitions. It is clear that our exponential kernel methods place more mass around the true location of the parameters, and that the shape of kNN-exp is slightly less regular than LRD-exp. Although LRD-exp induces a larger bias in estimating the drift $(\mu_1, \mu_2)$, it does a better job at capturing the correct location of the volatilities.

\begin{table}[!ht]
\centering
\scalebox{0.78}{
\begin{tabular}{c c |  *{10}{c}}
\toprule
                 &  & LRD   & LRD-exp & kNN   & kNN-exp & ES    & CA    & W2    & AL    & SS    & DNN    \\
                  \midrule
\multirow{2}{*}{$\mu_1=0$} & MSE$(\times 10^{-4})$   & 2.004 & 2.113   & 0.638 & \bf{0.159}   & 1.484 & 1.497 & 0.170 & 0.628 & 1.216 & 1.838  \\
  & 95\% CI width   & 1.760 & \bf{0.331}   & 1.480 & 0.442   & 0.740 & 1.660 & 0.831 & 1.240 & 1.734 & 3.079    \vspace{0.1in}\\
 \multirow{2}{*}{ $\mu_2=0$} & MSE$(\times 10^{-4})$  & 0.734 & 7.439   & 0.363 & 0.259   & 0.050 & 0.142 & 0.277 & 0.089 & \bf{0.021} & 7.029  \\
  & 95\% CI width   & 1.628 & \bf{0.431}   & 1.435 & 0.538   & 0.737 & 1.564 & 0.813 & 1.247 & 1.569 & 2.527  \vspace{0.1in}\\
 \multirow{2}{*}{$\sigma_{11}=1$} & MSE$(\times 10^{-3})$  & 0.311 & 0.111   & 0.167 & 0.460   & 0.834 & \bf{0.002} & 2.634 & 1.669 & 1.404 & 3.502  \\
  & 95\% CI width   & 0.963 & 0.371   & 0.954 & \bf{0.329}   & 1.231 & 0.896 & 1.140 & 0.910 & 0.763 & 6.233  \vspace{0.1in}\\
  \multirow{2}{*}{$\sigma_{12}=0.5$} & MSE$(\times 10^{-3})$ & 1.035 & \bf{0.103}   & 0.427 & 0.107   & 6.657 & 0.861 & 1.528 & 3.337 & 0.109 & 18.427 \\
  & 95\% CI width   & 1.193 & 0.355   & 1.092 & \bf{0.324}   & 1.700 & 1.122 & 0.988 & 1.114 & 0.628 & 5.530 \vspace{0.1in} \\
 \multirow{2}{*}{$\sigma_{22}=1$} & MSE$(\times 10^{-3})$  & 0.416 & 0.304   & 0.124 & 0.018   & 0.776 & \bf{0.007} & 4.244 & 1.679 & 3.687 & 1.882  \\
  & 95\% CI width   & 0.959 & \bf{0.252}   & 0.974 & 0.314   & 1.274 & 0.920 & 1.157 & 0.922 & 0.783 & 6.693 \\
                  \bottomrule
\end{tabular}}
\vspace{-0.1in}
\caption{\label{tab:brownian_simu} \footnotesize Performance on the stock volatility estimation example, averaged over 10 repetitions, with top 1\% selected. All 95\% CIs have full coverage of the true parameters. The bold fonts mark the best model in each row.}
\vspace{-0.2in}
\end{table}

\subsection{Real Data Analysis}
Following the example in \citet{rogers2008estimating}, we examine a small data set of stock prices focusing on two stocks: Boeing (BA) and Proctor \& Gamble (PG). The prices were obtained from NYSE (Yahoo Finance), starting from 3rd January 2011 and consisting  of $1\,000$ trading days. Since the off-market  trades  follow a different mechanism than the market trade, we only model  price changes from the opening price to the closing price each day where the log price differences $X(t)$ are all computed based on the opening prices of that day.

\begin{figure}[!ht]
\centering
\includegraphics[width=0.8\textwidth]{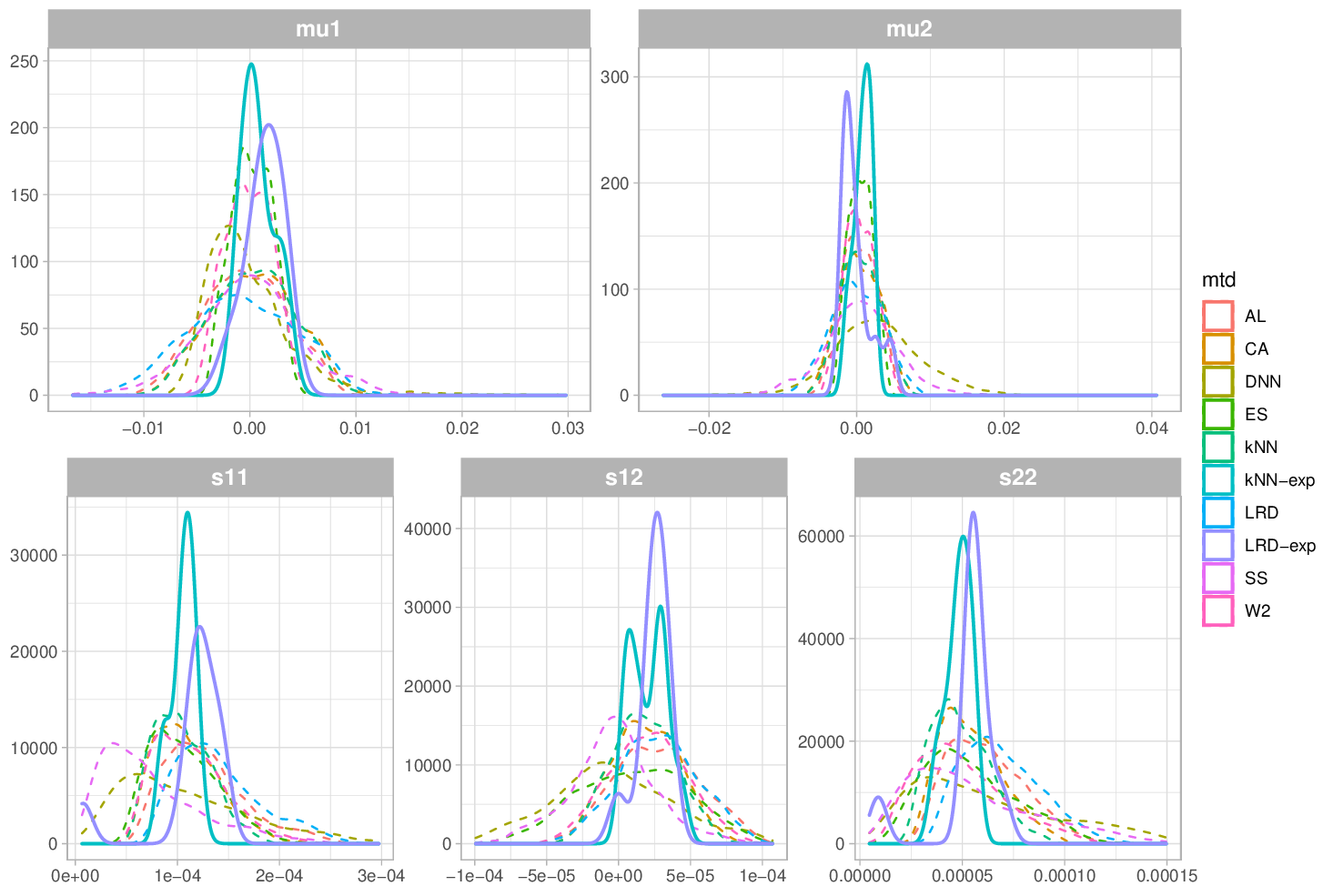}
\vspace{-0.1in}
\caption{ \footnotesize Posterior densities estimated from  log-prices of BA and PG}\label{fig:brownian_v2_d2}
\vspace{-0.1in}
\end{figure}

We observe that the fluctuations in the prices are much smaller than in our simulated time series and we thereby choose the hyperparameters based on the mean and covariance of the closing prices. \Cref{fig:brownian_v2_d2} gives the ABC posterior distributions estimated from different methods and the corresponding summaries of the distributions are provided in \Cref{tab:brownian_v2_d2}. We again observe that our methods with the exponential kernels return much narrower posterior distributions. The estimates on the volatilities are quite close except for the ABC with summary statistics (SS). The drifts of both BA and PG are not significantly different from zero.

\begin{table}[!ht]
\centering
\scalebox{0.79}{
\begin{tabular}{rr |  *{10}{c}  }
\toprule
                  &     & LRD        & LRD-exp & kNN    & kNN-exp & CA     & ES     &  W2     & AL     & DNN     & SS     \\
                  \midrule
\multirow{3}{*}{$\mu_1$
$(\times 10^{-3})$} & $\bar\mu_1$  &  -0.774  & 1.354   & 0.250  & 0.779   & 0.342  & 0.173  & -0.126 & -0.500 & -0.091  & -0.071 \\
&$l$& -10.380 & -1.981  & -6.816 & -1.145  & -7.099 & -3.201 & -3.987 & -7.603 & -5.415  & -9.532 \\
&$u$&  8.202   & 3.514   & 7.229  & 3.385   & 7.395  & 3.443  & 3.736  & 6.810  & 9.335   & 10.121 \\
     \vspace{-0.12in}      \\
\multirow{3}{*}{$\mu_2$ $(\times 10^{-3})$} & $\bar\mu_2$ & 0.009   & -0.039  & 0.491  & 0.822   & 0.445  & 0.545  & 0.453  & 0.316  & 2.793   & 0.450  \\
&$l$ & -6.591  & -1.718  & -4.329 & -1.406  & -4.604 & -2.364 & -3.491 & -4.103 & -8.913  & -9.831 \\
& $u$ &  6.531   & 4.563   & 5.503  & 2.166   & 5.538  & 3.616  & 4.450  & 4.690  & 14.701  & 10.171 \\
     \vspace{-0.12in}    \\
\multirow{3}{*}{$\sigma_{11}$$(\times 10^{-4})$} & $\bar \sigma_{11}$  & 1.406   & 1.142   & 1.049  & 1.054   & 1.101  & 1.070  & 1.130  & 1.273  & 1.224   & 0.738  \\
&$l$ & 0.824   & 1.142   & 0.633  & 0.844   & 0.634  & 0.577  & 0.626  & 0.689  & 0.243   & 0.155  \\
&$u$ &  2.362   & 1.142   & 1.640  & 1.202   & 1.819  & 1.818  & 1.938  & 2.259  & 3.616   & 1.976  \\
      \vspace{-0.12in}      \\
\multirow{3}{*}{$\sigma_{12}$$(\times 10^{-5})$}  & $\bar \sigma_{12}$ &  2.811   & 2.446   & 1.767  & 2.084   & 1.906  & 1.074  & 2.052  & 2.759  & -0.530  & -0.376 \\
&$l$ &  -2.379  & 2.446   & -2.359 & 2.084   & -2.390 & -6.152 & -3.137 & -2.656 & -25.243 & -6.660 \\
&$u$ &8.226   & 2.446   & 6.037  & 2.084   & 6.442  & 8.008  & 7.253  & 8.568  & 23.343  & 6.065  \\
      \vspace{-0.12in}     \\
\multirow{3}{*}{$\sigma_{22}$$(\times 10^{-4})$} & $\bar \sigma_{22}$ &  6.922   & 5.263   & 4.924  & 4.820   & 5.470  & 5.419  & 4.947  & 6.179  & 18.485  & 5.521  \\
&$l$ & 4.069   & 5.263   & 2.766  & 4.820   & 3.149  & 2.005  & 1.734  & 3.322  & 1.571   & 1.326  \\
&$u$ &11.049  & 5.263   & 8.020  & 4.820   & 8.826  & 10.258 & 9.844  & 10.223 & 93.218  & 12.852
  \\
                  \bottomrule
\end{tabular}}
\vspace{-0.1in}
\caption{\label{tab:brownian_v2_d2} \footnotesize Posterior estimates on analysis of BA and PG. For each parameter, we report three summary statistics, the posterior means, the lower limit of the 95\% CI intervals ($l$) and the upper limit of the 95\% CI intervals ($u$).}
\vspace{-0.2in}
\end{table}

\section{Discussion}
This paper develops an ABC variant using a classification-based KL estimator as a discrepancy measure. By deploying a flexible classifier, the empirical KL divergence can be estimated with a vanishing error.  In addition, inspired by the connection between the KL divergence and the log-likelihood ratio, we propose a scaled exponential kernel to aggregate ABC samples. This smoothing variant avoids the need  for  choosing the ad hoc threshold $\epsilon_n$ and fully utilizes  information returned from all ABC samples. Under mild conditions, we show that the posterior concentration rate of the accept-reject ABC depends on the estimation error $\delta_n$ and the accept-reject threshold $\epsilon_n$, while the rate of the smooth version depends on the estimation error $\delta_n$ and the contraction rate of the actual posterior distribution. Our methodology can also be related to many other likelihood-free inference methods, including ABC with Classification Accuracy \citep{gutmann2018likelihood}, Wasserstein distance ABC \citep{bernton2019approximate}, and Generalized Posteriors \citep{schmon2020generalized}. Our methods  coincide with the $c$-posterior \citep{miller2018robust}, which is robust to perturbations. In particular, the accept-reject ABC can be shown to be robust under model misspecification (see \Cref{sec:model_misspec}). In addition, the exponential kernel can be motivated as  an instantiation  of General Bayesian Inference (GBI) \citep{bissiri2016general} with  the KL estimator  as the  loss function. See \citet{thomas2019diagnosing} and \citet{thomas2020generalised} for examples of conducting robust inference using probabilistic classifiers under the generalized Bayes update setup. Along with our theoretical investigations, we demonstrate competitive performance of our methods on benchmark examples. Our theoretical analysis provides theoretical justifications for the method of \cite{gutmann2018likelihood}.

\section*{Acknowledgements}
The authors gratefully acknowledge the support from the James S.\ Kemper Faculty Fund, the Liew Family Junior Faculty Fellowship, and the Richard Rosett Faculty Fellowship at the University of Chicago Booth School of Business and the National Science Foundation (Grant No. NSF DMS-1944740).

\bibliography{abc-gan}

\clearpage

\appendix

\section{Convergence Rate of Estimation Errors with Neural Network Sieves} \label{sec:nn}

To develop a precise definition of the low underlying dimension $d^\ast$ described in Remark \ref{rmk:nn_sieve}, we borrow the smoothness notion of \citet{bauer2019deep}.

\begin{defn}[$(p,C)$-Smoothness]
Let $p=q+s$ for some $q\in\mathbb{N}_0$ and $0<s\leq 1$. A function $m:\mathbb{R}^d\to\mathbb{R}$ is called {\em $(p,C)$\hyp{}smooth} if for every $\alpha=(\alpha_1,\dots,\alpha_d)\in\mathbb{N}_0^d$ with $\sum_{j=1}^d\alpha_j=q$, the partial derivative $\frac{\partial^q m}{\partial x_1^{\alpha_1}\cdots\partial x_d^{\alpha_d}}$ exists and satisfies
\[
	\biggl|\frac{\partial^q m}{\partial x_1^{\alpha_1}\cdots\partial x_d^{\alpha_d}}(x)-\frac{\partial^q m}{\partial x_1^{\alpha_1}\cdots\partial x_d^{\alpha_d}}(z)\biggr|\leq C\|x-z\|^s
\]
for every $x,z\in\mathbb{R}^d$ where $\|\cdot\|$ denotes the Euclidean norm.
\end{defn}

With this, the nested composition structure is defined as follows.

\begin{defn}[Generalized Hierarchical Interaction Model] \label{defn:hier}
Let $d \in\mathbb{N}$, $d^\ast \in \{1,\dots,d\}$, and $m : \mathbb{R}^d \rightarrow \mathbb{R}$.
We say that the function $m$ satisfies a {\em generalized hierarchical interaction model of order $d^\ast$ and level $0$}, if there exist $a_1\in\mathbb{R}^d,\ldots, a_{d^\ast}\in\mathbb{R}^d$, and $f : \mathbb{R}^{d^\ast} \rightarrow \mathbb{R}$ such that
\[
m(x) = f(a_1' x,\ldots,a_{d^\ast}' x)
\]
for every $x \in \mathbb{R}^d$.
We say that $m$ satisfies a {\em generalized hierarchical interaction model of order $d^\ast$ and level $l + 1$ with $K$ components} if there exist $g_k: \mathbb{R}^{d^\ast} \rightarrow \mathbb{R}$ and $f_{1,k},\dots,f_{d^\ast,k}: \mathbb{R}^{d} \rightarrow \mathbb{R}$ $(k = 1,\ldots,K)$ such that $f_{1,k},\ldots,f_{d^\ast,k}$ $(k = 1,\dots,K)$ satisfy a generalized hierarchical model of order $d^\ast$ and level $l$ and 
\[
m(x) = \sum_{k=1}^K g_k(f_{1,k}(x),\dots,f_{d^\ast,k}(x))
\]
for every $x \in \mathbb{R}^d$.
We say that the generalized hierarchical interaction model is {\em $(p,C)$\hyp{}smooth} if all functions occurring in its definition are $(p,C)$\hyp{}smooth.
\end{defn}

For example, a conditional binary choice model satisfies a generalized hierarchical interaction model of order $d^\ast\leq 3$ and level $0$, irrespectively of the dimension of the covariates.

\begin{exa}[Binary Choice Model] \label{exa:binary}
Let $y_i=\mathbb{I}\{x_i'\alpha+\varepsilon_i>0\}$, $\varepsilon_i\sim P_\varepsilon$, be the true DGP and $y_i^\beta=\mathbb{I}\{x_i'\beta+\tilde{\varepsilon}_i>0\}$, $\tilde{\varepsilon}_i\sim\tilde{P}_\varepsilon$, be the generative model.
Then,
\[
	\log\frac{p_0(y,x)}{p_\theta(y,x)}=y\log\frac{1-P_\varepsilon(-x'\alpha)}{1-\tilde{P}_\varepsilon(-x'\beta)}+(1-y)\log\frac{P_\varepsilon(-x'\alpha)}{\tilde{P}_\varepsilon(-x'\beta)}.
\]
Therefore, we can write this as $f(a_1'z,a_2'z,a_3'z)$ where $z=(y,x')'$, $a_1=(1,0,\dots,0)'$, $a_2=(0,-\alpha')'$, $a_3=(0,-\beta')'$, and \(
	f(y,x_1,x_2)=y[\log(1-P_\varepsilon(x_1))-\log(1-\tilde{P}_\varepsilon(x_2))]+(1-y)[\log P_\varepsilon(x_1)-\log\tilde{P}_\varepsilon(x_2)]
\).
If $P_\varepsilon$ and $\tilde{P}_\varepsilon$ are logistic distributions, then $f$ is $(p,C)$\hyp{}smooth for $(p,C)=(8,1)$ or for $(p,C)=(11,10)$, for example.
Therefore, it satisfies the generalized hierarchical interaction model of order $d^\ast=3$ and level $l=0$.
\end{exa}

\begin{exa}[Diffusion Process] \label{exa:brownian}
Let $X_{i,t/d}$, $t=1,\dots,d$ be discretely sampled observations of a Brownian motion $\d X_{it}=\mu \d t+\sigma \d W_{it}$ with $X_{i0}=0$, where $W_{it}$ is a standard Brownian motion independent across $i$.
Let the generative model be $X_{i,t/d}^\theta$, $t=1,\dots,d$ from $X_{it}^\theta=m\d t+s\d \tilde{W}_{it}$ and $\theta=(m,s)$.
Then, the log likelihood ratio is
\begin{multline*}
	\log\frac{p_0(x_{1/d},\dots,x_1)}{p_\theta(x_{1/d},\dots,x_1)}
	=d\log\frac{s}{\sigma}-\frac{d}{2}\biggl(\frac{1}{\sigma^2}-\frac{1}{s^2}\biggr)\sum_{j=1}^d(x_{j/d}-x_{(j-1)/d})^2\\
	+\biggl(\frac{\mu}{\sigma^2}-\frac{m}{s^2}\biggr)x_{1}-\frac{1}{2}\biggl(\frac{\mu^2}{\sigma^2}-\frac{m^2}{s^2}\biggr).
\end{multline*}
Letting $z=((x_{1/d}-x_{0})^2,\dots,(x_{1}-x_{(d-1)/d})^2,x_{1})'$, we can write this as $f(a'z)$ where $a=(-\frac{d}{2}(\frac{1}{\sigma^2}-\frac{1}{s^2}),\dots,-\frac{d}{2}(\frac{1}{\sigma^2}-\frac{1}{s^2}),\frac{\mu}{\sigma^2}-\frac{m}{s^2})'$ and
\(
	f(y)=d\log\frac{s}{\sigma}+y-\frac{1}{2}\bigl(\frac{\mu^2}{\sigma^2}-\frac{m^2}{s^2}\bigr)
\).
Then, $f$ is $(p,C)$\hyp{}smooth with $(p,C)=(\infty,1)$, and the log likelihood ratio satisfies the hierarchical model with $d^\ast=1$ and level $l=0$.
\end{exa}

Next, we define the configuration of the neural network appropriate for estimating a generalized hierarchical interaction model.

\begin{defn}[Hierarchical Neural Network]
Let $\sigma:\mathbb{R}\to\mathbb{R}$ be a $q$\hyp{}admissible activation function.
For $M^\ast\in\mathbb{N}$, $d\in\mathbb{N}$, $d^\ast\in\{1,\dots,d\}$, and $\alpha>0$, let $\mathcal{F}_{M^\ast,d^\ast,d,\alpha}$ be the class of functions $f:\mathbb{R}^d\to\mathbb{R}$ such that
\[
	f(x)=\sum_{i=1}^{M^\ast}\mu_i\sigma\Biggl(\sum_{j=1}^{4d^\ast}\lambda_{i,j}\sigma\Biggl(\sum_{v=1}^d\theta_{i,j,v}x_v+\theta_{i,j,0}\Biggr)+\lambda_{i,0}\Biggr)+\mu_0
\]
for some $\mu_i,\lambda_{i,j},\theta_{i,j,v}\in\mathbb{R}$, where $|\mu_i|\leq\alpha$, $|\lambda_{i,j}|\leq\alpha$, and $|\theta_{i,j,v}|\leq\alpha$.
For $l=0$, define the set of neural networks with two hidden layers by $\mathcal{H}_{M^\ast,d^\ast,d,\alpha}^{(0)}=\mathcal{F}_{M^\ast,d^\ast,d,\alpha}$; for $l>0$, define the set of neural networks with $2l+2$ hidden layers by
\begin{multline*}
	\mathcal{H}_{M^\ast,d^\ast,d,\alpha}^{(l)}=\\
	\Biggl\{h:\mathbb{R}^d\to\mathbb{R}:h(x)=\sum_{k=1}^K g_k(f_{1,k}(x),\dots,f_{d^\ast,k}(x)),\
	g_k\in\mathcal{F}_{M^\ast,d^\ast,d^\ast,\alpha},\ f_{j,k}\in\mathcal{H}^{(l-1)}\Biggr\}.
\end{multline*}
\end{defn}

With these definitions, the convergence rate of the neural network sieve is characterized as follows.
If the log likelihood ratio satisfies the generalized hierarchical model of order $d^\ast$ and the corresponding neural network sieve is used, the convergence rate of the discriminator depends only on $d^\ast$, not on $d$.
The precise assumption is formulated as follows.

\begin{assumption}[Neural Network Discriminator] \label{asm:nn}
Let $P_0$ and $P_\theta$ have subexponential tails and finite first moments.\footnote{We say that $P$ on $\mathbb{R}^d$ has {\em subexponential tails} if $\log P(\|X\|_\infty>a)\lesssim-a$ for large $a$.}
Let $\log(p_0/p_\theta)$ satisfy a $(p,C)$\hyp{}smooth generalized hierarchical interaction model of order $d^\ast$ and finite level $l$ with $K$ components for $p=q+s$, $q\in\mathbb{N}_0$, and $s\in(0,1]$.
Let $\mathcal{H}_{M^\ast,d^\ast,d,\alpha}^{(l)}$ be the class of neural networks with the Lipschitz activation function with Lipschitz constant $1$
for
\begin{gather*}
	M_\ast=\biggl\lceil{d^\ast+q\choose d^\ast}(q+1)\biggl(\biggl[\frac{(\log\delta_n)^{2(2q+3)}}{\delta_n}\biggr]^{\frac{1}{p}}+1\biggr)^{d^\ast}\biggr\rceil,\\
	\alpha=\biggl[\frac{(\log\delta_n)^{2(2q+3)}}{\delta_n}\biggr]^{\frac{d^\ast+p(2q+3)+1}{p}}\frac{\log n}{\delta_n^2},
\end{gather*}
and
\(
	\delta_n=[(\log n)^{\frac{p+2d^\ast(2q+3)}{p}}/n]^{\frac{p}{2p+d^\ast}}
\).
Denote by $\mathcal{D}_n=\{\Lambda(f):f\in\mathcal{H}_{M^\ast,d^\ast,d,\alpha}^{(l)}\}$ the sieve of neural network discriminators for the standard logistic cdf $\Lambda$.
\end{assumption}

\Cref{asm:nn} gives a sufficient condition for the entropy condition in \Cref{ass:entropy}.
With this, we obtain the ``classification counterpart'' of \citet[Theorem 1]{bauer2019deep} as below.

\begin{theorem}[{\citealp[Proposition S.3]{kaji2020adversarial}}] \label{thm:nnrate}
Suppose \Cref{asm:nn} holds.
If $n/m$ converges and an estimator $\hat D^\theta_{n,m}$ exists that satisfies $\M_{n,m}(\hat D^\theta_{n,m})\geq  \M_{n,m}^\theta(D_\theta)-O_P(\delta_n^2)$ for the $\delta_n$ in \Cref{asm:nn},
then $d_\theta(\hat{D}^\theta_{n,m},D_\theta)=O_P^\ast(\delta_n)$.
\end{theorem}

This theorem, combined with the explicit expression of $\delta_n$ in \Cref{asm:nn}, tells that if $d^\ast<2p$, we have $\delta_n=o_P(n^{-1/4})$, which is often the desired rate for the nonparametric estimator of a nuisance parameter.
For the binary choice model in \Cref{exa:binary}, the discriminator converges much faster than $n^{-1/4}$.

\begin{exa}[continues=exa:binary]
The binary choice model with logistic errors satisfies a $(p,C)$\hyp{}smooth hierarchical model with $(p,C)=(8,1)$ and $d^\ast=3$.
We can substitute these numbers (and $q=p-1$) into $\delta_n$ in \Cref{asm:nn} and obtain
\(
	\delta_n=\bigl(\frac{(\log n)^{13.75}}{n}\bigr)^{8/19}\lesssim n^{-2/5}.
\)
\end{exa}

\begin{exa}[continues=exa:brownian]
The discretely sampled Brownian motion model satisfies a $(p,C)$\hyp{}smooth hierarchical model with $d^\ast=1$, $C=1$, and arbitrarily large $p$.
Therefore, $\delta_n$ can be arbitrarily close to $n^{-1/2}$, however large the sampling frequency is.
\end{exa}

When there is no low\hyp{}dimensional structure, \Cref{defn:hier} simply reduces to the smoothness of $m$ and $d^\ast$ is equal to $d$.
Therefore, the convergence rate in \Cref{thm:nnrate} reduces to the traditionally proven rate that deteriorates quickly with $d$.

We note that the hierarchical structure is not the only way to proving the superior adaptivity of a neural network discriminator.
Similar results can possibly be deduced with other smoothness assumptions and network configurations, such as \citet{schmidt2020nonparametric} and \citet{yarotsky2017error}.

\section{Frequentist's Analysis on the Exponential Kernel}\label{sec:exp_kernel_post}
We thus study the concentration in terms of a KL neighborhood around $Q^{(n)}_{\theta^*}$ defined as
\begin{equation}\label{eq:B_tildeP}
B(\epsilon, Q^{(n)}_{\theta^*}; P_0^{(n)}) = \{Q^{(n)}_{\theta^*} \in \mathcal{Q}^{(n)}: \tilde K(\theta^*,\theta)\leq n\epsilon^2, \tilde V(\theta^*,\theta)\leq n \epsilon^2\},
\end{equation}
where $\tilde K(\theta^*,\theta) \equiv P_0^{(n)}\log \frac{q_{\theta^*}^{(n)}}{q_{\theta}^{(n)}}$ and $\tilde V(\theta^*,\theta)\equiv P_0^{(n)} \abs{\log \frac{q_{\theta^*}^{(n)}}{q_{\theta}^{(n)}} -\tilde K(\theta^*,\theta)}^2$.
 
  The following corollary is directly adopted from {Theorem 5.1} of \citet{kaji2021mh}. 
 
\begin{corollary}
 Denote with $\wt Q_\theta^{(n)}$ a measure defined through $\d \wt Q_\theta^{(n)}= ({p_0^{(n)}}/{q_{\theta^*}^{(n)}})\d P_\theta^{(n)}$ and let $d(\cdot, \cdot)$  be a semi-metric on $ \mP^{(n)}$. Suppose that there exists a sequence $\epsilon_n>0$ satisfying $\epsilon_n \to 0$ and $n\epsilon_n^2\to \infty$ such that for every $\epsilon>\epsilon_n$ there exists a test $\phi_n$ (depending on $\epsilon$) such that for every $J \in \N_0$
\begin{equation}
P_0^{(n)}\phi_n \lesssim e^{-n\epsilon^2/4} \quad \text{ and } \sup_{Q_\theta^{(n)}: d(Q_\theta^{(n)}, P_{\theta^*}^{(n)})> J\epsilon} \wt Q_\theta^{(n)}(1-\phi_n)\leq e^{-n J^2\epsilon^2/4}.
\end{equation}
Let $B(\epsilon,Q^{(n)}_{\theta^*}; P_0^{(n)})$ be as in \eqref{eq:B_tildeP} and let $\tilde \Pi_n(\theta)$ be a prior distribution with a density $\tilde \pi(\theta)\propto C_\theta\pi(\theta)$ with $C_\theta$ as in \eqref{eq:tilde_p}. Assume that there exists a constant $L>0$ such that, for all $n$ and $j \in \N$,
\begin{equation}
\frac{\tilde \Pi_n\left( \theta\in \Theta: j\epsilon_n < d(Q_\theta^{(n)}, P_{\theta^*}^{(n)}) \leq (j+1)\epsilon_n \right)}{\tilde \Pi_n \left(B(\epsilon_n, Q^{(n)}_{\theta^*}; P_0^{(n)}) \right)}\leq e^{n\epsilon_n^2 j^2/8}.
\end{equation}
Then for every sufficiently large constant $M$, as $n\to \infty$,
\begin{equation}
P_0^{(n)}\Pi^{EK}\left(Q_\theta^{(n)}: d(Q_\theta^{(n)}, P_{\theta^*}^{(n)}) \geq M \epsilon_n\mid X^{(n)}\right)\to 0.
\end{equation} 
 \end{corollary}
 
 Next, we want to show the shape of the posterior is actually asymptotically gaussian around $\theta^*$. The following corollary follows from Theorem 2.1 of \citet{kleijn2012bernstein} and {Lemma 8.1} of \citet{kaji2021mh}. 

\begin{corollary} (Bernstein von-Mises) Assume that the posterior \eqref{eq:tilde_p} concentrates around $\theta^*$ at the rate $\epsilon_n^*$ and that for every compact $K\in \R^d$
\begin{equation}
\sup_{h\in K} \abs{\log \frac{q_{\theta^*+\epsilon_n^* h}^{(n)}(\X)}{q_{\theta^*}^{(n)}(\X)}- h' \tilde V_{\theta^*}\tilde\delta_{n, \theta^*}-\frac{1}{2}h'  \tilde V_{\theta^*} h}\to 0
\end{equation}
for some random vector $\tilde \delta_{n, \theta^*}$ and a non-singular matrix $\tilde V_{\theta^*}$. Then the approximated posterior $\Pi^{EK}(\cdot)$ converges to a sequence of normal distributions in total variation at the rate $\epsilon_n^*$, i.e.
\begin{equation}
\sup_B \abs{\Pi^{EK}\left({\epsilon_n^*}^{-1}(\theta-\theta^*\right)\in B\mid \X)- N_{\tilde \delta_{n, \theta^*},\tilde V_{\theta^*}^{-1}(B)}}\to 0 \quad \text{ in } P_0^{(n)}-probability.
\end{equation}
\end{corollary}

\section{ABC versus MHC}\label{sec:mhc_compare}
One may wonder how  the performance of the exponentially weighted ABC (Algorithm \ref{alg:abc_exp}) differs from the sequential MHC version \citet{kaji2021mh}. The target distribution is the same but one may expect convergence issues of the sequential version.  The major difference are: (1) ABC does not  need a proposal distribution for state changes; (2)  ABC is parallelizable while MHC can  only be computed in  a sequential fashion; (3)  ABC is  less sensitive to how the $\theta$'s
are initialized. For the scenarios where the likelihood is  spiky (as in \Cref{fig:lv_heatmap}),  priors can be adjusted for ABC focus on the most promising region \citep{blum2018regression}. Meanwhile, the performance of MHC can be largely influenced by the initializations. When the acceptance ratio is low (or the effective  sample size is small), tuning can be difficult. 

\begin{figure}[!ht]
\centering
\includegraphics[width=0.9\textwidth]{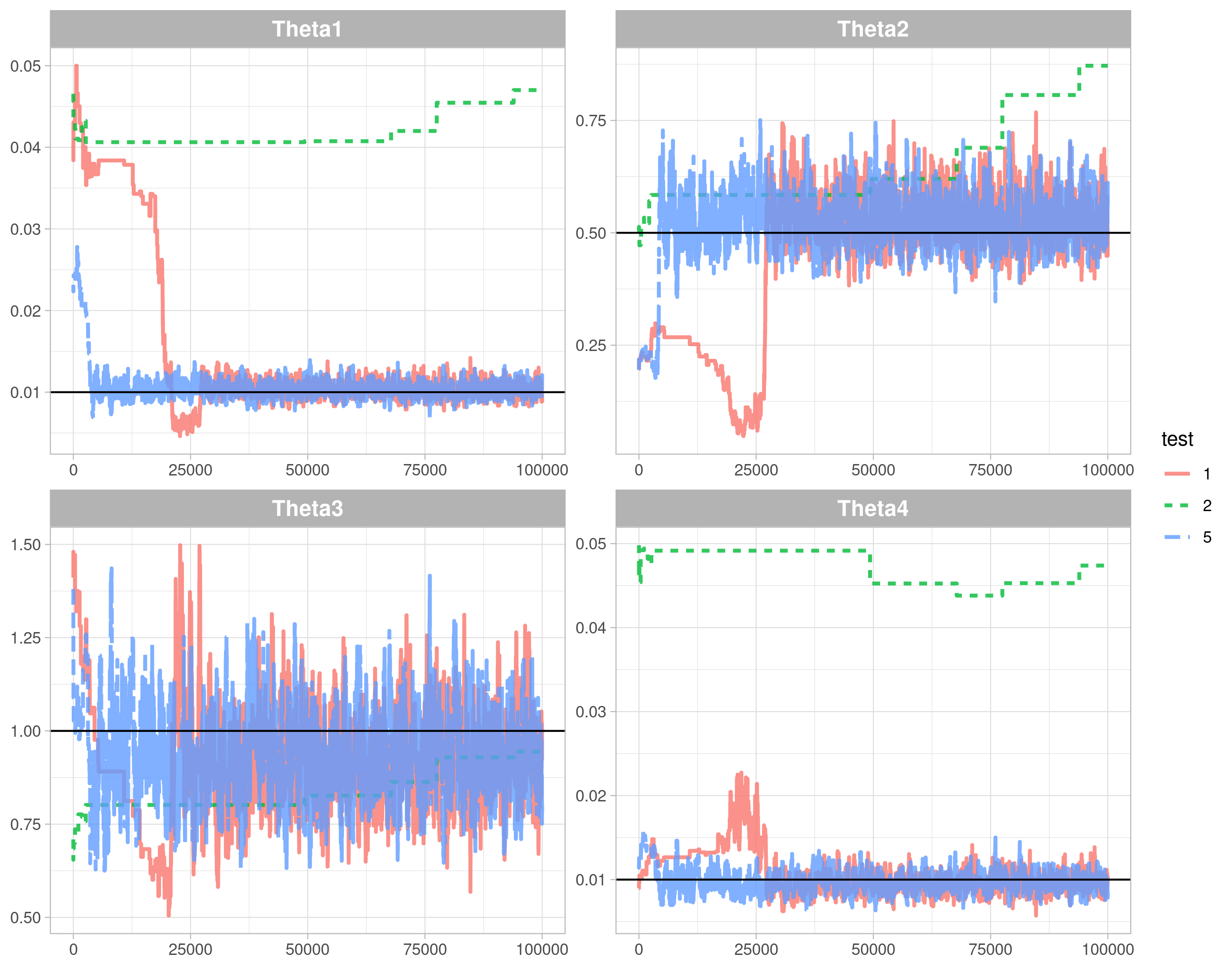}
\vspace{-0.1in}
\caption{Trace plots of MHC from different initializations.  The tests 1, 2 and 5 are corresponding  to the cases \#1, \#2 and \#5 in \Cref{tab:mhc_comp}. The black solid horizontal lines mark the locations of the true parameters.}\label{fig:mhc_comp}
\vspace{-0.2in}
\end{figure}

We use the Lotka-Volterra model to illustrate how MHC could suffer when the initialization point lands in a non-informative region. We randomly  sampled a few different initialization  points from the prior, and run the MHC for 100\,000 iterations with the same proposal distribution. The posterior means, effective sample sizes and the acceptance rates are reported in \Cref{tab:mhc_comp} and trace-plots are provided in \Cref{fig:mhc_comp}.

\begin{table}[!ht]
\centering
\begin{tabular}{l |c | l *{4}{r}}
\toprule
& \%accept && $\theta_1$ & $\theta_2$ &$\theta_3$ &$\theta_4$\\
\midrule
\multirow{3}{*}{\#1} & \multirow{3}{*}{11.934\%} & initial         & 0.041  & 0.197   & 1.411   & 0.009   \\
         &         & posterior mean & 0.013  & 0.463   & 0.900   & 0.011   \\
           &       & ESS            & 3.483  & 21.114  & 109.692 & 36.124  \\
                  \midrule
                  
\multirow{3}{*}{\#2} & \multirow{3}{*}{0.013\%} & initial        & 0.047  & 0.508   & 0.709   & 0.050   \\
           &       & posterior mean & 0.042  & 0.663   & 0.846   & 0.047   \\
             &     & ESS            & 2.980  & 2.474   & 4.074   & 2.508   \\                         \midrule
\multirow{3}{*}{\#3} & \multirow{3}{*}{0.019\%} & initial        & 0.016  & 0.355   & 0.761   & 0.042   \\
            &      & posterior mean & 0.016  & 0.417   & 0.582   & 0.041   \\
             &     & ESS            & 3.537  & 49.821  & 6.926   & 49.189  \\                         \midrule
\multirow{3}{*}{\#4} & \multirow{3}{*}{0.035\%} & initial        & 0.031  & 0.710   & 0.777   & 0.033   \\
            &      & posterior mean & 0.026  & 0.477   & 1.183   & 0.030   \\
              &    & ESS            & 2.012  & 1.861   & 1.069   & 4.323   \\
                         \midrule
\multirow{3}{*}{\#5} & \multirow{3}{*}{14.828\%} & initial        & 0.022  & 0.208   & 1.398   & 0.012   \\
             &     & posterior mean & 0.010  & 0.527   & 0.935   & 0.010   \\
               &   & ESS            & 45.167 & 123.381 & 236.331 & 164.153 \\
                  \bottomrule
\end{tabular}
\vspace{-0.1in}
\caption{\label{tab:mhc_comp}  \footnotesize MHC from different initializations for the Lotka-Volterra model with $\theta_0=(0.01, 0.5, 1, 0.01)'$. For each case, we report the initializations (sampled randomly from  the prior), the acceptance ratio  and the effective sample sizes (ESS) over a chain of 100\,000 iterations, the posterior mean with the first 10\,000 iterations treated as burn-ins.}
\vspace{-0.2in}
\end{table}

Obviously, the performance and mixing  speed of MHC is largely contingent on the starting point for this model. Even  for the case \#1, whose acceptance ratio is relatively high, it took more than 25\,000 iterations  for the chain to  slowly  navigate itself towards  the informative region,  and the effective sample sizes are still small since  the  MHC samples are highly correlated.  For the case \#3, the convergence of the  chain seems hopeless after 100\,000  iterations. The  ABC method,  on the contrary, is robust to the initializations since the parameters are sampled completely independently identically from the prior,  as shown in the  repetition experiment in \Cref{tab:lv_comp}.

\section{Other GAN-style Discrepancy Metrics}\label{sec:other_gans}

There are  many other divergences that can be estimated using different configurations of \eqref{eq:bce_obj}. Given  a set $\mF$ of functions  from $\mX$ to $\R$, we can define 
\begin{equation}\label{eq:ipm}
d_\mF(p_0, p_\theta)=\sup_{f\in\mF} \P_n f(X_i)-\P_m^\theta f(\tilde X^\theta_i),
\end{equation}
as the Integral Probability Metric (IPM) \citep{muller1997integral} associated with the function class $\mF$. \citet{arjovsky2017wasserstein} show that different choices of the function class $\mF$ can dramatically change the topology, as we discuss below,  and the regularity of $d_\mF(p_0, p_\theta)$ as a loss function. Several widely used discrepancy metrics  can be obtained from IPMs. For example, the 1-Wasserstein distance or the Earth-Mover distance \citep{villani2009optimal} are obtained when $\mF$ is the set of 1-Lipschitz functions. \citet{arjovsky2017wasserstein} proposed Wasserstein GANs, which approximate the 1-Wasserstein distance with gradient-clipped neural networks.  Another example is the  Total Variation (TV) distance,  obtained when  $\mF$ is the set of all measurable functions bounded between -1 and 1. Next, the Maximum Mean (MM) discrepancy is obtained  when $\mF=\{f\in \mH: \norm{f}_\infty\leq 1\}$ for $\mH$ some Reproducing Kernel Hilbert Space (RKHS) associated with a given kernel $k: \mX \times \mX \to \R$. Our classification-based GAN framework thus provides an alternative route towards  implementing ABCs with the Wasserstein distance \citep{bernton2019approximate} as well as MM \citep{park2016k2}.

One important aspect  of these various metrics is  that they induce topologies of different strengths. A divergence makes it easier for a sequence of distributions to converge when it induces a weaker topology. \citet{arjovsky2017wasserstein} show that KL induces the strongest one, followed by JS and TV,   with Wasserstein distance being the weakest. However, the computation costs of the Wasserstein distance are high, i.e. $O((n+m)^3\log (n+m))$ to calculate  it exactly \citep{burkard2012assignment}. The GAN version approximation also requires  weight clipping to enforce the Lipschitz constraint which can be problematic to implement. Our KL estimator, on the other hand, is more scalable and places no constraints on the discriminator. {The neural network discriminator can converge faster than traditional nonparametric methods if some low-dimensionality assumption is satisfied \citep{kaji2020adversarial, bauer2019deep}}. In addition, the KL estimator has a natural interpretation through its  link to  the log-likelihood ratio. 
The KL divergence appears in many theoretical results in Bayesian non-parametrics. For example, Schwartz's Theorem \citep{schwartz1965bayes} requires that, for posterior consistency,  the prior should assign positive probability to any KL neighborhood of $p_0$. By adopting the KL estimator within our framework, we can relate ABC to classical  posterior convergence rate results.

\section{Simulations: Univariate g-and-k Distributions}\label{sec:gk}
Another classical example in the ABC literature is the univariate $g$-and-$k$ distribution. It is defined implicitly by its inverse distribution function
\[
F^{-1}(x)=A+B\Big[1+0.8\frac{1-\e^{-gz_x}}{1+\e^{-gz_x}}\Big](1+z_x^2)^k z_x,
\]
where $z_x$ is the $x$-th quantile of the standard normal distribution, and parameters $A,B,g,k$ are related to location, scale, skewness and kurtosis. The probability density function has no analytical form but can be numerically calculated with high precision since it only involves one-dimensional inversions and differentiations of the quantile function. Therefore, Bayesian inference can be carried out.

We generate $n=500$ observations from the model by letting $A=3, B=1, g=2, k=0.5$.  Among the four parameters, $g$ is the hardest to identify, as observed in previous analyses. From a pilot run,
narrow the range of our uniform priors to  $A\sim U([2,4]), B\sim U([0,0.25]), g\sim U([0,10])$ and  $k\sim  U([0,2.5])$.
  
For our methods,  two classifiers are explored for the model, both built on observed data $\X$ and the  higher-order terms $\X^2$ and $\X^3$.\footnote{We are trying to manually create more features for this univariate data. We also explore even higher order terms and find the including up to $\X^3$ works the best in practice.} The first one is a random forests (RF) classifier, and the second one is a neural network (NND) with two hidden layers with 10 nodes each. The first layer is activated with the rectified linear unit (ReLU) function, while the second layer is equipped with a hyperbolic tangent function (tanh).

\begin{figure}[!ht]
\centering
\includegraphics[width=\textwidth]{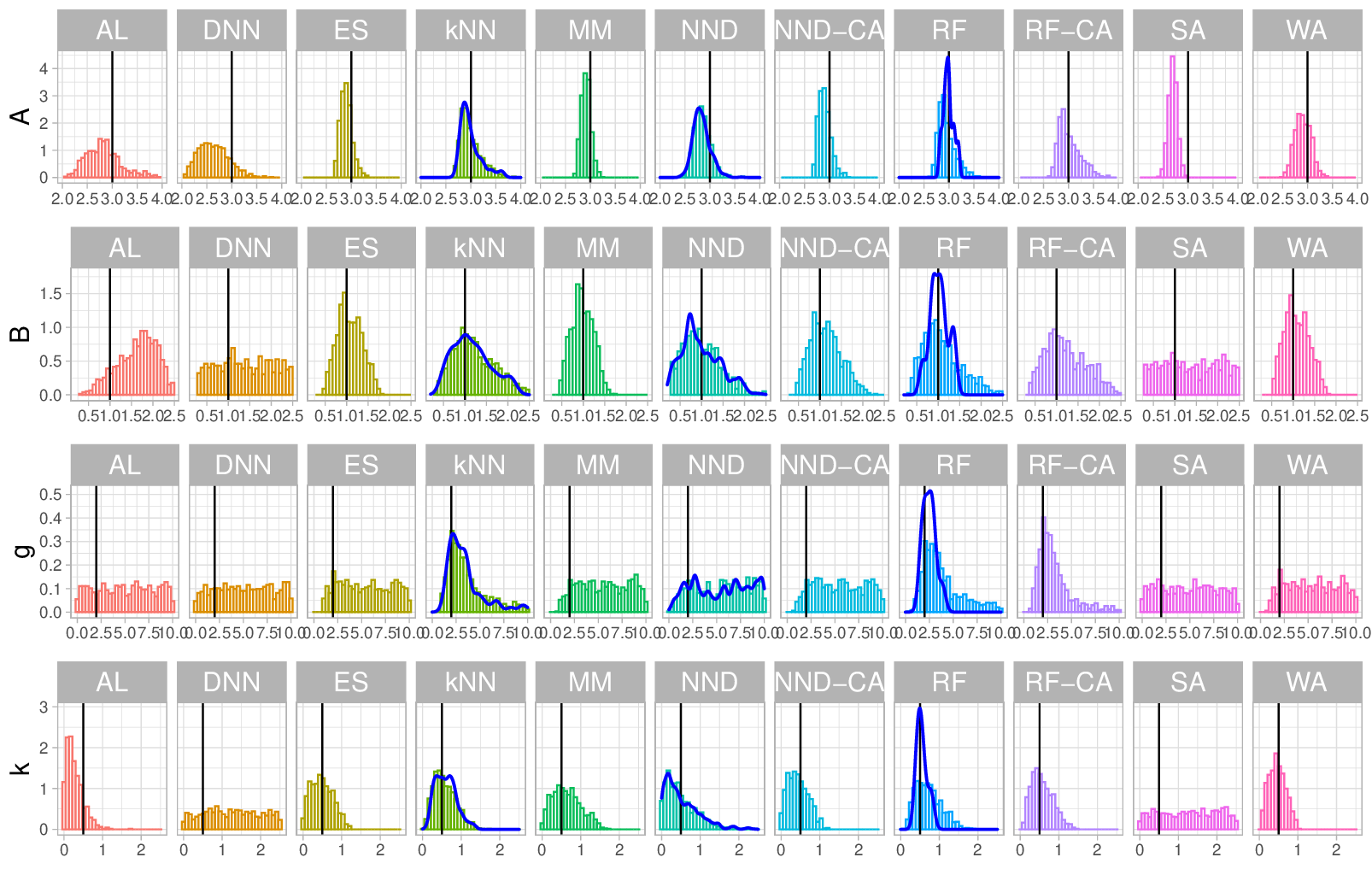}
\vspace{-0.1in}
\caption{Posterior densities for the $g$-and-$k$ distributions $(A=3, B=1, g=2, k=0.5)$.  The black vertical lines mark the true  parameter values. The blue curves in kNN, NND, RF boxes represent the weighted density calculated from the exponential kernel. The ABC posteriors were plotted with the top 1\% of $10^5$ samples.} \label{fig:gk_post}
\vspace{-0.2in}
\end{figure}

We compare the posteriors obtained from different ABC methods in  \Cref{fig:gk_post}. We can see that, indeed, the parameter $g$ is the most difficult one to estimate. While the majority of methods return a flat posterior,  KLC and kNN with the exponential kernel still place most of the posterior mass around the true value. AL and SA misidentify the location of one or two parameters.  While CA and KLC give relatively similar ranking on the ABC sampled parameters,   KLC with the exponential kernel  can utilize all the samples and re-adjust the shape according to weights. In addition, RF-exp  returns more spiky posteriors than kNN-exp. This is also reflected in the performance summaries in \Cref{tab:gk_comp} which show that RF-exp has a narrower CI  than kNN-exp on average. The NND method may not  ideal for this  univaraite $g$-and-$k$ distribution.

\begin{table}[!ht]
\centering
\scalebox{0.75}{
\begin{tabular}{l | *{2}{c} | *{2}{c}  | *{2}{c} | *{2}{c} }
             \toprule
             & \multicolumn{2}{c|}{$A=3$}                             & \multicolumn{2}{c|}{$B=1$}                             & \multicolumn{2}{c|}{$g=2$}                             & \multicolumn{2}{c}{$k=0.5$}         \\                    \
   {Method}     &{$(\hat A-A)^2$} & {95\% CI} width & {$(\hat B-B)^2$} & {95\% CI} width & {$(\hat g-g)^2$}  &{ 95\% CI} width & {$(\hat k-k)^2$}  &{95 \%CI} width \\ 
   \midrule
   RF     & 0.0036 & 0.758 & 0.0316 & 1.699 & 2.545  & 7.748 & 0.02550 & 1.170 \\
RF-exp & \bf{0.0016} & 0.394 & 0.0281 & \bf{0.913} & \bf{0.131}  & \bf{1.373} & 0.00541 & \bf{0.486 (0.9)}\\
NND   & 0.0143 &  0.535 & 0.0369 & 1.528 & 12.840 &  8.356 & 0.00341 & 1.044 \\
NND-exp & 0.0156 &  0.455 & 0.0339 & 1.229 & 12.516 & 8.191 & 0.00800 & 0.784 \\
kNN      & 0.0024 & 0.843 & 0.0486 &  1.770 & 2.695  &  7.558 & 0.00421 & 1.086 \\
kNN-exp  & 0.0032 &  0.610 & 0.0410 & 1.333 & 1.673  &  4.109 & 0.00480 & 0.805 \\
\vspace{-0.14in} &&&&&&&&\\
AL      & 0.0176 &  1.325 & 0.2211 &  1.593 & 8.704  & 9.480 & 0.04451 & 0.835 \\
NND-CA      & 0.0125 &  0.503 & 0.0466 & 1.568 & 13.077 & 8.162 & 0.00574 &  1.080 \\
RF-CA  & \bf{0.0016} & 0.843 & 0.0414 & 1.719 & 1.293  & 7.184 & \bf{0.00286} & 1.022 \\
SA      & 0.1191 &  0.424 (0) & 0.0598 &  2.379 & 8.800  &  9.478 & 0.56399 & 2.379 \\
WA      & 0.0145 & 0.627 & 0.0240 & 1.034 & 11.621 & 8.566 & 0.01060 & \bf{0.776}\\
DNN    & 0.0560 & 1.166 & 0.0622 & 2.373 & 8.976  & 9.456 & 0.56306 & 2.374 \\
ES     & 0.0140 & 0.487 & 0.0154 & 1.023 & 11.015 & 8.497 & 0.00298 & 0.957 \\
MM     & 0.0120 & \bf{0.386} & \bf{0.0102} & 0.916 & 13.379 & 8.340 & 0.01459 & 1.248 \\
\bottomrule
\end{tabular}}
\vspace{-0.1in}
\caption{\label{tab:gk_comp} \footnotesize Performance summaries on $g$-and-$k$ distributions over $10$ repetitions with  top 1\% ABC samples selected. Most of the 95\% CIs have full coverage with the rest  having their coverage marked after the CI width. The bold fonts mark the best model in each metric.}
\vspace{-0.2in}
\end{table}

\section{Discriminator Calibration}\label{sec:dhat_training}
We  provide  more intuition for how  the choice of the discriminator, including the  $m/n$ ratio and the number of sets of latent variables needed to approximate the KL divergence,  affects our results. We   also discuss  possible modifications we make to the KL estimation  to eliminate bias. We use the M/G/1-queuing model as an example.

\begin{figure}[!ht]
\begin{center}
\includegraphics[width=0.75\textwidth]{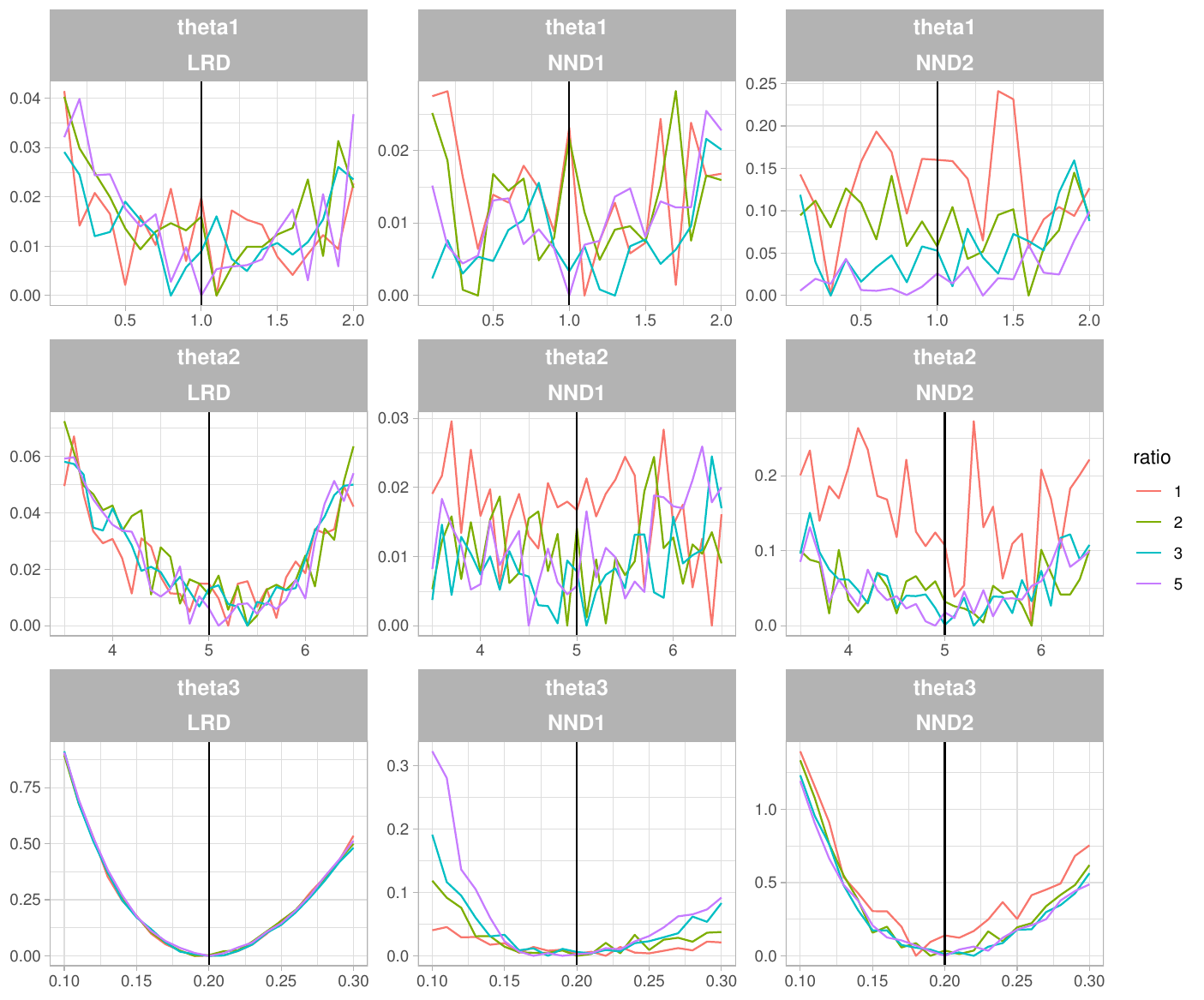}
\end{center}
\vspace{-0.2in}
\caption{Estimated KL divergence (rescaled) under different discriminators (nlatent=10). We plot the marginal changes in estimated KL divergence with respect to different parameters and different $m/n$ ratios here. The estimated values are rescaled by subtracting the minimum in each scenario so the changes are more comparable. }\label{fig:mg1_param1}
\vspace{-0.2in}
\end{figure}

First, we want to investigate how the choice of the discriminator and  the $m/n$ ratio could impact the estimation of the KL divergence. We consider three discriminators: (1)  logistic regression on degree-2 polynomial terms of $\X$ (LRD); (2) a neural network with  one layer of 10 ReLu nodes and two layers of 10 tanh nodes with input $\X$ (NND1); (3) a neural network with one layer of 10 ReLu nodes and one layer of 10 tanh nodes with input as the degree-2 polynomial terms of $\X$ (NND2). For each discriminator, we do a grid search over $m/n\in\{1,2,3,5\}$. We plot the estimated KL divergence (rescaled) in  \Cref{fig:mg1_param1}. From the plot we can see it is quite obvious that LRD best identifies the location of the true values, and the fluctuations/variance in its estimations are the smallest. It also worth noting that the difficulties in estimating different parameters are significantly different here. All discriminators are able to learn the location of $\theta_3$ with little variance, but only LRD is able to learn $\theta_1$ and $\theta_2$ well. The $m/n$ ratio does not seem to have significant impact for LRD, but we observe for NND1, NND2 and some other models that the variance in the estimation tends to get smaller when $m/n>1$ but not overly large  to avoid bias. We continue our experiments with LRD and $m/n=3$.

\begin{figure}[!ht]
\begin{center}
\includegraphics[width=0.75\textwidth]{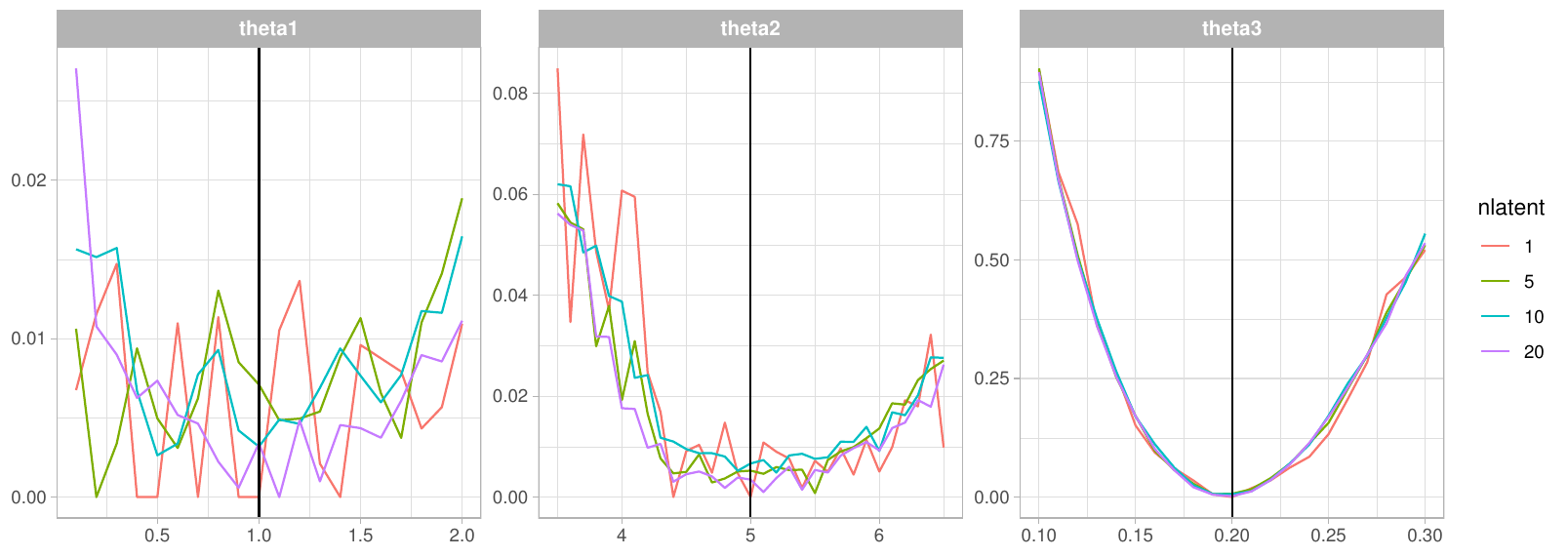}
\end{center}
\vspace{-0.2in}
\caption{Estimated KL divergence (rescaled) using LRD and $m/n=3$. We plot the marginal changes in estimated KL divergence with respect to different parameters and different number of sets of latent variables. The estimated values are rescaled by subtracting the minimum in each scenario so the changes are more comparable.}\label{fig:mg1_param2}
\vspace{-0.2in}
\end{figure}

Next, we want to know how many different sets of latent variables are needed to control the variance in estimation. We focus on the best discriminator we have previously found (i.e. LRD with $m/n=3$). We include the cases where $nlatent\in\{1,5,10\}$ here. The results are shown in  \Cref{fig:mg1_param2}. It is expected that with more sets of latent variables to average over, the variance in estimation gets smaller. The main challenge for the M/G/1 model lies in estimating $\theta_1$ correctly and stably. With only one or five sets of latent variable, it seems hopeless to identify $\theta_1$. To trade off variance and computation costs, we set $\text{nlatent}=10$ in our experiments.

\section{Computational Complexities}\label{sec:complex}

We list the computation complexity and the computation time (in hours) of the three simulation examples  in \Cref{tab:comp_cost}. The computation complexities are considered in the multivariate settings,  and the costs of ES, WA,  MM, kNN are adopted from \citet{jiang2018approximate, nguyen2020approximate}. For logistic regression, RF and DNN, the  cost depends on how fast the iterations actually converge and we provide only a rough approximation here. From the three  simulation examples, g-and-k examples and m/g/1-queuing examples has small $d$ but large $n$,  while the Lotka-Volterra example is large in $d$ but small in $n$.  We  see that  conventional  summary statistics calculation usually costs least  amount of time.  AL is also fast  but it becomes almost infeasible  when the dimension of $X$, i.e., $d$, is large and calculating the  density function of the multivariate normals has numerical issues. For the KL estimators, kNN is easiest to  compute and it has  the computation advantage when $d$ is small, but both its accuracy and complexity suffer  when $d$ is large. Though our classifier costs longer time on examples with small $d$, the computation costs are relatively similar for the LV example whose dimension $d$ is large.

\begin{table}[!ht]
\centering
\begin{tabular}{l c r r r}
\toprule
&\multirow{2}{*}{Complexity}& \multicolumn{3}{c}{Computation Time (in  hours)}\\
      &  & g-and-k & m/g/1  & LV  \\
      \midrule
kNN    &     $\mO ((n+m)\log(n+m))$       & 0.017   & 0.150  & 11.618 \\
Logistic   &   $\mO(n+m)$   & $\backslash$   & 12.052 & 12.214 \\
NN & $\mO((n+m)\times \# \rm{weight}\times \# \rm{epoch})$& 28.624& $\backslash$ &  $\backslash$ \\
RF    &    $\mO((n+m)\log(n+m))$        & 4.926   & 8.268  & 12.382 \\
\vspace{-0.1in}\\
WA    &   $\mO((n+m)^2)$         & 0.009   & 0.706  & 11.706 \\
MM    &  $\mO((n+m)^2)$           & 10.009  & 24.617 & 11.882 \\
ES    &   $\mO((n+m)^2)$         & 0.048   & 0.180  & 12.192 \\
SA/SS &            & 0.961   & 1.986  & 2.517  \\
DNN   &  $\mO(10^{6}\times 10^6 \times\abs{\Theta}\times \# \rm{epoch})$          & 1.968   & 2.378  & 34.556 \\
AL    &            & 0.009   & 0.120  &  $\backslash$      \\
\bottomrule
\end{tabular}
\vspace{-0.1in}
\caption{\label{tab:comp_cost}\footnotesize Computational Complexities/Runtime Comparisons}
\vspace{-0.2in}
\end{table}

\section{Proofs}
\subsection{Proof of \Cref{thm:est_error_conc}} \label{pf:est_error_conc}
Denote $K_n=\P_n\log\frac{p_0}{p_\theta}$. 
Using Chebyshev's inequality and \Cref{ass:ratio_bd} as
\begin{align*}
P_0^{(n)}[\abs{K_n-K(p_0, p_\theta)}>u]&=P_0^{(n)}\Big(\abs{(\P_n- P_0) \log \frac{p_0}{p_\theta} }>u \Big)  \leq \frac{1}{u^2}P_0^{(n)}\Big[\abs{(\P_n- P_0) \log \frac{p_0}{p_\theta}}^2\Big]\\
&=\frac{1}{u^2}P_0^{(n)}\Big[\abs{\P_n  \big(\log \frac{p_0}{p_\theta}-P_0\log\frac{p_0}{p_\theta}\big)}^2\Big] = \frac{1}{nu^2}P_0 \Big[\abs{  \log \frac{p_0}{p_\theta}-P_0\log\frac{p_0}{p_\theta}}^2\Big]  \\
&\leq \frac{16(2+\Lambda)h(p_0, p_\theta)}{nu^2},
\end{align*} 
where the last inequality follows from {Lemma 2.1} (iii) of \citet{kaji2021mh}.
Next, note
$$
\hat K(\X, \tilde \X^\theta)-K_n=- \P_n\Big(\log \frac{1-\hat D^\theta_{n,m}}{1-D_\theta}- \log\frac{\hat D^\theta_{n,m}}{D_\theta}\Big).
$$
{Recall the outer expectation $P^*$ in \Cref{lem:delta_n} and the set of classifiers $D^\theta_{n, \delta}$ defined in \Cref{ass:entropy},} we can bound 
\begin{align*}
& P\biggl(\abs{\P_n\Big(\log \frac{1-\hat D^\theta_{n,m}}{1-D_\theta}- \log\frac{\hat D^\theta_{n,m}}{D_\theta}\Big)}>u, d_\theta(\hat D^\theta_{n,m}, D_\theta)\leq C_n\delta_n\biggr) \\
&\qquad\qquad \leq P^*\Big( \sup_{D\in \mD^\theta_{C_n\delta_n}}\abs{\P_n\Big(\log \frac{1-D}{1-D_\theta}- \log\frac{D}{D_\theta}\Big)}>u \Big)\\
&\qquad \qquad \leq \frac{1}{u}\E^* \sup_{D\in \mD^\theta_{C_n\delta_n}} \abs{\P_n\Big(\log \frac{1-D}{1-D_\theta}- \log\frac{D}{D_\theta}\Big)}
\end{align*}
by Markov's inequality. The proof of {Theorem 4.1} of \citet{kaji2021mh} shows that the expectation is $O(C_n\delta_n)$.
Using the triangle inequality and the Bonferroni inequality, since $h(p_0, p_\theta)\leq 2$, we can then write
\begin{align*}
&P\biggl(\abs{\hat K(\X, \tilde \X^\theta)-K(p_0, p_\theta)} >2u, d_\theta(\hat D^\theta_{n,m}, D^\theta)\leq C_n\delta_n\biggr) \\
&\quad \leq P\biggl(\abs{\hat K(\X, \tilde \X^\theta)-K_n}+\abs{K_n-K(p_0, p_\theta)}>2u, d_\theta(\hat D^\theta_{n,m}, D_\theta)\leq C_n\delta_n \biggr)\\
&\quad  \leq P\biggl(\abs{\hat K(\X, \tilde \X^\theta)-K_n}>u,  d_\theta(\hat D^\theta_{n,m}, D_\theta)\leq C_n\delta_n \biggr) 
+ P_0^{(n)}[\abs{K_n-K(p_0, p_\theta)}>u]\\
&\quad  \leq O\Big(\frac{C_n\delta_u}{u}\Big)+\frac{32(2+\Lambda)}{nu^2}.
\end{align*}

 \subsection{Proof of  \Cref{thm:post_rate}}\label{pf:post_rate}

Throughout, we continue to assume that $\tilde \X^\theta=g_\theta(\tilde \X)$ and we denote with $P= P_0^{(n)}\otimes \tilde P^{(m)}$ the joint measure for $(\X,\tilde\X)$. Below, we will be using the notation $\Pi(\cdot)$ to denote the generic probability, i.e. for  $(\theta,\tilde\X)$ or for the conditional probability $\theta$ given $\tilde \X$. 
 Later, we will define a high-probability event $\Omega_n(C,\varepsilon_n)$ such that $P_0^{(n)}[\Omega_n(C,\varepsilon_n)^c]=o(1)$ for some $C\in(0,1)$.
Given  $\delta_n>0$ from our assumptions,  we can write for every $\lambda_n>0$ and $\epsilon_n>0$ and  for every arbitrarily slowly increasing sequence $C_n>0$
\begin{multline}\label{eq:conc_cond}
P_0^{(n)}\Pi\left(K(p_0, p_\theta)>\lambda_n\middle\vert \hat K(\X,\tilde\X^\theta) \leq \epsilon_n \right)\leq \Pi_1+o(1)+ \\
P_0^{(n)} \Pi\left(K(p_0, p_\theta)>\lambda_n\middle\vert \hat K(\X,\tilde\X^\theta) \leq \epsilon_n, d(\hat D^\theta_{n,m}, D_\theta) \leq C_n\delta_n\right)\1[\Omega_n(C,\varepsilon_n)],
\end{multline}
where, using Lemma \ref{lem:delta_n} and the fact that the rate $\delta_n$ is uniform (see Remark \ref{rmk:worst_rate}),  
$$
\Pi_1\equiv  P_0^{(n)} \Pi(d(\hat D^\theta_{n,m}, D_\theta)> C_n\delta_n)\leq \sup_{\theta\in\Theta}P(d(\hat D^\theta_{n,m}, D_\theta)> C_n\delta_n)=o(1).
$$
Consider the joint event, for some $\delta'>0$, 
 $$
 A_{\epsilon_n}(\delta')=\{(\tilde \X ,\theta):\hat K[\X,  g_\theta(\tilde \X)] \leq \epsilon_n\}\cap\{ K(p_0, p_\theta)>\delta'\}.
$$ 
For every $(\tilde \X, \theta)\in A_{\epsilon_n}(\delta')$ we have  
\begin{align*}
K(p_0,p_\theta)\leq\hat K(\X, g_\theta(\tilde \X)) +\abs{\hat K(\X, g_\theta(\tilde \X))-K(p_0, p_\theta)} \leq\epsilon_n+ \abs{\hat K(\X, g_\theta(\tilde \X))-K(p_0, p_\theta)}.
\end{align*}
Hence $(\tilde \X , \theta)\in A_{\epsilon_n}(\delta')$ implies that
\[
\abs{\hat K(\X, \tilde \X^\theta)-K(p_0, p_\theta)} > \delta'-\epsilon_n,
\]
and choosing $\delta'\geq \epsilon_n+t_\epsilon$ leads to
\[
\Pi [A_{\epsilon_n}(\delta')]\leq \int_{\Theta} \tilde P^{(m)} \left[ \abs{\hat K(\X,\tilde\X^\theta)-K(p_0, p_\theta)} >t_\epsilon \right] d\Pi(\theta).
\]
Using  \eqref{eq:conc_cond}, we now focus on the conditional probability, given $d(\hat D^\theta_{n,m}, D_\theta) \leq C_n\delta_n$, 
\begin{multline}\label{eq:conc_frac}
\Pi\left(K(p_0, p_\theta)>\epsilon_n+t_\epsilon \middle\vert \hat K(\X,\tilde\X^\theta) \leq \epsilon_n, d(\hat D^\theta_{n,m}, D_\theta) \leq C_n\delta_n \right) \\
\leq \frac{\int_\Theta \tilde P^{(m)} \left[\abs{\hat K(\X,\tilde\X^\theta)-K(p_0, p_\theta)} >t_\epsilon\mid  d(\hat D^\theta_{n,m}, D_\theta) \leq C_n\delta_n \right] d\Pi(\theta)}{\int_\Theta \tilde P^{(m)} \left[\hat K(\X,\tilde\X^\theta) \leq \epsilon_n \mid d(\hat D^\theta_{n,m}, D_\theta) \leq C_n\delta_n \right]d\Pi(\theta)}.
\end{multline}
We now find a lower bound for the denominator. 
Recall the KL neighborhood $B_2(p_0,\epsilon_n)$ defined in \eqref{eq:KL_ball}. Since
\[
\hat K(\X, \tilde \X^\theta) \leq K(p_0,p_\theta)+\abs{\hat K(\X, \tilde \X^\theta)- K(p_0, p_\theta)}\leq \epsilon_n/2+ K(p_0, p_\theta), 
\]
provided that $\abs{\hat K(\X, \tilde \X^\theta) -K(p_0, p_\theta)} \leq \epsilon_n/2$.  The denominator can be then bounded by
\begin{align*}
& \int_\Theta \tilde P^{(m)} [\hat K(\X, \tilde \X^\theta) \leq \epsilon_n \mid d(\hat D^\theta_{n,m}, D_\theta) \leq C_n\delta_n ]d\Pi(\theta)\\
&\qquad \geq \int_{B_2(p_0, \epsilon_n/{2})} \tilde P^{(m)} \left[\abs{\hat K(\X, \tilde \X^\theta)-K(p_0, p_\theta)} \leq   \epsilon_n/2 \mid d(\hat D^\theta_{n,m}, D_\theta) \leq C_n\delta_n \right] d\Pi(\theta) \\
& \qquad \geq \Pi[{B_2(p_0, \epsilon_n/{2})}  ] -\int_{B_2(p_0, \epsilon_n/{2})} \tilde P^{(m)} \left[\abs{\hat K(\X, \tilde \X^\theta)-K(p_0, p_\theta)} >   \epsilon_n/2 \mid d(\hat D^\theta_{n,m}, D_\theta) \leq C_n\delta_n \right] d\Pi(\theta).
\end{align*}
Denoting 
$$
Z(\X)\equiv \int_{B_2(p_0, \epsilon_n/{2})} \tilde P^{(m)} \left[\abs{\hat K(\X, \tilde \X^\theta)-K(p_0, p_\theta)} >   \epsilon_n/2 \mid d(\hat D^\theta_{n,m}, D_\theta) \leq C_n\delta_n \right] d\Pi(\theta)
$$ 
we can write, for every $C>0$, using Fubini's theorem and Markov's inequality
\begin{align*}
P_0^{(n)}\left(Z(\X)> C\right)&\leq \frac{1}{C} \int_{B_2(p_0, \epsilon_n/{2})} P \left[\abs{\hat K(\X, \tilde \X^\theta)-K(p_0, p_\theta)} >   \epsilon_n/2 \mid d(\hat D^\theta_{n,m}, D_\theta) \leq C_n\delta_n \right]d\Pi(\theta)\\
&=\frac{\Pi(B_2(p_0, \epsilon_n/{2}))}{C}\frac{\sup\limits_{\theta\in\Theta}  P \left[\abs{\hat K(\X, \tilde \X^\theta)-K(p_0, p_\theta)} >   \epsilon_n/2, d(\hat D^\theta_{n,m}, D_\theta) \leq C_n\delta_n \right]}{\sup\limits_{\theta\in\Theta}  P \left[ d(\hat D^\theta_{n,m}, D_\theta) \leq C_n\delta_n \right]}\\
&=\frac{\Pi(B_2(p_0, \epsilon_n/{2}))}{C(1+o(1))}\sup\limits_{\theta\in\Theta}\rho_{n,\theta}(\varepsilon_n/2;C_n;\delta_n),
\end{align*}
where we have used Theorem \ref{thm:est_error_conc} and the fact that $P(d(\hat D^\theta_{n,m}, D_\theta)\leq C_n\delta_n)=1+o(1)$ for every $\theta\in\Theta$ from Lemma \ref{lem:delta_n}.
We now define an event, for some $0<C<1$ and $\epsilon_n>0$,
$$
\Omega_n(C,\epsilon_n)=\left\{\X: Z(\X)  \leq C  \times  \Pi(B_2(p_0, \epsilon_n/{2}))\right\}.
$$
Using Theorem \ref{thm:est_error_conc}, we have 
$$
\rho_{n,\theta}(\epsilon_n/2;C_n;\delta_n)=O\Big(\frac{C_n\delta_n}{\epsilon_n} +\frac{1}{n\epsilon_n^2}\Big)\quad\text{  for every}\quad \theta\in\Theta.
$$
Choosing $\epsilon_n>0$ such that  $\epsilon_n=o(1)$ and $n\epsilon_n^2\rightarrow\infty$ and  $C_n\delta_n=o(\epsilon_n)$ we have $P_0^{(n)}[\Omega_n(C,\epsilon_n)^c]=o(1)$ for every $C\in(0,1)$.
On the event  $\Omega_n(C,\epsilon_n)$, for some $0<C<1$, we can lower-bound the denominator with 
$$
 \int_\Theta \tilde P^{(m)} [\hat K(\X, \tilde \X^\theta) \leq \epsilon_n \mid d(\hat D^\theta_{n,m}, D_\theta) \leq C_n\delta_n ]d\Pi(\theta)> (1-C)\times  \Pi(B_2(p_0, \epsilon_n/{2}).
$$
Using this bound and applying  Fubini's theorem, we  can further write
\begin{multline}\label{eq:frac_exp}
 P_0^{(n)} \Pi\left(K(p_0, p_\theta)>\epsilon_n+t_\epsilon \mid \hat K(\X,\tilde\X^\theta) \leq \epsilon_n, d(\hat D^\theta_{n,m}, D_\theta) \leq  C_n\delta_n \right)\1[\Omega_n(C,\varepsilon_n)] \\
\leq  \frac{\int_\Theta P\left[\abs{\hat K(\X,\tilde\X^\theta)-K(p_0, p_\theta)} >t_\epsilon \mid d(\hat D^\theta_{n,m}, D_\theta) \leq C_n\delta_n \right] d\Pi(\theta)}{(1-C)\times  \Pi(B_2(p_0, \epsilon_n/{2})}  
\end{multline}
Using  Theorem \ref{thm:est_error_conc} again,   we obtain an upper bound for the display above with
$$
  \frac{ \rho_n(t_\epsilon;C_n;\delta_n)}{(1-C)\times  \Pi(B_2(p_0, \epsilon_n/{2})(1+o(1))}.
$$
Using the prior Assumption \ref{ass:prior_mass}, we can choose $t_\epsilon$ such that  
$$
\Big(\frac{C_n\delta_n}{t_\epsilon} +\frac{1}{nt_\epsilon^2}\Big)/\epsilon_n^{\kappa}=1/M_n
$$
 for some arbitrarily slowly increasing sequence $M_n>0$.
We choose $t_\epsilon=M_n C_n\delta_n/\epsilon_n^{\kappa}+\sqrt M_n n^{-1/2}/\epsilon_n^{\kappa/2}$. Since $\delta_n\gtrsim n^{-1/2}$ and $\epsilon_n^{-\kappa}\geq \epsilon_n^{-\kappa/2}$, the overall rate is then driven by 
$\epsilon_n+t_\epsilon=\epsilon_n+\wt M_n\delta_n \epsilon_n^{-\kappa}$, where $\wt M_n=M_nC_n$.

\subsection{Proof of \Cref{thm:unif_ci}}\label{sec:proof_unif_ci}
Recall from \Cref{thm:post_rate} that with the accept-reject strategy, the true KL divergence $K(p_0, p_\theta)$ is contracting at the rate $\lambda_n=\epsilon_n+ \wt M_n \delta_n \epsilon_n^{-\kappa}$, where $\wt M_n$ is a slowly increasing sequence that diverges faster than $C_n$.
Consider the case when $\epsilon_n \gg \wt M_n \delta_n \epsilon_n^{-\kappa}$ or, equivalently, $\epsilon_n \gg \delta_n^{1/(\kappa+1)}$. Denote
 \[
   x(\theta)=\epsilon_n^{-1}K(p_0,p_\theta)\quad\text{ and}\quad f_n(\theta-\theta_0)=f\big(\epsilon_n^{-1/2}(\theta-\theta_0)\big).
 \]

 We express the ABC posterior expectation of $f_n (\theta-\theta_0)$ for  a non-negative and bounded function $f_n(\cdot)$ by
\begin{align*}
& P_0^{(n)}E_{\hat \Pi^{AR}_{\epsilon_n}} \Big[f_n(\theta-\theta_0)\Big] =P_0^{(n)} \int f_n(\theta-\theta_0) \d \hat \Pi^{AR}_{\epsilon_n}(\theta\mid \X)\\
&\quad =\underbrace{P_0^{(n)}  \int f_n(\theta-\theta_0) \1[K(p_0,p_\theta)\leq \lambda_n, d(\hat D^\theta_{n,m}, D_\theta )\leq C_n\delta_n)] \d \hat \Pi^{AR}_{\epsilon_n}(\theta\mid \X)}_{(\text{I})}\\
&\qquad  + \underbrace{P_0^{(n)}  \int f_n(\theta-\theta_0) \1[K(p_0,p_\theta)\leq \lambda_n, d(\hat D^\theta_{n,m}, D_\theta )> C_n\delta_n)] \d \hat \Pi^{AR}_{\epsilon_n}(\theta\mid \X)}_{(\text{II})}  \\
& \qquad +
\underbrace{P_0^{(n)}  \int f_n(\theta-\theta_0) \1[K(p_0,p_\theta)> \lambda_n] \d \hat \Pi^{AR}_{\epsilon_n}(\theta\mid \X)}_{(\text{III})}
\end{align*}
where the term $(\text{III})$ can be controlled using Fubini's theorem and the concentration result in \Cref{thm:post_rate} as follows
\begin{align*}
(\text{III})& =   \int f_n(\theta-\theta_0)  P_0^{(n)}  \1[K(p_0,p_\theta)> \lambda_n] \d\hat \Pi^{AR}_{\epsilon_n}(\theta\mid \X)\\
& \leq \norm{f}_\infty P_0^{(n)}\Pi\Big(K(p_0, p_\theta)>\lambda_n\mid \hat K(\X, \tilde \X^\theta)\leq \epsilon_n\Big)=o(1).
\end{align*}
The second term $(\text{II})$ can be  bounded similarly as
\begin{align*}
(\text{II})&= P_0^{(n)} \int_{K(p_0, p_\theta)\leq \lambda_n } f_n(\theta-\theta_0) \1 [d(\hat D^\theta_{n,m}, D_\theta )> C_n\delta_n)]\d  \hat \Pi^{AR}_{\epsilon_n}(\theta\mid \X) \\
& \leq \norm{f}_\infty \int_{K(p_0, p_\theta)\leq \lambda_n}P(d(\hat D^\theta_{n,m}, D_\theta )> C_n\delta_n)\d\Pi(\theta) \\
&\leq  \norm{f}_\infty   \sup_{\theta \in \Theta} P(d(\hat D^\theta_{n,m}, D_\theta )> C_n\delta_n) = o(1)
\end{align*}
where we use the fact that $\sup_{\theta\in \Theta} P(d(\hat D^\theta_{n,m}, D_\theta )> C_n\delta_n) = o(1)$ from \Cref{lem:delta_n}.

Thus, the asymptotic behavior is mainly determined by  the term $(\text{I})$. Combined with the continuity of $\pi(\theta)$ at $\theta_0$   we can re-write $(\text{I})$  as
\begin{align*}
(\text{I})&= P_0^{(n)} \frac{\int_{K(p_0, p_\theta)\leq \lambda_n} \pi(\theta) f_n(\theta-\theta_0)\tilde P^{(m)}\big[\hat K(\X, \tilde \X^\theta) \leq \epsilon_n , d(\hat D^\theta_{n,m}, D_\theta )\leq C_n\delta_n  \big]\d\theta}{\int_{K(p_0, p_\theta)\leq \lambda_n}\pi(\theta) \tilde P^{(m)}\big[\hat K(\X, \tilde \X^\theta) \leq \epsilon_n,d(\hat D^\theta_{n,m}, D_\theta )\leq C_n\delta_n  \big]\d\theta}\\
&= P_0^{(n)} \frac{\int_{K(p_0, p_\theta)\leq \lambda_n}  f_n(\theta-\theta_0)\tilde P^{(m)}\big[\hat K(\X, \tilde \X^\theta) \leq \epsilon_n , d(\hat D^\theta_{n,m}, D_\theta )\leq C_n\delta_n  \big]\d\theta}{\int_{K(p_0, p_\theta)\leq \lambda_n}\tilde P^{(m)}\big[\hat K(\X, \tilde \X^\theta) \leq \epsilon_n,d(\hat D^\theta_{n,m}, D_\theta )\leq C_n\delta_n  \big]\d\theta} (1+o(1))
\end{align*}

We have $x(\theta)\geq 0$ for all $\theta\in \Theta$ and since
$$
\hat K(\X, \tilde \X^\theta)=\hat K(\X, \tilde \X^\theta)-K(p_0,p_\theta)+\epsilon_n x(\theta),
$$ 
 we can write 
\begin{multline}\label{eq:N_khat}
\tilde P^{(m)} \big[\hat K(\X, \tilde \X^\theta )\leq \epsilon_n, d(\hat D^\theta_{n,m}, D_\theta )\leq C_n\delta_n \big]\\
 \geq \tilde P^{(m)} \left[\abs{\hat K(\X, \tilde \X^\theta)-K(p_0,p_\theta)}\leq \epsilon_n(1-x(\theta)), d(\hat D^\theta_{n,m}, D_\theta )\leq C_n\delta_n  \right]\\
 = 1- \tilde P^{(m)} \left[\abs{\hat K(\X, \tilde \X^\theta)-K(p_0,p_\theta)}> \epsilon_n(1-x(\theta)), d(\hat D^\theta_{n,m}, D_\theta )\leq C_n\delta_n  \right].
\end{multline}
Denoting with
\[
\tilde Z(\X )=\int_{x(\theta)\leq 1- \frac{M_nC_n\delta_n}{\epsilon_n}} \tilde P^{(m)} \left[\abs{\hat K(\X, \tilde \X^\theta)-K(p_0,p_\theta)}>\epsilon_n\big(1-x(\theta)\big), d(\hat D^\theta_{n,m}, D_\theta )\leq C_n\delta_n \right] \d \theta.
\]
Using Markov's inequality and Fubini's theorem we have, for every $\tilde C>0$,
\begin{align*}
&P_0^{(n)}(\tilde Z(\X)>\tilde C)\\
&\quad \leq \frac{1}{\tilde C}\int\limits_{x(\theta)\leq 1- \frac{M_nC_n\delta_n}{\epsilon_n}} P \left[\abs{\hat K(\X, \tilde \X^\theta)-K(p_0,p_\theta)}>\epsilon_n\big(1-x(\theta)\big), d(\hat D^\theta_{n,m}, D_\theta )\leq C_n\delta_n \right] \d\theta \\
&\quad \leq \frac{1}{\tilde C} \int_{x(\theta)\leq 1- \frac{M_nC_n\delta_n}{\epsilon_n}} \rho_n\big(\epsilon_n(1-x(\theta));C_n;\delta_n\big) \d\theta \\
&\quad \leq  \frac{1}{\tilde C} \times\int_{x(\theta)\leq 1- \frac{M_nC_n\delta_n}{\epsilon_n}} \d\theta\times  \sup_{x(\theta)\leq 1- \frac{M_nC_n\delta_n}{\epsilon_n} } \rho_n\big(\epsilon_n(1-x(\theta));C_n;\delta_n\big)  \\
&\quad \leq  \frac{\int_{x(\theta)\leq 1- \frac{M_nC_n\delta_n}{\epsilon_n}} \d\theta}{\tilde C} \times \rho_n(M_nC_n\delta_n; C_n; \delta_n) \\
&\quad= \frac{\int_{x(\theta)\leq 1- \frac{M_nC_n\delta_n}{\epsilon_n}} \d\theta}{\tilde C}\times O\left(\frac{1}{M_n}+\frac{1}{n M_n^2C_n^2\delta_n^2}\right). \end{align*}
where $\rho_n(\cdot; C_n; \delta_n)$ is defined in \Cref{thm:est_error_conc}. 
Thus, we can define a set $\tilde\Omega_{n}(\tilde C_n)$, for some
for some arbitrarily slowly increasing sequence $\tilde C_n>0$, and $\tilde C_n =O(1/M_n)$, as
\[
\tilde\Omega_{n}(\tilde C_n)=\left\{\X:  \tilde Z(\X)\leq \tilde C_n\times \int_{x(\theta)\leq 1- \frac{M_nC_n\delta_n}{\epsilon_n}} \d\theta \right\}.
\] 
Then we have that $P_0^{(n)}(\tilde \Omega_n(\tilde C_n)^c)=o(1)$. Recall the inequality in \eqref{eq:N_khat}, on $\tilde \Omega_n(\tilde C_n)$, we have that 
\begin{multline*}
\int_{x(\theta)\leq 1- \frac{M_nC_n\delta_n}{\epsilon_n}} \tilde P^{(m)} \big[\hat K(\X, \tilde \X^\theta )\leq \epsilon_n, d(\hat D^\theta_{n,m}, D_\theta )\leq C_n\delta_n \big] \d\theta \\
 \geq \int_{x(\theta)\leq 1- \frac{M_nC_n\delta_n}{\epsilon_n}}  \d\theta - \tilde Z(\X)=  \int_{x(\theta)\leq 1- \frac{M_nC_n\delta_n}{\epsilon_n}}  \d\theta \times (1-o(1))
\end{multline*}
And this quantity is upper-bounded by $\int_{x(\theta)\leq 1- \frac{M_nC_n\delta_n}{\epsilon_n}}  \d\theta$. Therefore, we can conclude that 
\[
\int_{x(\theta)\leq 1- \frac{M_nC_n\delta_n}{\epsilon_n}} \tilde P^{(m)} \big[\hat K(\X, \tilde \X^\theta )\leq \epsilon_n, d(\hat D^\theta_{n,m}, D_\theta )\leq C_n\delta_n \big] \d\theta = \int_{x(\theta)\leq 1- \frac{M_nC_n\delta_n}{\epsilon_n}}  \d\theta \big(1+o(1)\big).
\]

Note that $x(\theta)\leq 1- M_nC_n\delta_n/\epsilon_n$ implies that 
$K(p_0, p_\theta)\leq \epsilon_n -M_nC_n\delta_n$. On this set $\tilde \Omega_n(\tilde C_n)$, we can further lower bound the denominator as
\begin{align*}
&\int_{K(p_0, p_\theta)\leq \lambda_n} \tilde P^{(m)}\big[\hat K(\X, \tilde \X^\theta) \leq \epsilon_n, d(\hat D^\theta_{n,m}, D_\theta )\leq C_n\delta_n   \big]\d\theta \\
&\qquad = \underbrace{\int_{K(p_0, p_\theta)\leq \epsilon_n-M_nC_n\delta_n}  \d \theta}_{D_1}(1+o(1))\\
&\qquad \quad +\underbrace{\int_{\epsilon_n-M_nC_n\delta_n<K(p_0, p_\theta)\leq \lambda_n}\tilde P^{(m)}\big[\hat K(\X, \tilde \X^\theta) \leq \epsilon_n,  d(\hat D^\theta_{n,m}, D_\theta )\leq C_n\delta_n    \big]\d\theta}_{D_2}.
\end{align*}

Next, we show that the second term $D_2$  is $o(D_1)$. Let $u=\epsilon_n^{-1/2}(\theta-\theta_0)$. Under Assumption \ref{ass:KL_taylor}, we have 
$K(p_0,p_\theta)=\frac{1}{2} (\theta-\theta_0)'I(\theta_0) (\theta-\theta_0)\{1+o(1)\}$, which is $x(\theta_0+\sqrt \epsilon_n u)=\frac{1}{2}u'I(\theta_0)u$.  Since $I(\theta_0)$ is positive definite we can write 
\begin{align}\label{eq:D1D2}
\frac{D_2}{D_1}& \leq \frac{\int_{1-M_nC_n\delta_n/\epsilon_n<x(\theta)\leq 1+\wt M_n\delta_n/\epsilon_n^{\kappa+1} } \d \theta}{\int_{x(\theta)\leq 1-M_nC_n\delta_n/\epsilon_n} \d \theta}\\
&\leq  \frac{\int_{2(1-M_nC_n\delta_n/\epsilon_n)\leq u'I(\theta_0)u
\leq 2(1+\wt M_n\delta_n/\epsilon_n^{\kappa+1})}\d\theta}{\int_{u'I(\theta_0)u \leq 2(1-M_nC_n\delta_n/\epsilon_n)}\d\theta}
\lesssim \frac{M_nC_n\delta_n}{\epsilon_n}+\frac{\wt M_n\delta_n}{\epsilon_n^{\kappa+1}} = o(1).
\end{align}
{where the approximation follows from the fact that $\epsilon_n \gg \delta_n^{1/(\kappa+1)}$ and because the denominator is approximating the integral $\int_{u'I(\theta_0)u \leq 2}\d\theta$ and the numerator is the length of shrinking intervals.}
Combining the above results, we find that, on the set $\tilde \Omega_n(\tilde C_n)$, the denominator can be lower-bounded by
$
\int_{x(\theta)\leq 1-\frac{M_nC_n\delta_n}{\epsilon_n}}  \d \theta \{1+o(1)\}.
$

This implies 

\begin{align}
(\text{I})&= \frac{\int_{K(p_0, p_\theta)\leq \lambda_n}  f_n(\theta-\theta_0) P\big[\hat K(\X, \tilde \X^\theta) \leq \epsilon_n , d(\hat D^\theta_{n,m}, D_\theta )\leq C_n\delta_n  \big]\d\theta}{(1+o(1))\int_{x(\theta)\leq 1-\frac{M_nC_n\delta_n}{\epsilon_n}}  \d \theta} (1+o(1))+ o(1)\nonumber\\
&=(N_1+N_2)\{1+o(1)\}+o(1).
\end{align}
with
\begin{equation}
N_1\equiv{\frac{\int\limits_{x(\theta)\leq 1-\frac{M_nC_n\delta_n}{\epsilon_n}} f\Big(\epsilon_n^{-1/2}(\theta-\theta_0)\Big) \d \theta}{\int\limits_{x(\theta)\leq 1-\frac{M_nC_n\delta_n}{\epsilon_n}}  \d \theta} }
\end{equation}
and
\begin{equation}
N_2\equiv\frac{\int\limits_{K(p_0, p_\theta)\leq \lambda_n} \1\left[{x(\theta)> 1-\frac{M_nC_n\delta_n}{\epsilon_n}}\right] f\Big(\epsilon_n^{-1/2}(\theta-\theta_0)\Big)   P \Big[\hat K(\X, \tilde \X^\theta) \leq \epsilon_n, d(\hat D^\theta_{n,m}, D_\theta )\leq C_n\delta_n\Big] \d \theta}{\int\limits_{x(\theta)\leq 1-\frac{M_nC_n\delta_n}{\epsilon_n}}  \d \theta}\label{eq:N2},
\end{equation}
where the second equality follows from the fact that $x(\theta)\leq 1-\frac{M_nC_n\delta_n}{\epsilon_n}$ leads to $K(p_0, p_\theta)\leq \epsilon_n -M_nC_n\delta_n$ and then $K(p_0, p_\theta)\leq \lambda_n$ is trivially satisfied and where the last $o(1)$ comes from  the set $\tilde \Omega_n(\tilde C_n)^c$.

Since we have $\epsilon_n\gg \delta_n$, with $u=\epsilon_n^{-1/2}(\theta-\theta_0)$, the first term is approximately equal to
\[
N_1=\frac{\int_{K(p_0,p_{\theta_0+\sqrt{\epsilon_n} u})\leq \epsilon_n} f(u) \d  u}{ \int_{K(p_0,p_{\theta_0+\sqrt{\epsilon_n} u})\leq \epsilon_n}   \d u}\left(1+o(1)\right).
\]
Using  Assumption \ref{ass:KL_taylor} again, we have
\[
\int_{K(p_0, p_{\theta_0+\sqrt{\epsilon_n} u})\leq \epsilon_n} \d u = \int_{\frac{1}{2}\sqrt{\epsilon_n} u' I(\theta)\sqrt{\epsilon_n} u \leq \epsilon_n} \d u+o(1)=\int_{u'I(\theta_0)u\leq 2}\d u+o(1).
\]
This leads to
\[
N_1=\frac{\int_{u'I(\theta_0)u \leq 2}f(u) \d u}{\int_{u'I(\theta_0)u\leq 2} \d u } \left(1+o(1)\right).
\]
Next, we show that the fraction $N_2$ converges to 0. The numerator can be simplified as an integral over $1-\frac{M_nC_n\delta_n}{\epsilon_n}<x(\theta)\leq 1+\frac{\wt M_n\delta_n}{\epsilon_n^{\kappa+1}}$, which can be  bounded by \eqref{eq:D1D2} as
\begin{align*}\label{eq:N2}
N_2&\leq \frac{\norm{f}_\infty \int_{1-\frac{M_nC_n\delta_n}{\epsilon_n}<x(\theta)\leq 1+\frac{\wt M_n\delta_n}{\epsilon_n^{\kappa+1}}} \d \theta }{
\int_{x(\theta)\leq 1-\frac{M_nC_n\delta_n}{\epsilon_n}} \d \theta}\\
& \leq \norm{f}_\infty \frac{D_2}{D_1}=o(1).
\end{align*}

Since we have $\epsilon_n^{\kappa+1}\gg \delta_n$, putting all the terms together, we obtain that the $P_0^{(n)}$-averaged ABC posterior distribution of $\epsilon_n^{-1/2}(\theta-\theta_0)$ is asymptotically uniform over the ellipsoid $\{u: u'I(\theta_0)u\leq 2\}$.

\end{document}